%% file: BFKLsub2.tex
\documentclass[11pt, reqno,preprint]{article}
\pdfoutput=1

\usepackage{jheppub}
\usepackage{epsfig}
\usepackage{amssymb}
\usepackage{amsmath}
\usepackage{mathrsfs}
\usepackage{hyperref}
\usepackage{multirow}
\usepackage{wasysym}

\definecolor{applegreen}{rgb}{0, 0.71, 0.0}

\def\be{\begin{equation}}
\def\ee{\end{equation}}
\def\ba{\begin{eqnarray}}
\def\ea{\end{eqnarray}}

\def\a{\alpha}

\def\MM{\mathcal{M}}

\def\Fdown{F^{\downarrow}}
\def\Fdisc{F^{\updownarrow}}

\def\OO{O}

\newcommand{\lsim}{\mathrel{\hbox{\rlap{\lower.55ex \hbox{$\sim$}} \kern-.3em \raise.4ex \hbox{$<$}}}}
\newcommand{\gsim}{\mathrel{\hbox{\rlap{\lower.55ex \hbox{$\sim$}} \kern-.3em \raise.4ex \hbox{$>$}}}}

\def\Gammacusp{\Gamma_{\rm cusp}}

\def\sister#1{\check{#1}}

\def\({\left(}
\def\){\right)}
\def\[{\left[}
\def\]{\right]}

\def\<{\langle}
\def\>{\rangle}

\newcommand{\beq}{\begin{equation}}
\newcommand{\eeq}{\end{equation}}
\newcommand{\beqq}{\begin{equation*}}
\newcommand{\eeqq}{\end{equation*}}
\newcommand\beqa{\begin{eqnarray}}
\newcommand\eeqa{\end{eqnarray}}
\newcommand\beqaa{\begin{eqnarray*}}
\newcommand\eeqaa{\end{eqnarray*}}
\newcommand\bea{\begin{array}}
\newcommand\eea{\end{array}}
\newcommand{\la}[1]{\label{#1}} 
\newcommand{\Blue}[1]{{\color{blue}#1\color{black}}}
\newcommand{\Red}[1]{{\color{red}#1\color{black}}}

\usepackage{color}
\definecolor{darkblue}{cmyk}{0.9,0.9,0,0}
\newcommand{\DBlue}[1]{{\color{darkblue}#1\color{black}}}

\title{\centering
Adjoint BFKL at finite coupling:\\a short-cut from the collinear limit}

\author{${}^{a}$Benjamin Basso, ${}^{b,c}$Simon Caron-Huot and ${}^c$Amit Sever}
\affiliation[a]{Laboratoire de Physique Th\'eorique, \'Ecole Normale Sup\'erieure, Paris 75005, France}
\affiliation[b]{Niels Bohr International Academy and Discovery Center, Blegdamsvej 17, 
Copenhagen 2100, Denmark}
\affiliation[c]{School of Natural Sciences, Institute for Advanced Study, Princeton, NJ 08540, USA}
\emailAdd{basso@lpt.ens.fr, schuot@nbi.dk, asever@ias.edu}

\bigskip\bigskip\bigskip\bigskip\bigskip

\abstract{In the high energy Regge limit, the six gluons scattering amplitude is controlled by the adjoint BFKL eigenvalue and impact factor. In this paper we determine these two building blocks at any value of the 't Hooft coupling in planar ${\cal N}=4$ SYM theory. This is achieved by means of analytic continuations from the collinear limit, where similar all loops expressions were recently established. We check our predictions against all available data at weak and strong coupling.
}

\notoc

\begin{document}
\maketitle

\bigskip\bigskip\bigskip\bigskip\bigskip\bigskip\bigskip\bigskip\bigskip\bigskip\bigskip\bigskip\bigskip\bigskip\bigskip\bigskip\bigskip\bigskip\bigskip
\newpage
\tableofcontents
\newpage

\section{Introduction}

A large part of our understanding of quantum field theory comes from perturbative expansions around weakly coupled limits.
This is especially true for complicated objects such as scattering amplitudes, which depend on many kinematic variables.
There is thus a general interest in methods which remain applicable exactly at finite values of the coupling.
A prominent example is the Renormalization Group.  It allows to make exact statement about the behavior of the amplitudes in certain regimes,
effectively resumming infinitely many terms in the perturbative series.



These sorts of resummations provide an excellent way to get a qualitative and quantitative grasp of the amplitudes at finite coupling.
They have been worked out recently to an exquisite level of detail in the planar ${\cal N}=4$ SYM theory, partly due to integrability in this theory~\cite{Beisert:2010jr}. The hexagon amplitude in this {theory}, for example, admits the following expansion~\cite{Alday:2010ku}  in the collinear (a.k.a.~OPE) limit {(large $\tau$)}:
\be
\mathcal{W}_{\rm hex}
 =1+ 2\sum_{\ell\geq 1} {(-1)^\ell} \cos(\ell\phi) \int\limits_{-\infty}^{+\infty} \frac{dp}{2\pi}\, \hat\mu_\ell (p)\, e^{ip\sigma-\tau E_\ell(p)} + \ldots\, ,
 \label{Collinear}
\ee
where ${\cal W}_\text{hex}$ is a regularized conformal invariant amplitude \cite{Basso:2013vsa}, $\{\tau,\sigma,\phi\}$ represent the conformal invariant data encoded in the external momenta and the dots stand for multi-particle excitations.
The form of eq.~(\ref{Collinear}) is largely controlled by symmetries,
but what makes this formula especially remarkable
is that the functions $E_m(p)$ and $\hat\mu_m(p)$ are now known exactly at all values of the `t Hooft coupling $\lambda$ \cite{Basso:2013vsa}.

Integrability made arguably its first appearance in four-dimensional quantum field theory through the Regge limit, where
the relevant BFKL evolution kernel \cite{Kuraev:1977fs,Balitsky:1978ic} was found to possess hidden symmetries \cite{Lipatov:1993yb,Faddeev:1994zg,Lipatov:2009nt}.
In this limit, the preceding hexagon amplitude admits an expansion of the form \cite{Lipatov:2010qf,Lipatov:2010ad,Bartels:2010tx}:
\be
 \mathcal{W}_{\rm hex}^{\,\circlearrowleft} e^{-i\pi \delta} =
 -2\pi i\sum_{m=-\infty}^\infty (-1)^m e^{im\phi} \int\limits_{-\infty}^{+\infty} \frac{d\nu}{2\pi} \,\hat\mu_{\!\,_\text{BFKL}}(\nu,m)\, e^{i(\sigma-\tau)\nu+(\sigma+\tau)\omega(\nu,m)}
 +\ldots \, ,
 \label{BFKL}
\ee
when written in terms of our variables. Here $\circlearrowleft$ stands for the relative analytic continuation from the kinematic regime of (\ref{Collinear}), $\delta$ is a known phase (see eq.~(\ref{BFKL_2a}))
and in the Regge limit $\sigma$ and $\tau$ are both large with a finite difference.

In spite of its historical significance, integrability of the BFKL Hamiltonian has not yet been successfully integrated into the all-loop $\mathcal{N}=4$ story,
and the building blocks of eq.~(\ref{BFKL}) are much less well-understood than their collinear counterparts (\ref{Collinear}).
The main aim of this paper will be to bridge this gap for this particular observable and obtain expressions for $\omega(\nu,m)$ and $\hat\mu_{\!\,_\text{BFKL}}(\nu,m)$
that are valid at all values of the coupling, extending the three-loop perturbative results of refs.~\cite{Bartels:2008ce,Bartels:2008sc,Fadin:2011we,Dixon:2012yy,Dixon:2014voa}
and the strong coupling results of refs.~\cite{Bartels:2013dja,Bartels:2010ej}.

The two preceding expansions resum completely different physics. As one approaches the Regge limit,
an infinite number of higher-twist contributions to the collinear expansion (\ref{Collinear}) become enhanced, and need to be resumed.  Conversely, an infinite number of terms in the
BFKL expansion (\ref{BFKL}) would be required to account for just the leading term of the collinear expansion.

Nevertheless, as we will demonstrate in this paper, following earlier suggestions from refs.~\cite{Bartels:2011xy,Hatsuda:2014oza},
the leading term in the collinear expansion \emph{completely} determines the leading term in the BFKL expansion!
Since the functions in the collinear expansion are known exactly, we will be able to determine the color-adjoint BFKL eigenvalue and impact factors exactly at all values of the coupling.

There is a rich history of interplay between collinear and high-energy expansions.
The BFKL equation was used early on to predict the behavior of the DGLAP  kernel
at small $x$ \cite{Jaroszewicz:1982gr,Catani:1989sg,Ball:1995vc}.  Conversely, the DGLAP equations yields an infinite number of predictions
regarding the collinear limit of the BFKL Hamiltonian, which turn out to be important to stabilize its perturbative expansion (see for instance \cite{Salam:1998tj,Vera:2005jt}).
These connections have been extended to higher orders in perturbation theory, strong coupling,
and to structure functions in conformal field theory \cite{Kotikov:2007cy,Brower:2006ea,Hatta:2007he,Cornalba:2008qf,Costa:2012cb}.
Generally, one uses one expansion to predict the leading logarithmic terms in the other expansion in a certain limit.

In this paper we will carry out, for the first time, this matching procedure exactly at all values of the coupling.

Conceptually, the key idea exploits the existence of a physical region 
in which both
the expansions (\ref{Collinear}) and (\ref{BFKL}) can be simultaneously truncated to the leading term, allowing
the relevant analytic functions to be measured to arbitrary accuracy.
Even though this is a rather restricted region, their agreement in an open set
will be sufficient to fully reconstruct one from the other.
This reconstruction, via analytic continuation, is fundamentally nonperturbative
and we will make full use of the finite-coupling expressions which are available on the OPE side.

The paper is organized as follows. In section \ref{sec:overview} we briefly review the
concept of Reggeization and its realization in gauge theory.
We introduce the concept of ``sister'' trajectory, whose
contribution to physical scattering amplitudes cancels in all but specific kinematic regions.
In section \ref{analyticcontinuationsec} we discuss the OPE expansion of the 6-particle amplitude,
and we demonstrate that in a certain Lorentzian region it is governed not by the usual operators,
but by ``sister'' ones related in a specific way.
In section \ref{goingthroughcutsec} we analyze the ``sister'' contribution in the high-energy limit and
perform the necessary analytic continuation which converts it to
finite-coupling expressions for the BFKL kernel. In section \ref{strongC:sec} we analyze the strong coupling limit of our finite coupling result. We construct a new semi-classical string solution and make contact
with existing literature. Finally, in section \ref{discussion}, we
elaborate on various important subtleties and in section \ref{Conclusion} we suggest directions for future work. Four appendices support the main text.

\section{Reggeization: a brief overview}\label{sec:overview}

High-energy forward scattering is a way to measure the transverse structure of a target.
This structure exhibits some energy dependence, which also tells us about the excitations of the theory.
This dependence is captured by the action of a Lorentz boost between the projectile and target.
As the classic work of Regge has demonstrated, the spectrum of the boost operator is intimately linked,
by a process of analytic continuation away from integer spin, to that of spatial rotations.
This has led to a rather successful phenomenology of the strong interactions wherein hadronic bound states are
organized into ``trajectories'' $\alpha(t)$, expressing their spin as a function of energy squared.
The trajectories $\alpha(t)$ are then analytically continued to the spacelike region and compared with
the energy dependence of various high-energy processes~\cite{Donnachie:2002en}.
As we will see, the continuation between the collinear and high-energy expansions is of precisely this type.

To illustrate {one important} subtlety we will {need to deal with}, it will be helpful to first discuss a {comparatively} simple example
involving open strings in flat space-time.
This example dates from the days where this subject was called
dual resonance models. As was the case then, {we find that} the dual resonance model usefully
illustrates certain features of gauge theory.

The example is $2\to 4$ scattering of open strings at tree level. We consider the high-energy limit,
where three particles ($2, 3$ and 4) form a cluster
of fast-moving particles, separated by a large rapidity gap from the other three (see figure \ref{fig:foldedstring}-a below).
Because of the Regge factorization built into string theory, one might {naively} expect an energy dependence of the type
\be
 A_6 \propto s^{\alpha(t_{234})}, \quad\mbox{with}\quad \alpha(t) = 1+\alpha' t\,.  \label{A6simple}
\ee
The exponent is the leading open-string Regge trajectory, exchanged between the two clusters.
\begin{figure}
\centering
\def\svgwidth{15cm}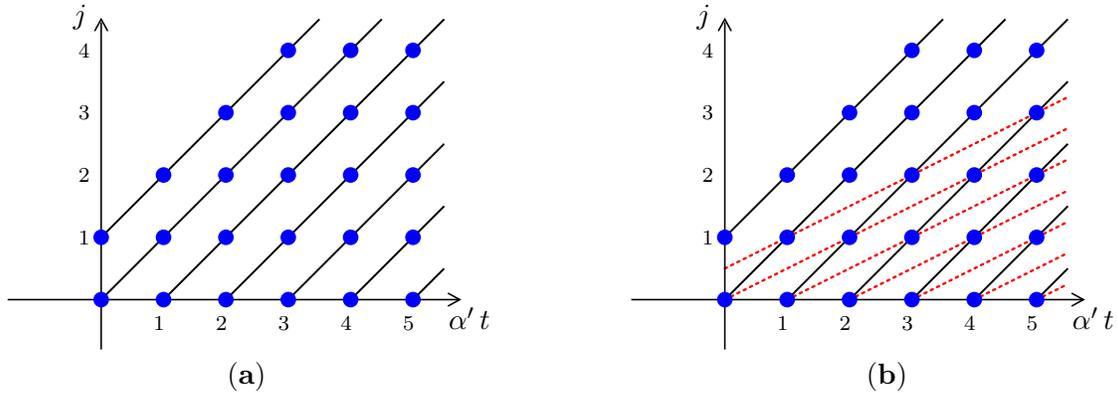
\caption{
(a) The most familiar continuation, away from integer spin, of the spectrum of the superstring in flat space. (b) {The familiar and sister trajectories},
which both enter the partial wave decomposition of the six-particle amplitude.\label{fig:stringspectrum}}
\end{figure}

Eq.~(\ref{A6simple}) is indeed the correct answer {for the $2\to 4$ amplitude} when the two incoming particles are adjacent along the color trace.
However it is incorrect when the two incoming particles are 3 and 6.
{In keeping with the literature} we will refer to the latter region as the {Mandelstam region of the $2\to4$ amplitude}.
{We stress that it is a perfectly physical configuration, it is just a different ordering of the same external momenta.}
A computation, however, reveals that in this region the asymptotic behavior for large $s$ and negative enough $t$ is controlled by a different trajectory:
\be
 A^{\rm folded}_6 \propto s^{\sister{\alpha}(t_{234})}\,,\quad \sister{\alpha}(t) = \,{1\over2} +\frac12\alpha' t\,.  \label{sister_traj_string}
\ee
This new, ``sister'' trajectory was first observed in the bosonic string amplitude in refs.~\cite{Hoyer:1976xw,Hoyer:1976xj,Sarbishaei:1976ag},
and subsequently also in the superstring \cite{Barratt:1976ha} (as well as for closed strings, where it appears in all kinematic regions \cite{Barratt:1977qb}).
It is part of an infinite sequence of trajectories translated by half-integer spin. Figure~\ref{fig:stringspectrum}
shows these trajectories, where the spin in the vertical axis being defined with respect to some SO(2) or SO(1,1) subgroup.
These trajectories are essential in order to correctly account for the high-energy behavior of the 6-particle amplitude.
It is important to stress that they are \emph{not} associated with any new state which would lie outside of the familiar lattice.
They simply represent a different way to group the states, away from integer spin, into analytic families.
\footnote{
The need for this alternative grouping may be seen from the multiplicities of SO(N) representations which occur
at a given mass level \cite{Nahm:1976es,Curtright:1986di}.  In particular, we have used table 6c of ref.~\cite{Curtright:1986di}
in order to obtain eq.~(\ref{sister_traj_string}) for the superstring, since the original amplitude calculation
in ref.~\cite{Barratt:1977qb} did not implement the GSO projection (which was invented later).}

\begin{figure}
\centering
\def\svgwidth{15cm}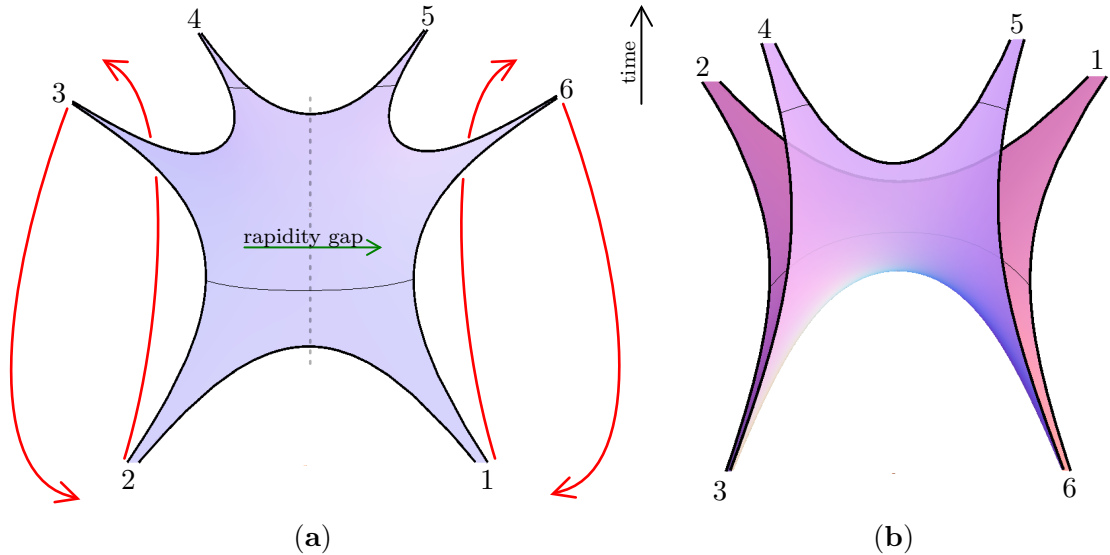
\caption{(a) Sketch of the string worldsheet exchanged between two sets of highly boosted particles (2, 3 and 4 vs. 5, 6 and 1) in the classical limit. Here, particles 1 and 2 are in-coming while the rest are out-going. (b) In the Mandelstam region where $3$ and $6$ are in-coming while the rest are out-going, the exchanged state is obtained by folding the picture in (a). These states lie on the ``sister'' trajectories of fig.~\ref{fig:stringspectrum}.
\label{fig:foldedstring}}
\end{figure}

The interpretation of the {sister} trajectories is most transparent in the classical regime $\alpha'|t_{234}|\gg 1$, where the exchanged string can be given a classical description
\`a la Gross-Mende-Manes \cite{Gross:1987kza,Gross:1989ge}.  To see this we have evaluated this classical solution for a sample choice of external momenta
taken both in the planar and Mandelstam kinematics. A sketch of  some representative solutions are shown in fig.~\ref{fig:foldedstring}. One sees that in the Mandelstam region
the exchanged string {can be} ``folded''.
This naturally doubles its tension, explaining the unusual slope in eq.~(\ref{sister_traj_string}).

{Although sufficiently simple to be worked out in full detail, }
this example turns out to be a good model for a certain gauge theory phenomenon.
In this case the elementary trajectory will be provided by the Reggeized gluon, while 
composite states containing several Reggeized gluons will be the analog of the higher-tension {or ``sister"} trajectories.

Exchanges of multiple Reggeized gluons are suppressed in perturbation theory by powers of the coupling,
but are necessarily important in precision calculations. Most importantly,
the amplitude for exchanging the Pomeron (a color-singlet pair of Reggeized gluons) has the property that it increases with energy.
Hence for {high} enough energies, multi-Reggeon exchange in principle dominates, regardless of the value of the coupling.
It is {possible} to consider situations where this phenomenon is
present, yet is kept under control by an adjustable parameter.
This happens, for example, when either the projectile or target is a color-singlet with a sufficiently small size. In this case one will always be in the so-called linear or dilute regime (a phenomenon sometimes {referred to} as ``color transparency'').\footnote{
This {can be seen as} analogous to the large impact parameter limit in classical Regge theory, which is similarly invoked to justify a picture based on exchanging a small number of reggeons, irrespective of their coupling --- in a gapped theory, interactions are exponentially suppressed at large distances. This analogy is made very precise in the AdS/CFT correspondence, see ref.~\cite{Brower:2006ea}.}
Another possibility is to consider theoretical limits such as large $N_c$, which we will do in this paper.

The large-$N_c$ suppression of multiple Reggeized gluons exchange (not only multi-Pomeron)
is a nontrivial aspect of BFKL dynamics.
It was (we presume) understood already in the early days of the BFKL approach.
It can be seen rather explicitly, for example, in the multi-Regge limit, where one has several rapidity gaps.
Technically, the formalism contains both a signature-odd and a signature-even Reggeized gluon, with otherwise the same quantum numbers (``signature'' is the eigenvalue under the flip in fig.~\ref{fig:foldedstring}).
Depending upon the kinematic region, these can thus interfere constructively or destructively when exchanged in a peripheral channel ($t_{23}$).
When one computes the contribution from two Reggeized gluons in the central channels ($t_{234}$), one finds that the color factors
cancel out exactly in the planar limit due to such interference, in all regions except for the Mandelstam region.
This in close analogy to what we have just discussed for tree-level open strings.
This two-Reggeon contribution was brought to attention in
a remarkable series of paper starting from \cite{Bartels:2008ce,Bartels:2008sc}, on which we will build.
It is worth mentioning however that, as far as we know, the fundamental ingredients have been around for a long time
(see, for example, the first, third and fourth diagram of fig.~17 in ref.~\cite{Bartels:1980pe}).

\begin{figure}
\centering
\def\svgwidth{13cm}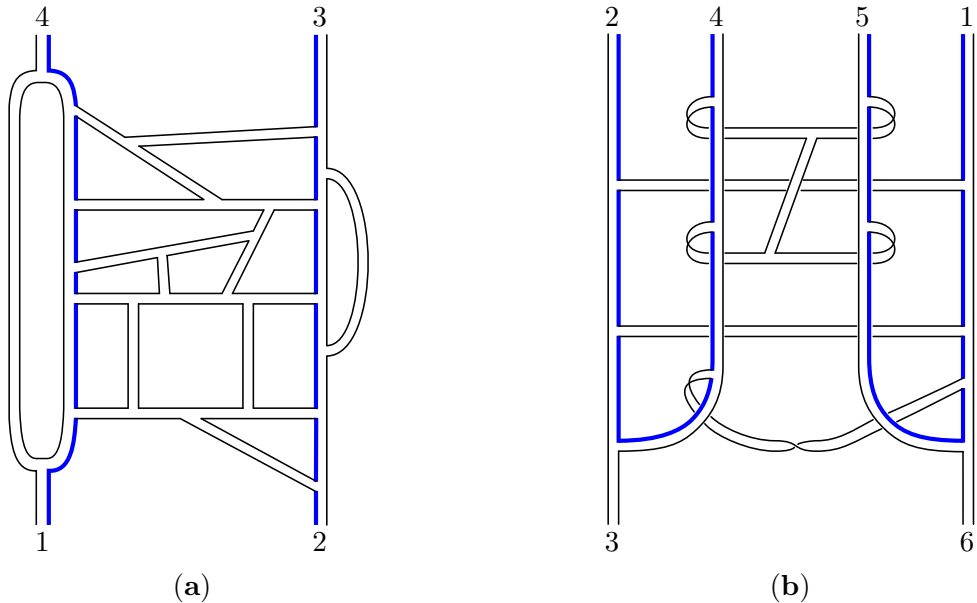
\caption{(a) Gluon Reggeization in the planar limit.  The multiple gluons exchanged between the target and projectile
are sourced by a single color source on each side (heavy blue line), whose location in the transverse plane is the dynamical variable
in the BFKL evolution equation. This equation originates from (factorized) high-energy
sub-loops as shown for example on the left.
(b) The amplitude in the Mandelstam region (compare with fig.~\ref{fig:foldedstring}(b)). At high energies it {factorizes} into dipole-dipole scattering.
\label{fig:reggeization}}
\end{figure}

A conceptually distinct approach to high-energy scattering exploits,  in gauge theory,  the dynamics of null Wilson lines.
This was pioneered notably in the work of Balitsky \cite{Balitsky:1995ub}.
The Wilson lines track the color charges of fast partons in the projectile.
In this formalism the Reggeized gluon can be identified with the logarithm of a null infinite Wilson line.
For more details, we refer to ref.~\cite{Caron-Huot:2013fea}.
In the planar limit the formalism simplifies dramatically: as argued in the last reference, the number of Wilson lines
required to describe a given process, regardless of the value of the t` Hooft coupling, should be a fixed number
which depends only on the kinematics, and determined through simple rules.
For planar scattering amplitudes,
by considering the available color charges during the (effectively instantaneous)
scattering process,
one easily sees that the six-particle amplitude in the Mandelstam region
is the first time where dipole-dipole scattering can occur (see fig.~\ref{fig:reggeization}).

One may think that these subtleties are of a rather technical nature, affecting {mainly} high-point amplitudes in the planar limit.
Our point of view is that exchange of multiple Reggeons is a
general phenomenon which affects all amplitudes away from the planar limit.
Planar high-point amplitudes allow us to study it in a controlled way.

\section{Continuation of collinear expansion and sister dispersion relation}\la{analyticcontinuationsec}
\begin{figure}[t]
\centering
\def\svgwidth{15cm}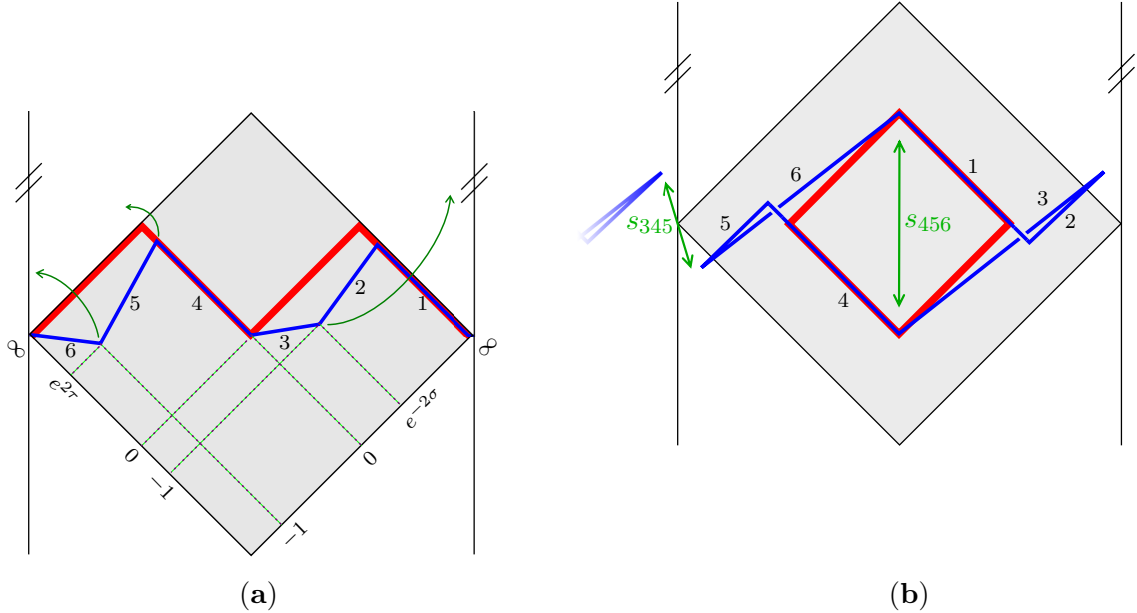
\caption{a). The hexagon representing the six-gluon amplitude in Euclidean kinematic, where all non-adjacent cusps are space-like separated. Cusps 5-6 and 2-3 in this figure are extended in the transverse space while all other cusps stand in the plotted ${\mathbb R}^{1,1}$ plane. In the collinear limit, these two cusps are flattened along the red square. After the analytic continuation $\{\sigma,\tau\}\to\{\sigma+i\pi/2,\tau-i\pi/2\}$ we arrive at the $3\to3$ kinematical configuration in (b). To be more precise, for the momentum transfers $s_{345}$ and $s_{456}$ to become time-like (while all other kept space-like), we also have to drag the cusp 5-6 to its antipodal point (at the top in (b)). This flip is not seen at the level of the conformal cross ratios  (\ref{crossratios}). In the multi-Regge limit, cusp 5-6 (2-3) approaches the antipodal point of cusp 1-6 (3-4).}
\label{3to3kinematics}
\end{figure}

We consider {the collinear expansion for} the process of $3\to 3$ scattering in the kinematics shown through the dual null polygon in fig.~\ref{3to3kinematics}-b.
This region contains the same information as the $2\to 4$ {Mandelstam} region that we have discussed so far,
but it will turn out to be more convenient for our analysis.

The 6-particle amplitude depends \emph{a-priori} on 9 Mandelstam invariants, but in the {theory} we will consider, planar $\mathcal{N}=4$ super Yang-Mills (SYM), dual conformal symmetry
ensures that the nontrivial dependence factors through 3 conformal cross-ratios: 
\be
 u_1 = \frac{s_{12}s_{45}}{s_{345}s_{456}}\approx \frac{1}{1+e^{-2\sigma}}\,,\qquad
 u_2 = \frac{s_{23}s_{56}}{s_{234}s_{456}}\approx e^{-2\tau}\,,\qquad
 u_3 = \frac{s_{34}s_{61}}{s_{345}s_{234}}\approx  \frac{1}{1+e^{2\sigma}} \,.
\label{crossratios}
\ee
These expressions are accurate up to corrections suppressed by powers of $e^{-\tau}$, which will not be important in the regimes we will consider.

The collinear limit $\tau\to\infty$ is now extremely well-understood. In this limit the amplitude organizes in a convergent
OPE series, whose relevant anomalous dimensions as well as the vacuum energy are known exactly from integrability \cite{Basso:2010in}, \cite{Beisert:2006ez}. Furthermore, a conjecture for the form factors at finite value of the coupling has been recently constructed \cite{Basso:2013aha}. This data govern the 6-particle amplitude (more precisely,
an infrared-finite combination of it and known four- and five-point amplitudes),
through an expansion of the form\footnote{
Relative to ref.~\cite{Basso:2013aha} we have moved a factor $(-1)^\ell$ from the measure to the outside.}
\be\la{collinear_2}
 \mathcal{W}_{\rm hex} -1 = 2\sum_{\ell\geq 1} (-1)^\ell \cos(\ell\phi) \int\limits_{-\infty}^{+\infty} \frac{dp}{2\pi} \hat{\mu}_\ell(p) e^{i\sigma p-\tau E_\ell(p)}+\ldots
 \equiv 2\sum_{\ell\geq 1}(-1)^\ell\cos(\ell\phi)\, F_\ell(\sigma,\tau)+\ldots
\ee
Here, we have retained only the leading-twist single-particle contribution for each angular momentum $\ell\geq 1$,
for reasons which will be explained below. These contributions as well as their multi-particle states were recently constructed in \cite{Basso:2014koa,Basso:2014nra}.

Importantly, the expansion (\ref{collinear_2}) is derived in the so-called Euclidean region, where all invariants
in eq.~(\ref{crossratios}) are spacelike and $\sigma,\tau$ are real.
We are interested in a Lorentzian region where the invariants $s_{345}$ and $s_{456}$ are time-like while all other invariants are space-like. This requires an analytic continuation, under which
the conformal cross ratios in (\ref{crossratios}) transform as $u_1\to e^{2\pi i}|u_1|$ and $u_{2,3}\to e^{i\pi}|u_{2,3}|$, see \cite{Bartels:2010tx}.
We will thus require a nontrivial continuation of the $\sigma$ and $\tau$ variables:
\be\la{continuation}
 \tau\to \tau-i\pi/2\ ,\qquad \sigma\xrightarrow[{\gamma}]{} \sigma+i\pi/2{+i0}\ , \qquad \cos\phi\to \cos\phi\,.
\ee
The contour $\gamma$ amounts for taking $\sigma$ large and negative, then shifting it by $i{\pi\over2}$ and finally taking its real part back to its original positive value. We keep $\phi$ unchanged through the continuation.\footnote{The continuation path $\gamma$ differs from the one used in e.g.~ref.~\cite{Bartels:2010tx} in that the norms $|u_i|$ are not kept constant.
As far as we can tell the two paths are nevertheless physically equivalent.}
 As depicted in fig.~\ref{3to3kinematics}, the cusp whose location is labeled by $\sigma$ (2-3) needs to be dragged around another cusp (1-2), which corresponds to adding the imaginary part to $\sigma$ while its real part is large and negative.

To see the effect of the continuation on eq.~(\ref{collinear_2}), it is useful to record the leading order expressions at weak coupling for the $\ell=1$ mode,
\be\begin{aligned}
 \hat\mu_1(p) &= 2\pi g^2\frac{1}{\cosh(\frac{\pi p}{2})(p^2+1)} + \mathcal{O}(g^4)\,,\\
 E_1(p) &=  1 + 2g^2 \left[ \psi\left(\frac32+\frac{ip}{2}\right)+\psi\left(\frac32-\frac{ip}{2}\right)-2\psi(1)\right] +\mathcal{O}(g^4)\,,
\end{aligned}\label{leadingorder_fE}
\ee
where $g^2=\lambda/(4\pi)^2$. We see that when ${\rm Im\,}\sigma=\pi/2$ the Fourier transform becomes marginally convergent.
This reflects the presence of a branch point at $\sigma=i\pi/2$, corresponding to the point where the cusp (2-3) becomes null separated from cusp (5-6) in figure \ref{3to3kinematics}-a, so that $u_3=\infty$.
Depending on which side of the branch cut one is, one can get two different functions.  Let us thus define the analytic continuation of the function $F_l(\sigma,\tau)$ in (\ref{collinear_2}) below the cut, and its discontinuity by:
\be \begin{aligned}
\Fdown_\ell(\sigma,\tau) &\equiv F_\ell\left(\sigma+i\frac{\pi}{2}-i0,\tau-i\frac{\pi}{2}\right)\,,\\
\Fdisc_\ell(\sigma,\tau) &\equiv F_\ell\left(\sigma+i\frac{\pi}{2}+i0,\tau-i\frac{\pi}{2}\right)-\Fdown_
 \ell(\sigma,\tau)\,.\end{aligned}\label{def_disc}
\ee
From now on we will absorb the imaginary shifts in the definitions of $\Fdown_\ell$ and $\Fdisc_\ell$, so that their arguments will be real.
The first function is defined directly by its Fourier representation, while the function we will be interested in is the second one, which is defined only by analytic continuation.  Physically, $\Fdisc_\ell$ represents the discontinuity of the amplitude in the $s_{345}$ channel. It vanishes,
by construction, for $\sigma<0$.

At leading order, for example,
\ba
 \Fdown_1(\sigma,\tau) &=& 2\pi g^2 \int\limits_{-\infty}^{+\infty} \frac{dp}{2\pi} \frac{i e^{-\pi p/2}\,e^{i\sigma p-\tau}}{\cosh(\frac{\pi p}{2})(p^2+1)} +\OO(g^4)
\\
&=& g^2e^{-\tau} \left\{ 2e^\sigma\sigma +i\pi e^\sigma-\big(e^{\sigma}-e^{-\sigma}\big)\log\left(1-e^{2\sigma}+i0\right)\right\} + \OO(g^4)\,.
\ea
The discontinuity comes just from the logarithm,
\be
\Fdisc_1(\sigma,\tau) = 2\pi i g^2 e^{-\tau} \left( e^{\sigma}-e^{-\sigma}\right)\theta(\sigma)+ \OO(g^4)\,.
\label{LO_Fdown}\ee
A crucial feature is that the two functions have dramatically different behavior in the Regge limit $\sigma,\tau\to\infty$:
$ \Fdown_1$ vanishes like $e^{-\tau-\sigma}$, while $\Fdisc_1$ goes like $e^{\sigma-\tau}$ and thus survives
in the Regge limit, as expected. We stress, that our analytic continuation path is such that the OPE twist expansion does not need to be resummed (not until $\sigma$ becomes so large that one enters
the BFKL regime so deeply that $e^{\sigma-\tau}\ll 1$ no longer holds, which we will not need to do in our calculations),
so that the continuation can be performed for each $F_\ell$ independently.

It is possible to calculate such discontinuities to rather high orders in perturbation theory, as initiated in ref.~\cite{Bartels:2011xy}
and pursued further in ref.~\cite{Hatsuda:2014oza}.  These references considered the $2\to 4$ kinematics but our discussion so far
has been essentially equivalent.
The expressions for $\hat\mu_\ell(p)$ and $E_\ell(p)$ are such that, to any loop order, $\Fdown_\ell(\sigma)$ can be expressed
in terms of harmonic polylogarithms with argument $e^{-2\sigma}$ \cite{Papathanasiou:2013uoa}; this allows for a systematic
treatment.


We found a simple systematic procedure to organize the result, which will allow us to pass directly to finite coupling.
In this procedure {a} ``sister'' dispersion relation will naturally appear.

\subsection{Analytic continuation of Fourier transform: general framework}

We illustrate the idea by working to leading-logarithm accuracy, e.g. we compute
$F_1$ in eq.~(\ref{collinear_2}) exactly in $g^2$ but using the functions $\hat\mu(p)$ and $E(p)$ truncated to order $g^2$.
For conciseness in this subsection we discuss the mode $\ell=1$ and temporarily omit the subscript.
The generalization to higher orders and to other values of $\ell$ will be immediate.

We begin by writing the evolution in $\tau$ as a convolution in $\sigma$-space, omitting temporarily the trivial $\mathcal{O}(g^0)$ piece:
\be
 -\frac{d}{d\tau} \Fdown(\sigma,\tau) = \int\limits_{-\infty}^{\infty} dt\, K(t) \Fdown(\sigma-t,\tau) \,,\qquad
\label{Kaction1}
 K(t) = -4g^2\frac{e^{-|t|}}{(e^{2|t|}-1)_+} + \mathcal{O}(g^4)\,.
\ee
The kernel $K(t)$ is the Fourier transform of the dispersion relation in eq.~(\ref{leadingorder_fE}).
The first observation is that it is not analytic, because of the absolute value.\footnote{For any test function $G(t)$, the $+$ prescription is defined as:
$\int\limits_0^\infty \frac{dt}{(e^{2t}-1)_+}G(t) \equiv \int\limits_0^\infty \frac{dt}{e^{2t}-1}(G(t)-G(0))$ {and in (\ref{Kaction1}) we have $G(t)=e^{-|t|}\Fdown(\sigma-t,\tau)$}\,.}
This is a problem because to analytically continue $\sigma$ we also  need to continue $t$, in order to prevent
$\sigma-t$ from hitting the cut of $\Fdown$ starting at $t=\sigma$.
To remedy this, we split the kernel into two analytic pieces $K_\pm$, which agree with $K$ respectively for positive and negative arguments.
Then
\be
 -\frac{d}{d\tau}\Fdown(\sigma,\tau) =
   \int\limits_{-\infty}^0 dt\,K_-(t) \Fdown(\sigma-t,\tau)
+ \int\limits_{0}^{\infty} dt\, K_+(t) \Fdown(\sigma-t,\tau)\,.
\label{Kaction2}
\ee
Both integrals now define analytic functions of $\sigma$.

\begin{figure}[t]
\centering
\def\svgwidth{15cm}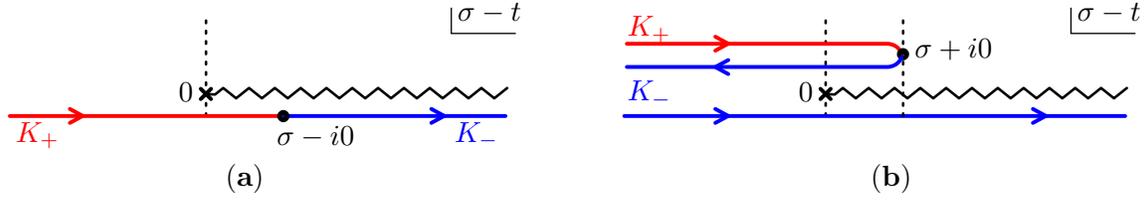
\caption{(a) The function $\Fdown$ is originally defined on a real line below the cut. (b) Contour for $(\sigma-t)$ in the convolution (\ref{Kaction2}) after $\sigma$ has been continued above the cut, passing to the left of the branch point. Only the range of integration between the two dashed lines contributes to the discontinuity [(b)-(a)].
\label{fig:contour}}
\end{figure}

The crucial step in our procedure is to properly drag the contour of integration around the cut
as we analytically continue the evolution equation, as a function of the argument $\sigma$ on the left-hand side.
By collecting the various pieces of the contour shown in fig.~\ref{fig:contour}, and using the appropriate kernel in each component, we obtain a simple closed evolution equation for the discontinuity:
\be
\boxed{-\frac{d}{d\tau} \Fdisc(\sigma,\tau) = \int\limits_{0}^\infty dt\,\big(K_+(t)-K_-(t+i0)\big) \Fdisc(\sigma-t,\tau)}\,.
\label{twotermcontinuation}
\ee
This is much like the original kernel except that the piece $K_-$ has been analytically continued clockwise to positive $t$.

The interest of eq.~(\ref{twotermcontinuation}) is that it disentangles the continuation of the form factor from that of the energy;
more generally this trick would help analytically continue any convolution.

At the lowest order, one can compute from eq.~(\ref{Kaction1}) that
\be
 K_+(t) - K_-(t+i0) = -4g^2\left(\frac{e^{-t}+e^{3t}}{(e^{2t}-1)_+} -\frac{i\pi}{2}\delta(t)\right)\,. \label{Ktspace}
\ee
{The first term arises trivially but} the $i\pi$ term is a slightly subtle anomaly which originates from the continuation of the $+$ prescription.
It will be important below so let us elaborate on it. It can be derived simply by expanding out the action of $K_-$ on a test function:
\be
 \int\limits_{-\infty}^0 \frac{dt\, G(t)}{(e^{-2t}-1)_+} \equiv \int\limits_{-\infty}^{-\epsilon} \frac{dt\,G(t)}{e^{-2t}-1} - G(0) \int\limits_{-\infty}^{-\epsilon} \frac{dt}{e^{-2t}-1}\,.
\nonumber\ee
The second integral is just a constant and so does not participate in the analytic continuation.
The first integral continues nontrivially when the $t$ contour is rotated clockwise to positive $t$:
\be
\int\limits_{-\infty}^{-\epsilon} \frac{dt\,G(t)}{e^{-2t}-1} \longrightarrow \int\limits_{-\epsilon}^\infty \frac{-dt\,G(t)}{e^{-2t}-1-i0} = -\frac{i\pi}{2}G(0)+ \int\limits_{\epsilon}^\infty \frac{-dt\,G(t)}{e^{-2t}-1}.
\label{anomaly}
\ee
{The integral re-combines with the constant to form the $+$ prescription, leaving the additional anomaly.}

Performing the Fourier transform of (\ref{Ktspace}) and reinstating the $\mathcal{O}(g^0)$ term in the energy
yields a new dispersion relation, {which we will call ``sister''}:
\be
 \sister{E}_1(p) = 1+ 2g^2\left[\psi\left(\frac32+\frac{ip}{2}\right)+\psi\left(-\frac12+\frac{ip}{2}\right)-2\psi(1)+i\pi\right] +\mathcal{O}(g^4)\,. \label{sister_1loop}
\ee
{It essentially differs from~(\ref{leadingorder_fE}) by the flipping of the argument of the second $\psi$-function, which makes it} analytic for ${\rm Im\,}p<-1$.
The discontinuity defined in (\ref{def_disc}) can now be written in the form (when $\ell = 1$)
\be
\boxed{\Fdisc_\ell(\sigma,\tau) =
\int\limits_{-\infty}^\infty \frac{dp}{2\pi}\,\hat\mu_\ell^\updownarrow(p) e^{ip\sigma}e^{-\tau \sister{E}_\ell(p)}}\, ,
\label{LLog_disc}
\ee
where the $p$ contour in the integral runs below all singularities, in accordance with the vanishing of the discontinuity for $\sigma<0$.
Equation (\ref{LLog_disc}) is the main result of this subsection.
It expresses that the discontinuity {in the Mandelstam region}
is not controlled by the original dispersion relation, but rather by a sister one.
This is quite reminiscent of our earlier discussion of strings in flat space.

The lowest order ``sister'' form factor can be obtained by Fourier transforming eq.~(\ref{LO_Fdown}):
\be
 \hat\mu_1^\updownarrow(p) =- 4\pi i g^2 \frac{1}{p^2+1} + \OO(g^4)\,.
\ee
Substituting into eq.~(\ref{LLog_disc}) yields a prediction for the discontinuity to leading-log accuracy and to any loop order.
With the help of the HPL package \cite{Maitre:2005uu}, we have verified this prediction against the direct analytic continuation of $\Fdown$
up to order $g^{10}$, and found perfect agreement.

\subsection{Sister dispersion relation at finite coupling}
\label{sec:sister_finite_coupling}

We are ready to go directly to finite coupling.
The dispersion relations $E_\ell(p)$, for $\ell\geq 1$, have been obtained
in ref.~\cite{Basso:2010in}. They are expressed in terms of an auxiliary spectral parameter $u$:
\be\begin{aligned}
 E_\ell(u) &= \ell+\int\limits_0^{\infty}\frac{dt}{t} K(t)\left(\cos(ut)e^{-\ell t/2}-1\right)\,,
   \\
 p_\ell(u) &= 2u +\int\limits_0^\infty \frac{dt}{t} K(-t)\sin(ut)e^{-\ell t/2}\,.
\end{aligned}\label{gauge_dispersion}
\ee
The function $K(t)$ admits the convergent expansion%
\footnote{In the notation of ref.~\cite{Basso:2010in}, $K(t)=\frac{\gamma_+(2gt)}{1-e^{-t}} - \frac{\gamma_-(2gt)}{e^t-1}$.\label{gamma-footnote}}
\be
 K(t) = \frac{1}{1-e^{-t}}\sum_{n\geq 1}2(2n)\gamma_{2n} J_{2n}(2gt)
 - \frac{1}{e^t-1}\sum_{n\geq 1}2(2n-1)\gamma_{2n-1}J_{2n-1}(2gt)\,. \label{Bessel_expansion}
\ee
Assuming this expansion, the BES equation, which determines the coefficients $\gamma_i$, takes the simple form
\be
 E_{0}(u) = p_{0}(u) =0 \quad \mbox{for }\quad  u\in [-2g,2g]\,. \label{BES}
\ee
Importantly, the $t\to 0$ limit of $K(t)$ is controlled by $\gamma_1$,
which, up to a factor, is the cusp anomalous dimension.
\be
-K(0)=2g\gamma_1
= 4g^2\left\{1-\frac{\pi^2}{3}g^2+\frac{11\pi^4}{45}g^4-g^6\left(\frac{73\pi^6}{315}+8\zeta_3^2\right)\right\} +\OO(g^{10}) \equiv \Gammacusp\,. \label{cusp}
\ee
Further coefficients are recorded in appendix \ref{app:expansion}.

In principle we should now try to apply the preceding analytic continuation procedure to the functions $E_\ell(p)$.
However, because the procedure
applies to any convolution, it is easy to see that we can simply treat $E_\ell(u)$ and $p_\ell(u)$ separately.
The procedure is then straightforward: we take all the terms in (\ref{gauge_dispersion}) which contain $e^{iut}$ (these are the terms which would go into $K_-$),
and we rotate the $t$ contour clockwise 180${}^\circ$ in them.  This yields
\be
\boxed{\begin{aligned}
 \sister{E}_\ell(u) &= \ell
 +\frac{i\pi}{2}\Gammacusp +\int\limits_0^{\infty}\frac{dt}{t}\left[ K(t)\frac{e^{-iut-\ell t/2}-2}{2} + K(-t) \frac{e^{-iut+\ell t/2}}{2}\right]
 \\
 \sister{p}_\ell(u) &=
 2u +\frac{\pi}{2}\Gammacusp -i \int\limits_0^\infty \frac{dt}{t}\left[ K(t)\frac{e^{-iut+\ell t/2}}{2}-K(-t)\frac{e^{-iut-\ell t/2}}{2}\right]
\end{aligned}}\label{gauge_sister_dispersion}
\ee
The $\Gammacusp$ terms originate from the pole at $t=0$, precisely as in eq.~(\ref{anomaly}); this pole is controlled by the cusp anomalous dimension thanks to eq.~(\ref{cusp}).
The integrals define analytic functions of $u$ in the lower-half plane ${\rm Im} \,u< -\ell/2$.

The analytic continuation of the form factor is similar and is discussed in appendix \ref{app:impact}.  Using the analytic expressions in ref.~\cite{Papathanasiou:2013uoa} together with the HPL package,
we have verified that eq.~(\ref{gauge_sister_dispersion}) reproduces the correct discontinuity up to four loops.
We stress, however, that the correctness of eq.~(\ref{gauge_sister_dispersion}) can be tested without knowledge of the correct form factor.
The concept of the sister dispersion relation allows to completely disentangle the continuation of the energy
from that of the form factor.

Finally, let us comment on the amplitude in the $2\to 4$ region, as was previously considered in refs.~\cite{Bartels:2011xy,Hatsuda:2014oza}.
The relevant continuation is then $\sigma\to \sigma-i\pi$, passing again to the left of the cut at $-i\pi/2$. Our derivation of $\check E_l(\check p_l)$ (\ref{gauge_sister_dispersion}) will not be effected by this change. However, the prediction for $F^{2\to4}_l$ is more involved. It can be obtained by adding, to the original amplitude, our result for the discontinuity.
Considering that this continuation is the complex conjugate of the case $\sigma\to\sigma+i\pi$, which is closer to our definitions, the prediction can be written:
\be
 \big(F_\ell^{2\to 4}\big)^* = \int\limits_{-\infty}^\infty \frac{dp}{2\pi} e^{i\sigma p} \left[
   \hat\mu_\ell(p)e^{-\tau E_\ell(p)-\pi p} 
 + \hat\mu^{\updownarrow}_\ell e^{-\tau\sister{E}_\ell(p)-\pi p/2}\right]
 + \oint_{\mathcal{C}} \frac{dp}{2\pi} e^{i\sigma p -\pi p/2}\hat\mu_\ell^{\updownarrow} e^{-\tau\sister{E}_\ell(p)} \,. \label{expansion24}
\ee
This formula should be compared against eq.~(\ref{LLog_disc}), which expresses the discontinuity of the OPE in the $3\to 3$ region.
The first integral runs along the real axis while $\mathcal{C}$ encloses all lower-half plane singularities of the second integrand.
The first integral is convergent as the behavior of $\hat\mu^\updownarrow$ is such
that the square bracket behaves at large negative $p$ like $e^{\pi |p|}\hat\mu_\ell(p)\big[e^{-\tau E_\ell(p)}-e^{-\tau \sister{E}_\ell(p)}\big]$. However
$(E_\ell(p)-\sister{E}_\ell(p))\sim e^{-\pi |p|/2}$ in this region.
In the Regge limit $\sigma\to\infty$ the second integral dominates and we reproduce the leading-log formula proposed in ref.~(\ref{LLog_disc}).
More generally, we have also verified eq.~(\ref{expansion24}) against the analytic expressions of ref.~\cite{Papathanasiou:2013uoa} up to order $g^{10}$.

\section{Going through the cut: finite coupling BFKL eigenvalue}\la{goingthroughcutsec}

Before we identify the sister dispersion relation (\ref{gauge_sister_dispersion}) with the leading Regge trajectory
in the adjoint-dipole sector with transverse angular momentum $|m|=\ell\geq 1$, let us briefly discuss the identification of the quantum numbers between the two expansions.
The Regge factorization formula of ref.~\cite{Lipatov:2010ad,Bartels:2010tx} gives the BDS remainder function in the form\footnote{
This formula of refs.~\cite{Lipatov:2010ad,Bartels:2010tx} is expressed in terms of variables
which are related to our variables through: $\left(\frac{w}{w^*}\right)^{\frac12}\mapsto e^{i\phi}$;
$|w|\mapsto e^{\sigma-\tau}$; $\frac{1}{\sqrt{|u_2u_3|}}\mapsto e^{\tau+\sigma}$; $\nu_{\rm here}=2\nu_{\rm there}$. Also we have omitted a ``$\cos\omega_{ab}$'' term as a consequence using a slightly different contour of integration to avoid real-axis poles,
as discussed at the end of subsection \ref{ssec:result_impact_factor}.}
\be
\mathcal{R}_{\rm hex}^{3\to 3}e^{-i\pi \delta} = 
-2\pi i\sum_{m=-\infty}^{+\infty} (-1)^m e^{im\phi} \int\limits_{-\infty}^{+\infty} \frac{d\nu}{2\pi}
 \hat\mu_{\rm BFKL}(\nu,m) e^{i(\sigma-\tau)\nu} e^{(\sigma+\tau)\omega(m,\nu)}\,. \label{BFKL_2a}
\ee
The phase $\delta=\frac{\Gammacusp}{2} \log \frac{|w|}{|1+w|^2}$, where $w=e^{\sigma-\tau+i\phi}$, essentially cancels a non-factorizing
phase present in the BDS Ansatz.  On the other hand, the OPE prediction (\ref{collinear_2}) is given for
the ratio  $\mathcal{W}_{\rm hex}$, which is associated with a specific propagation channel. The conversion,
in the multi-Regge limit, can be obtained readily using eq.~(119) of ref.~\cite{Basso:2013aha}:
$\mathcal{W}_{\rm hex}^{3\to 3}=\mathcal{R}_{\rm hex}^{3\to 3}e^{-i\pi {\Gamma_\text{cusp}\over2} \log |1+w|^2}$.
This extra phase will turn out to be irrelevant in the saddle point region that we will consider shortly,
but we find pleasing that it neatly cancels a piece of the phase $-i\pi\delta$.  The part which remains, however,
is the one which will be important for the analysis. It gives that
\be
   \mathcal{W}^{3\to 3}_{\rm hex} e^{-\frac12 i\pi(\sigma-\tau)\Gammacusp} = \mbox{right-hand side of eq.~(\ref{BFKL_2a})}. \label{BFKL_2}
\ee
Comparing with the collinear expansion (\ref{collinear_2}) and (\ref{LLog_disc}) then leads to the following identification of quantum numbers:
\be
  \nu=\frac12(\sister{p}-i \sister{E})-\frac{\pi}{2}\Gammacusp\,,\qquad
  -\omega=\frac12(\sister{E}-i \sister{p})\,. \label{BFKLvariables}
\ee
The physical interpretation of these formulas is simple: the OPE and BFKL expansions resum the exchanged states
in the same $s_{234}$ channel, both organized in terms of the symmetries of the same associated null square, see figure \ref{3to3kinematics}. However, for the two limits one chooses to diagonalize different symmetries (dilatations and boosts, as opposed to $\sigma$-translations and twist), which differ by a $45^\circ$ rotation.

It seems likely that the simple form of eq.~(\ref{BFKL_2}) can be generalized to higher-point amplitudes, using a sequence of null
squares and pentagons appropriate to the multi-Regge limit; we also believe that the extra phases, responsible
for the $\Gammacusp$ shift in $\nu$, can be predicted in a simple and systematic way using the anomalous Ward identities for dual conformal symmetry as done in ref.~\cite{Caron-Huot:2013fea}. We hope to discuss this in a future publication.

Returning to the interplay between the collinear and Regge limits, we must address the question of whether there exists some
physical observable for which we can simultaneously apply
(\ref{collinear_2}) and (\ref{BFKL_2}), thereby justifying equating them.
Clearly this observable will involve complex momenta, due to the map (\ref{BFKLvariables}).
This arises naturally when evaluating real configuration space amplitudes using the saddle point method.

If we take both $\sigma$ and $\tau$ sufficiently large, regardless of the value of the coupling, the integrals will develop narrow saddle points
at the points $p_*$ and $\nu_*$:
\be
\frac{\sigma}{\tau}= \frac{1}{i}\frac{d\sister{E}}{dp}(p_*) \,\quad\mbox{and}\quad
\frac{\tau-\sigma}{\tau+\sigma} = \frac{1}{i}\frac{d\omega}{d\nu}(\nu_*) \,.  \label{saddle_points}
\ee
For physical observables the left-hand sides are real.  As can be seen by taking $u$ to be pure imaginary
in eq.~(\ref{gauge_sister_dispersion}), the saddle point will then be at pure imaginary $u_*$, corresponding to
($\nu_*,\omega_*$) pure imaginary and real, respectively, and ($\sister{p}_*,\sister{E}_*$) also pure imaginary and real (up to the
$\Gammacusp$ shifts).

As we begin from $\sigma/\tau=0$ and increase it, the OPE saddle point will move upward from $u=-i\infty$
along the imaginary axis, eventually approaching the first branch cut at $u=-i\ell/2\pm 2g$.
A natural hypothesis is that if we keep increasing $\sigma/\tau$, the saddle point $u_*$ will traverse the cut
where it will meet with the BFKL saddle point.
As we further increase $\sigma/\tau$ towards unity, corresponding to the Regge limit with fixed transverse size,
the BFKL $\nu_*$ will approach its final value of zero, where we will make contact with the usual BFKL eigenvalue in which $\nu$ is real.

This hypothesis will indeed turn out to be correct. We will justify it in the discussion section by showing that the two expansions have a non-empty overlapping
regime of validity, in some open neighborhood of the cut, where the corresponding analytic functions
can be matched with each other to arbitrary precision.
For the moment, we will simply proceed assuming the hypothesis, for which
there will be several nontrivial checks.

\subsection{Going through the cut}

Let us now explain how one can perform the analytic continuation across the cut at ${\rm Im}\,u=-\ell/2$, where the integral representation (\ref{gauge_sister_dispersion}) diverges.
Since the cut is very narrow at weak coupling,
this is the one fundamentally nonperturbative step in our procedure.

The necessary input is that $K(t)$ solves the BES equations (\ref{BES}).
These equations hold exactly \emph{on} the cut.
Thus, depending on whether it is the term with $K(t)$ or $K(-t)$ which becomes singular on the cut,
we add one or the other of the BES equations. In this way, right inside the cut at ${\rm Im\,}u=-\ell/2+i0$,
the dispersion relations (\ref{gauge_sister_dispersion})
are also equal to
\be\begin{aligned}
\sister{E}_\ell (u)&= \sister{E}_\ell(u) + i p_0(u+i\ell/2) \\
 &=
 2iu +i\frac{\pi}{2}\Gammacusp +\int\limits_0^{\infty}\frac{dt}{t}\left[ K(t)\frac{e^{-iut-\ell t/2}-2}{2} + K(-t) \frac{e^{iut-\ell t/2}}{2}\right]\,, \\
 \sister{p}_\ell(u) &= \sister{p}_\ell(u) +iE_0(u+i\ell/2) \\
&=
 2u +\frac{\pi}{2}\Gammacusp +i \int\limits_0^\infty \frac{dt}{t}\left[ K(t)\frac{e^{iut-\ell t/2}-2}{2}+K(-t)\frac{e^{-iut-\ell t/2}}{2}\right]\,. \label{goingthrough}
\end{aligned}
\ee
The integrals now converge throughout the strip $|{\rm Im}\,u|< \ell/2$, providing the desired analytic continuation.
Using the identification (\ref{BFKLvariables}) and setting $\ell=|m|$, we obtain the desired result:
\be\boxed{\begin{aligned}
 -\omega(u,m)&= \int\limits_0^\infty\frac{dt}{t}\left(\frac{K(-t)+K(t)}{2}\cos(ut)e^{-|m|t/2}-K(t)\right)\\
 \nu(u,m) &= 2u+ \int\limits_0^\infty \frac{dt}{t}\frac{K(-t)-K(t)}{2}\sin(ut)e^{-|m|t/2}
\end{aligned}}\label{BFKL_exact}\ee
This formula is our main result: the color-adjoint BFKL eigenvalue, expressed to all loop orders in the planar limit in terms of the solution $K(t)$ to the BES integral equation (\ref{BES}).

It is extremely nontrivial that for real $u$ the combinations $\omega$ and $\nu$ are both real and define a symmetric trajectory $\omega(\nu)=\omega(-\nu)$,
let alone that they reproduce the known perturbative expansion.
For example, at leading order we have $u\approx \nu/2$ and $K\approx -4g^2 t/(e^t-1)$, and the Fourier transform easily produces
\be\label{loBFKL}
 -\omega(\nu,m) = 2g^2\left\{
 -\frac{2|m|}{\nu^2+m^2}+\psi\left(1+\frac{|m|+i\nu}{2}\right) +\psi\left(1+\frac{|m|-i\nu}{2}\right)-2\psi(1)\right\} + \OO(g^4)\,.
\ee
This is precisely the standard BFKL result.\footnote{We recall that our normalization for $\nu$, which agrees with refs.~\cite{Brower:2006ea,Costa:2012cb} and
is such that $i\nu$ has the physical interpretation of dimension, departs from that of refs.~\cite{Bartels:2008ce,Bartels:2008sc,Fadin:2011we,Dixon:2012yy,Dixon:2014voa}: $\nu_{\rm here} = 2\nu_{\rm there}$.}

The trajectory (\ref{BFKL_exact}) can be easily expanded out to any desired order at weak coupling;
the expansion involves always the same building blocks: $\psi$ functions or their derivatives, or rational factors.
This is detailed in appendix \ref{app:expansion}, where the reader will find a comparison against the three-loop
results of refs.~\cite{Bartels:2008ce,Bartels:2008sc,Fadin:2011we,Dixon:2012yy,Dixon:2014voa} as well as our four-loop prediction.

\subsection{Comments on the $m=0$ mode}

Our main result (\ref{BFKL_exact}) was derived for all $m\neq 0$. The $m=0$
case is in principle more difficult for us because the flux tube, which underpins the collinear expansion~(\ref{collinear_2}), does not have any elementary excitation with $m=0$ and vanishing $R$-charge. From the perspective of how integrability organizes the states of the theory,
this means that the $m=0$ states which dominate in the Regge limit must be composites.  This was analyzed in ref.~\cite{Hatsuda:2014oza} where these composites were identified as a pair of scalar flux tube excitations. In principle, to obtain the BFKL eigenvalue with $m=0$ one should thus study the energy of this scalars pair.

The formula (\ref{BFKL_exact}), however,
displays such a compellingly simple dependence on $m$ that it is very tempting to simply set $m=0$ into it. Evidence that this procedure yields the correct eigenvalue can be found by specializing to the points $\nu = \pm \frac{\pi}{2}\Gamma_{\textrm{cusp}}$. At these points the $m=0$ eigenvalue was predicted in ref. \cite{Caron-Huot:2013fea} to vanish identically
\be
\omega(\nu =  \pm\frac{\pi}{2}\Gamma_{\textrm{cusp}}, m=0) = 0\, , \label{coll-eq}
\ee
giving us a sharp test for our exact formula~(\ref{BFKL_exact}).
There is not much to verify to three loops, actually, since our formula reproduces all known results to this order, which were shown in ref.~\cite{Caron-Huot:2013fea} to obey eq.~(\ref{coll-eq}).
The representation~(\ref{BFKL_exact}), however, does not make eq.~(\ref{coll-eq}) obvious at finite coupling. A careful analysis, carried in appendix \ref{meq0appendix}, reveals that this is nonetheless true.
This particular but nontrivial check makes us confident that the formula~(\ref{BFKL_exact}) works properly in the special case $m=0$ as well.

\section{Strong coupling and string theory}\label{strongC:sec}

The integral representation~(\ref{BFKL_exact}) determines the color-adjoint BFKL eigenvalues all the way from weak to strong coupling.
In the latter regime we should therefore be able to make contact with the semiclassical string description and, most notably, with the analysis of refs.~\cite{Bartels:2013dja,Bartels:2010ej}. These papers studied the Regge limit of scattering amplitudes using the Thermodynamical Bethe Ansatz (TBA) equations derived in~\cite{Alday:2009dv,Alday:2010vh}. This is what we shall do in this section.

\subsection{Strong coupling intercept}

\begin{figure}[t]
\centering
\def\svgwidth{16cm}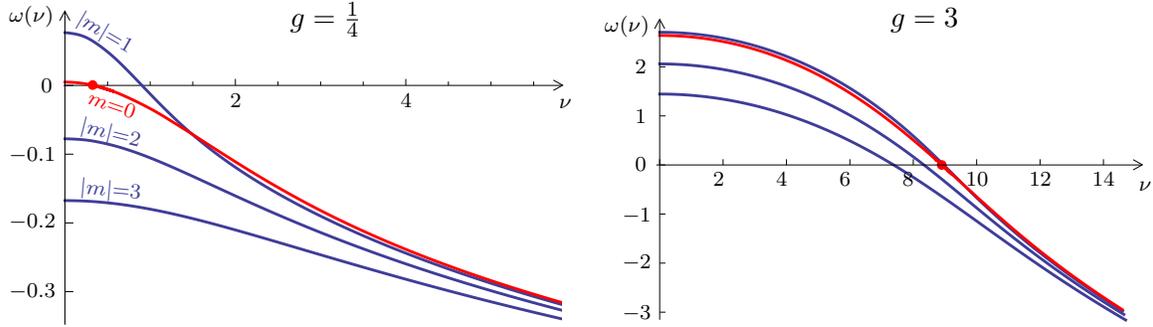
\caption{Plots of the eigenvalues $\omega_{m}(\nu)$ for $m=0$ (in red) and $|m|=1, 2, 3$ (in blue and from top to bottom) at weak coupling $g = 1/4$ (left) and strong coupling $g=3$ (right). The red dot indicates the point $(\omega, \nu) = (0, \frac{\pi}{2}\Gamma_{\textrm{cusp}})$ which lies on the $m=0$ trajectory. The trajectories are sparse in appropriate unit and mostly negative at weak coupling. They get closer and wider at strong coupling where they approach some universal curve. The latter is given parametrically in eq.~(\ref{bfklstrong}) for $|\nu| < \frac{\pi}{2}\Gamma_{\textrm{cusp}}$.}\label{flattening}
\end{figure}

Expanding our integrals at strong coupling is relatively easy since the latter are expressed in terms of the solution to the BES equation and thus already constructed, see~\cite{Alday:2007qf,Basso:2007wd,Kostov:2008ax}.
For convenience, the required expressions are reproduced in appendix \ref{app:strong}.
To the leading order at strong coupling $\sqrt{\lambda}\gg 1$, for instance, we immediately find that
\be\la{wmax}
\omega(\nu) = \frac{\sqrt{\lambda}}{2\pi}(\sqrt{2}-\log(1+\sqrt{2})) + O(1)\, ,
\ee
which applies to all values of $m$ and $\nu$ --- as long as these are not of order $O(g)$. The BFKL trajectories are thus becoming flat and identical at strong coupling, as illustrated in figure~\ref{flattening}. Their intercept~(\ref{wmax}) turns into a prediction for the energy growth of the remainder function at strong coupling which is in perfect agreement with the string theory results derived in refs.~\cite{Bartels:2013dja,Bartels:2010ej}.

It is important here to recall that this growth is only relative to a background which is itself exponentially small;
$\omega$ is really the dipole Regge trajectory minus the gluon one, which is infinitely large and negative due to infrared divergences.

It is quite remarkable that all trajectories share the same positive intercept~(\ref{wmax}) at strong coupling. It hints at the fact that they all pertain to the same semiclassical string saddle point, on which we shall come back shortly. The situation contrasts, however, with weak coupling where the intercepts are mostly negative and sparse in appropriate unit. As we crank up the coupling all the negative intercepts eventually turn around and start growing as depicted in figure~\ref{BFKLintercepts}. The observed degeneracy among levels is of course an artifact of the strict strong coupling limit and, with a little more work, one could actually demonstrates that  for $|m|>0$
\be
\omega_m(0)  = \omega_0(0) - \frac{|m|-1}{\sqrt{2}} + O(1/\sqrt{\lambda})\, .
\ee
This is showing, in agreement with figure~\ref{BFKLintercepts}, that trajectories with $|m|>1$ have remained below the $m=0$ and $m=|1|$ ones, while the degeneracy among the latter is not yet resolved at this loop order.

\begin{figure}[h]
\centering
\def\svgwidth{11cm}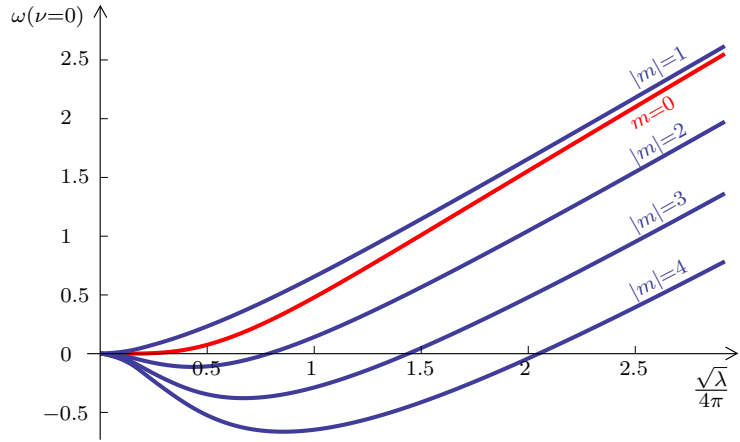
\caption{Plot of the intercept $\omega(\nu=0, m)$ as function of the coupling $g \equiv \sqrt{\lambda}/(4\pi)$ for trajectories $|m|=1, 2, 3, 4$ (in blue) and $m=0$ (in red). The upper curve corresponds to $|m|=1$ and it dominates the Regge limit at all couplings. The following one is associated to the $m=0$ trajectory. Its intercept is extremely small at weak coupling (it kicks in at three loops, $\omega(0, 0) =  4\pi^2\zeta_{3}g^6 +\ldots )$ but is stricly positive at finite $\lambda$. We note that the $m=|1|$ and $m=0$ curves gets closer to one another as the coupling grows, with the splitting among them being $o(1)$ at strong coupling. The remaining curves correspond to $|m|=2, 3, 4$. (The $|m|>4$ trajectories are not represented but would be sitting below them.)
The spacing among the $|m|=1, 2, 3, 4, ...$ curves is of order $O(1)$ at strong coupling.}
\la{BFKLintercepts}
\end{figure}

\subsection{Giant hole, giant fold and Wilson lines}\label{giant}

The strong coupling eigenvalue becomes more interesting if we scale both $\omega$ and $\nu$ with $\sqrt\lambda$. In this regime we find that
\be\begin{aligned}
\omega(\theta)&={\sqrt\lambda\over4\pi}\[{2\sqrt2\cosh{\theta}\over\cosh{(2\theta)}}-\log\({\sqrt2\cosh{\theta}+1\over\sqrt2\cosh{\theta}-1}\)\] \, ,\\
\nu(\theta)&={\sqrt\lambda\over4\pi}\[{2\sqrt2\sinh{\theta}\over\cosh{(2\theta)}}-i\log\({1+i\sqrt2\sinh{\theta}\over1-i\sqrt2\sinh{\theta}}\)\]\,,
\end{aligned}\la{bfklstrong}
\ee
as explained in appendix~\ref{app:strong}. The large value of the intercept (\ref{wmax}) as well as the non-trivial dispersion relation of the rescaled charges (\ref{bfklstrong}) indicate that there should exist a corresponding semi-classical string worldsheet. We can construct it via analytic continuation as follows. 

First, we note that the eigenvalue (\ref{bfklstrong}) is reminiscent of the dispersion relation of the so-called giant hole,
which represents a classical macroscopic spike on top of the GKP string: see equation (1) in ref.~\cite{Dorey:2010iy} with $v_\text{there}=\tanh\theta_\text{here}$.
In fact, the two are related as%
\footnote{
This particular analytic continuation proceeds to the left of the $\theta=0$ cut in the giant hole dispersion relation. 
}
\beq\la{gianttoBFKL}
\begin{aligned}
-\omega(\theta)&=
{1\over2}\[E_\text{giant}(\theta+i\pi/4)-iP_\text{giant}(\theta+i\pi/4)\]\,,  \\
\nu(\theta)&=
 {1\over2}\[P_\text{giant}(\theta+i\pi/4)-iE_\text{giant}(\theta+i\pi/4)\] - \frac{\sqrt{\lambda}}{4}\,. 
 \end{aligned}
\eeq
This is precisely of the form (\ref{BFKLvariables}), revealing that in addition to being analytically related to BFKL the giant hole
is nothing but the ``sister'' of the usual gluonic excitation.

We can be more precise.
The gluonic excitation at strong coupling admits a relativistic
form $E=\sqrt{p^2+2}$ up to momenta $p\sim \sqrt{\lambda}$, where it flattens out and becomes $E_\text{giant}(P_\text{giant})$.
This means that the saddle-point equation $dE/dp = i \frac{\sigma}{\tau}$, for small $\sigma/\tau$, has \emph{two} solutions: one with $p\to 0$ and one with $p\to\infty$.
For the amplitude in the Euclidean region only the first saddle point contributes, while the identification with the sister trajectory
shows that in the Mandelstam region for $3\to 3$ scattering it is the giant hole saddle point which takes over.

The shift of $\theta$ by $i\pi/4$ also has a simple physical interpretation.
It is closely related to the so-called mirror transformation, $\theta\to\theta+i\pi/2$,
which is known to rotate the giant hole solution by 90$^\circ$. This transformation amounts to moving the excitation from one side of the square in fig.~\ref{3to3kinematics} to the next.
The shift by $i\pi/4$ may thus be interpreted as a 45$^\circ$ rotation or ``half mirror transformation,"
reflecting the fact that the BFKL time runs diagonally within the square (c.f. eq.~(\ref{BFKLvariables})).

Since the giant hole classical string solution is known explicitly \cite{Dorey:2010iy}, the analytic continuation in $\theta$ provides us with a simple way of constructing the sister classical solution (or equivalently, the BFKL solution at complex momentum). We simply take $\theta$ imaginary and analytically continue the solution from Lorentzian to Euclidean worldsheet.
In this way we arrive at the solution \cite{Sakai:2009ut,Sakai:2010eh}
\begin{figure}[t]
\centering
\def\svgwidth{16cm}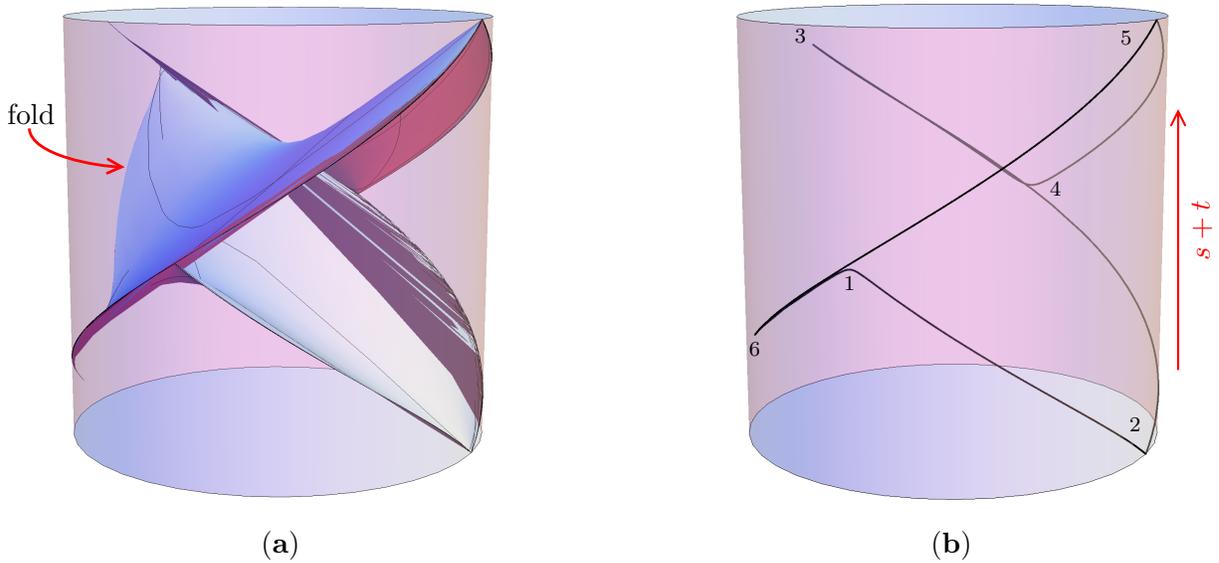
\caption{(a) The strong coupling sister solution (\ref{BFKLstrong}) or, equivalently, the BFKL solution with imaginary momenta $v=1/20$. The classical open string extends in AdS$_3$, represented by the solid cylinder. It ends on a boundary null polygon curve made of six edges
(b). The solution has a single fold in the bulk that is extended between cusps 3 and 6 at the boundary. 
Remarkably the worldsheet is real and Euclidean.
The four short edges can be viewed as part of the reference square, see figure \ref{3to3kinematics}. The two long edges, 5-6 and 2-3, connects antipodal points on the cylinder.
}
\la{stroncouplingsol}
\end{figure}

%
\be\begin{aligned}
X_{-1}\pm X_2&=\pm e^{\pm t}{e^{\Sigma}(\sqrt{1+ v^2}\sinh s-\cosh s)+e^{-\Sigma}( v\mp\sqrt{1+ v^2})\cosh s\over  v\,e^{\Sigma}+e^{-\Sigma}}\\
X_1\pm X_0&=\pm e^{\pm t}{e^{\Sigma}(\sqrt{1+ v^2}\cosh s-\sinh s)+e^{-\Sigma}( v\mp\sqrt{1+ v^2})\sinh s\over  v\,e^{\Sigma}+e^{-\Sigma}}\end{aligned}\la{BFKLstrong}
\ee
where $X_{-1}^2+X_0^2-X_1^2-X_2^2=1$ are the AdS$_3$ embedding coordinates, $\Sigma={ s- v\, t\over\sqrt{1+ v^2}}$ and $ v=-i{\partial E_\text{giant}\over\partial P_\text{giant}}=-i\tanh\theta$ is the (complexified) velocity which we take to be real. (Note that $ v\to-1/ v$ is the mirror transformation prescribed above.) This solution is plotted in figure \ref{stroncouplingsol}~a. It is an Euclidean surface in AdS$_3$ that ends on the boundary null polygon potted in figure \ref{stroncouplingsol}~b. The string has a fold in the bulk, which reminds of the flat space solution shown already in fig.~\ref{fig:foldedstring}~b.
The BFKL time $(s+ t\,)$ translates points between cusps 2 and 5, indicated in figure \ref{stroncouplingsol}~b. At any fixed BFKL time, the open string stretches between cusps 1 and 4 while going through the fold in between. One manifestation of the fold is the phase shift $-\frac{i}{2} \pi(\sigma-\tau)\Gammacusp$ in (\ref{BFKL_2}). Here, $\Gamma_\text{cusp}$ is the string tension pushing the fold in the $(s- t\,)$ direction. In accordance with that factor, we expect the fold to persist at the quantum level, meaning at any value of $\lambda$.


The reader may wonder in what sense this solution describes the expected dipole-dipole scattering:
after all the boundary curve contains only a single null-infinite Wilson line going in each direction (such as the one between cusps $2$ and $3$).
The resolution is simply that we are working on the Wilson loop side of the scattering amplitude -- Wilson loop duality.
In fact the boundary curve has precisely the expected structure shown in fig.~12~b of ref.~\cite{Caron-Huot:2013fea} for that side of the duality,
where two semi-infinite Wilson lines border each infinite one.
We thus expect that the T-dual solution to (\ref{BFKLstrong}) will describe precisely the scattering of two null-infinite adjoint dipoles.

Another surprising feature is that the semi-classical solution resides within an AdS${}_3$ subspace associated with the gauge theory $x^\pm$ plane.
The transverse plane, whose dynamics BFKL is concerned with, appears to be invisible!
The explanation is simply that solutions with imaginary $\nu\neq 0$ and $m\sim 1$ (as we are plotting) arise physically from a saddle point evaluation of the amplitude with fixed $\frac{\tau-\sigma}{\tau+\sigma}=v\sim -i\nu/\sqrt{\lambda}$ (see eq.~(\ref{saddle_points})).
The scaling limit thus implies that $\tau-\sigma\gg 1$ so that the solution unavoidably describes a collinear configuration.
Therefore, the transverse plane is still present, but only as an infinitesimal perturbation of the classical solution.

This interpretation is further supported by the fact that for ${\rm Im\,}\nu>0$, the line 5-6 turns out to become collinear with 4-5 instead of 6-1,
reflecting collinearity in a different channel.
At the special point $\nu=0$,
the solution ``cannot know'' where it should end at the boundary.
Instead of spontaneously breaking the symmetry we find that eq.~(\ref{BFKLstrong}) develops a line of singularity inside the bulk,
but,
as far as we can tell, this is not associated with any singularity in any observable quantity. Another interesting locus is when $\nu$ diverge (corresponding to $v=\theta=0$). At this point the fold reach the AdS boundary and the solution factorize into two squares overlapping along one edge -- corresponding to the fold.

A classical solution related to BFKL in the color-singlet sector (that is, the Pomeron with large Mellin moment $\nu\sim\sqrt{\lambda}$) was described in ref.~\cite{Janik:2013pxa}.
In that reference the case of real $\nu$ was considered, and the solutions are correspondingly complex.
The same would be true in our case; but as the saddle-point equation
 $\frac{\tau-\sigma}{\tau+\sigma}= -i\frac{d\omega}{d\nu}$ shows,
it is the the classical wordsheets with imaginary $\nu$ which correspond to physically observable amplitudes. This is why we have plotted  in fig.~\ref{stroncouplingsol}
the (real) solution for imaginary $\nu$.  One similarity with
ref.~\cite{Janik:2013pxa} is that for $\nu$ real and large enough the solution changes non-analytically.
For us the transition occurs at $\theta=\pm\infty$ where $\nu=\pm\frac{\pi\Gammacusp}{2}$ and eq.~(\ref{bfklstrong}) ceases to apply.
For larger values $|\nu|>\frac{\pi \Gammacusp}{2}$, the trajectory defined by eq.~(\ref{BFKL_exact})
is negative and made of two disconnected branches whose form is identical to (minus) the energy $E_\text{giant}$ of a giant hole, see equation (\ref{matching}). It is depicted in figure~\ref{SCTrajectory} together with the BFKL/sister branch in the form of the Chew-Frautschi plot.

\begin{figure}
\begin{center}
\def\svgwidth{13cm}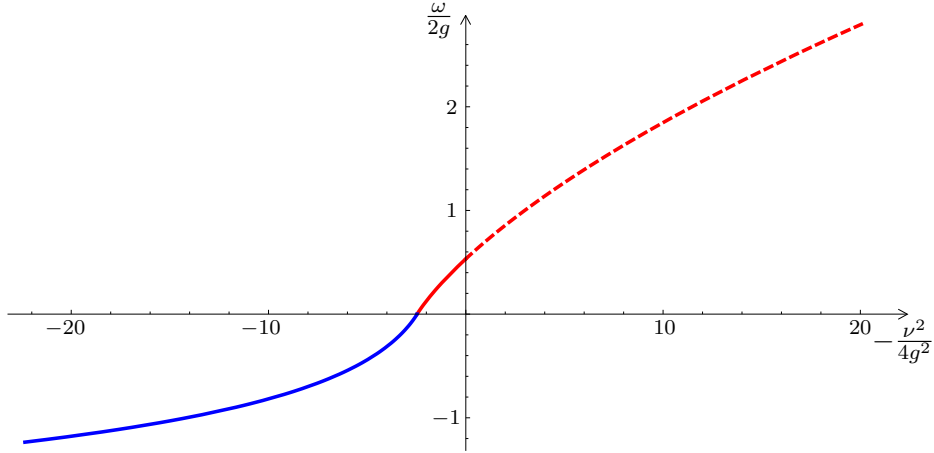
\end{center}
\caption{Chew-Frautschi plot of the universal BFKL/sister trajectory at strong coupling for $\nu \propto \sqrt{\lambda} = 4\pi g$. The sister trajectory (dashed red line) is defined for $\nu^2 < 0$ and stands on the right of this plot. At strong coupling it is described by the giant hole dispersion relation with imaginary $\theta$ between $0$ and $i\frac{\pi}{4}$. The continuation of the sister curve to positive $\nu^2$ defines the BFKL trajectory (solid line). The region that dominates in the Regge limit is described by the positive branch (solid red line) which ends at $\nu^2/(4g^2) = \pi^2/4\simeq 2.47$. Outside of this domain the trajectory is negative (solid blue line) and happens to coincide with (minus) the energy $E$ of a giant hole with momentum $p = \nu\pm \tfrac{\pi}{2}\Gamma_{\textrm{cusp}}$. The color-adjoint trajectory shares three qualitative features with the BFKL Pomeron (a.k.a Reggeized graviton): a linear behavior at small $\nu^2$, a logarithmic behavior $\omega \sim -\Gamma_{\textrm{cusp}}\log{\nu}$ at large positive $\nu^2$, and a square root scaling $\omega^2 \sim -\nu^2$ at large negative $\nu^2$.
}\la{SCTrajectory}
\end{figure}

It is interesting to ask what the solution of ref.~\cite{Janik:2013pxa}, itself an analytic continuation of the GKP string, would look like for imaginary $\nu$.
In fact this was plotted already in fig.~11 of ref.~\cite{Alday:2010ku}, where the connection with BFKL was however not discussed.
With hindsight, we can give a simple interpretation to that solution: in the gauge theory it represents the scattering of two fast dipoles,
exactly as one would expect by naively extrapolating the weak coupling description of BFKL.  More precisely, the classical solution with $\nu\sim\sqrt{\lambda}$
describes the scattering of one very small dipole against a very large one.

\section{Discussion}\la{discussion}

In this paper we have related the collinear OPE and Regge expansion of the 6-particle amplitude through a sequence of two analytic continuations.  Using known expressions for the collinear expansion
this allowed us to derive all-loop expressions for the BFKL eigenvalue (\ref{BFKL_exact}) and impact factor (\ref{bfkl_measure}).

The first continuation, described in section \ref{analyticcontinuationsec}, takes place in the momentum space of the gauge theory scattering amplitude.
Its role is to reach a specific Lorentzian region of high-energy scattering, in which color dipoles can be exchanged (the ``Mandelstam region'').
This continuation has a rather radical effect on the OPE: it replaces the anomalous dimensions which control it $E(p)$, with new, ``sister'' versions $\check E(p)$.

The second continuation, carried out in section \ref{goingthroughcutsec}, is somewhat more magical.
It takes place in the momentum space of the flux tube, which is essentially the Mellin space of the gauge theory. It involves passing through a cut,
where the collinear OPE resides on one side and the BFKL expansion resides on the other side.  We would like to elaborate on this step here.

We should stress that this step is fundamentally nonperturbative. At weak coupling, the cut gets very narrow, and passing through it
entails a resummation of perturbation theory.  It could be carried out explicitly because
$E(p)$ and $\sister{E}(p)$ are governed by the BES equation, which as a matter of fact does resum all orders in perturbation theory.

A good way to phrase the discussion is in terms of the Chew-Frautschi plot shown in fig.~\ref{CFplot}.
On the real axis we place ``energy'' squared, which in conformal field theory is identified with dimension squared: $\Delta^2\equiv(i\nu)^2=-\nu^2$.
On the vertical axis we place the spin $\omega$.
The far right of fig.~\ref{CFplot} then shows the sister trajectory $\omega(\nu=-i\Delta,|m|=1)$, which is just the function (\ref{gauge_sister_dispersion}) for $\ell=1$ and with the change of variables (\ref{BFKLvariables}).
In the left-hand side of the plot, the ``momentum transfer'' $\nu=-i\Delta$ is real and the figure shows the BFKL eigenvalue in its natural domain.\footnote{
The right-hand-side of a Chew-Frautschi plot normally shows discrete physical states
located at positive integer values of the spin. This discreteness is lost here as a result of the subtraction of the
gluon Regge trajectory, which adds a large negative offset to $\omega$.
}

\begin{figure}
\begin{center}
\def\svgwidth{13cm}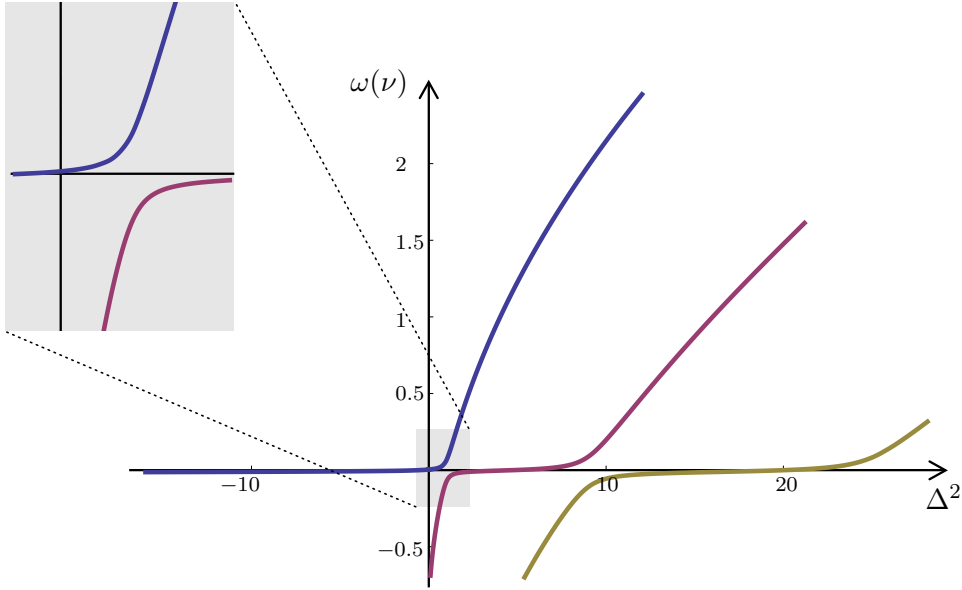
\end{center}
\caption{Chew-Frautschi plot of the BFLK eigenvalue $|m|=1$ and its daughter branches at weak coupling ($g=1/20$). The leftmost curve (painted in blue) is the principal branch of the BFKL eigenvalue.
It is of order $O(g^2)$ on the BFKL side (left) and approximated by $\omega\approx \Delta-|m|$ for $\Delta>|m|$.
In the cross-over region, which is magnified in the top left corner, the eigenvalue is controlled (to leading order at weak coupling) by the square-root formula~(\ref{Zhukowski_like}).
The second curve (painted in purple) is the first of an infinite sequence of daughter branches,
which represent the same trajectory but evaluated on different Riemann sheets.
It is related to energy- or twist-suppressed contributions (see eq.~(\ref{OPE_minus_BFKL})).}\la{CFplot}
\end{figure}

The two branches meet near the point $\Delta\approx |m|$, see inset of fig.~\ref{CFplot}.
It turns out that one can reliably describe the curve near this point using perturbation theory, without knowing anything
about integrability.  This becomes apparent upon rewriting the perturbation theory for the sister dispersion relation $\sister{E}(p)$, in the form
of a spin-dependent dimension $\Delta(\omega)$. These are the variables one conventionally uses to discuss twist operators.
The perturbative expansion develops singularities as $\omega\to 0$, but with only a \emph{single} pole $(g^2/\omega)$ at each loop order.  For example,
\be
  \Delta(\omega) = |m|+\omega - \frac{2g^2}{\omega}+ \OO(g^2\omega^0,g^4/\omega^2,g^6/\omega^3,\ldots)\,. \label{single_log}
\ee
This property can be verified from the explicit expressions in sections~\ref{sec:sister_finite_coupling} and~\ref{goingthroughcutsec}.
Here we would just like to mention that it is analogous to the familiar single-log behavior
of color-singlet twist-two operators near $j\to 1$, which controls small-$x$ parton distribution functions, see for example \cite{Jaroszewicz:1982gr,Catani:1989sg}.  In that context it is well-known that one gets at most one additional power of $\log x$ for each loop order.  For this reason we expect the discussion to be presented here to apply
almost unchanged to the color-singlet sector as well.

The implication of eq.~(\ref{single_log}) is that the joint series in $g^2$ and $1/\omega$ is reliable throughout the region $g^2\ll \omega\ll 1$.
This is interesting because this region \emph{includes} the cusp where the two curves meet. The omitted terms in (\ref{single_log}) are then subdominant,
and the relation can be inverted to obtain \cite{Mueller:1981ex}\footnote{
The paper \cite{Mueller:1981ex} did not explicitly consider the BFKL trajectory $\omega(\Delta)$ but rather the essentially equivalent
time-like anomalous dimension $\gamma_T(\Delta) \equiv \Delta-\omega(\Delta)$.}
\be
 \omega(\Delta) = \frac{\Delta-|m| + \sqrt{(\Delta-|m|)^2+8g^2}}{2} + \OO(g^2)\,. \label{Zhukowski_like}
\ee
The error remains under control for all real $\Delta$ in the range $|\Delta-|m||\ll 1$, and, in particular, the conclusion that
the branch point is of the square-root type is robust.
For $\Delta<|m|$ the two terms mostly cancel and the trajectory is perturbatively small, $\omega(\Delta)\approx 2g^2/(|m|-\Delta)$.
This is in perfect agreement with the first pole of the one-loop BFKL result (see eq.~(\ref{loBFKL})).

This well-known perturbative matching is, strictly speaking, sufficient to unambiguously identify the two trajectories $\sister{E}(p)$ and $\omega_{\rm BFKL}$ as being one and the same. For a more detailed discussion along these lines we refer to \cite{Jaroszewicz:1982gr,Kotikov:2002ab,Korchemsky:2003rc,Costa:2013zra}.

To better understand the matching at the nonperturbative level, we find it useful to consider a physical
observable which can probe the trajectory.
Schematically, the BFKL and OPE expansions of the hexagon can be written in a similar form
\be\begin{aligned}
 \mathcal{W}_{\rm hex}^{\,\circlearrowleft} &\sim
  \int_{\mathcal{C}_{\rm OPE}} d\Delta~ F_{\rm OPE}(\Delta)~e^{(\tau+\sigma)\omega_{\rm OPE}(\Delta)+(\sigma-\tau)\Delta} +\ldots\,
\\   \mbox{and}\quad \mathcal{W}_{\rm hex}^{\,\circlearrowleft} &\sim
  \int_{\mathcal{C}_{\rm BFKL}} d\Delta~ F_{\rm BFKL}(\Delta)~e^{(\tau+\sigma)\omega_{\rm BFKL}(\Delta)+(\sigma-\tau)\Delta}+\ldots\,.
\end{aligned}\label{OPE_BFKL_cartoon}
\ee
The essence of what we have done is that we have equated the two integrands.
This requires explanation: the integration contours and omitted terms are certainly different.
The OPE contour runs vertically to the right of the large cut at $\Delta\approx |m|$, as described below eq.~(\ref{LLog_disc})
The BFKL contour runs vertically also but in a different sheet of the same function.
Therefore the two contours differ by a cut contribution. A useful way to rewrite the difference is to exploit the fact that the cut is of the square-root type.
This leads to the following identity among integration contours:
\be
\def\svgwidth{15.5cm}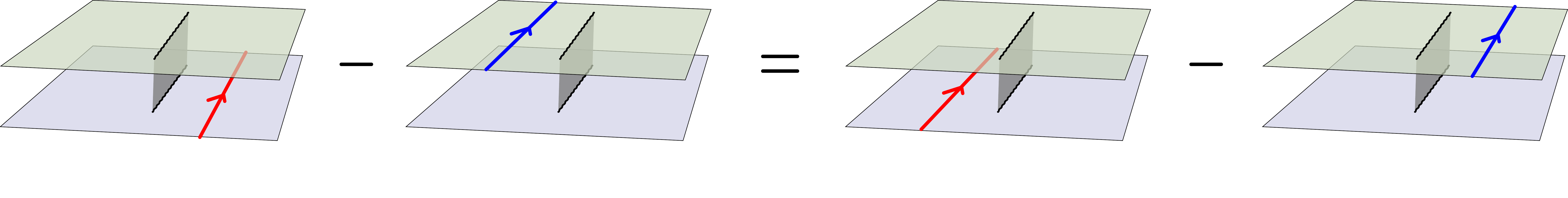
\label{OPE_minus_BFKL}
\ee
Here each term can be associated with one of the four semi-infinite segments appearing in the inset of fig.~\ref{CFplot},  the two terms on the left-hand-side belonging to the blue curve and the other two to the purple curve.
The first integral on the right-hand side is the OPE integral after we move it to the left of the $\Delta$-plane cut;
because the next singularity on this trajectory is displaced by an amount $\delta\omega=\delta \Delta\approx-1$ (see fig.~\ref{CFplot}),
this contribution is suppressed in the high-energy limit by $\sim e^{-2\sigma}$.
The second term is the BFKL integral after moving it to the right of the cut; since the next singularity occurs two units to the right,
$\delta\Delta\approx 2$, this term is suppressed by $e^{2\sigma-2\tau}$. (These estimates hold for small coupling, in general the exponents will vary with the coupling.)
The identity (\ref{OPE_minus_BFKL}) together with the Chew-Frautschi plot thus make explicit that the difference between the two integration contours in eq.~(\ref{OPE_BFKL_cartoon}) is of the same order as the expected size of the omitted terms.

To show that the integrands agree exactly, it thus suffices to show that the omitted terms can be made arbitrarily small.
This is done simply by taking $\sigma,\tau\to\infty$ where the integral will be saturated by a saddle point, as discussed around eq.~(\ref{saddle_points}).
As long as $g^2 \ll \frac{\tau-\sigma}{\tau+\sigma} < 1$, at weak coupling for example, one can check both that the saddle point
will not cross other singularities and that the error terms will be small.
Since the position of the saddle point can be varied freely within an open set, this ensures that the analytic functions entering (\ref{OPE_BFKL_cartoon})
agree within that set, hence that they are analytic continuations of each other.

We wish to stress that this argument does \emph{not} constitute a derivation of the BFKL expansion (\ref{BFKL_2a}) starting from the OPE expansion (\ref{collinear_2}).
What we derived is the \emph{integrand} of eq.~(\ref{BFKL_2a}), \emph{assuming} that the high-energy expansion indeed takes this form.
We can see however what would constitute a derivation of BFKL from the OPE.
For the BFKL formula (\ref{BFKL_2a}) to be correct, as we expect to be the case,
the second integral in eq.~(\ref{OPE_minus_BFKL}) (suppressed by $e^{2\sigma-2\tau}$) must approximately cancel against the first subleading twist correction to the OPE,
that is the next branch to the right in the Chew-Frautschi plot in fig.~\ref{CFplot}.
The remaining $\sim e^{4\sigma-4\tau}$ error must then cancel against the next twist, and so on.
Demonstrating this sequence of cancelations would constitute, in our view, a
rigorous derivation of the BFKL formula from the OPE. It would also be an extremely nontrivial cross-check on the OPE predictions.
Given recent progress on the OPE side, this would appear to be realistically feasible.

We would like to comment also on so-called wrapping corrections, which are known to play a crucial role in the relation between color-singlet twist-two operators and the BFKL Pomeron \cite{Kotikov:2007cy}.
These arise from (planar) Feynman diagrams with the topology of the cylinder, and complicate considerably the integrability
approach, although powerful methods are being developed to deal with this, see refs.~\cite{Gromov:2013pga,Gromov:2014caa}.
These effects are known to disappear in the asymptotic limit of large spin,
\be
 \lim_{\omega_s\to\infty} E^{\textrm{twist-two}}(\omega_s) = 2\Gammacusp \log\omega_s + C + \OO(1/\omega_s)\,. \label{GKP_background}
\ee
This is true as well for small excitations of this background.
For example, the dispersion relation $E(p)$ originates from the anomalous dimension of local twist-three operators,
which are low-lying excitations of the background (\ref{GKP_background}):
\be
 \lim_{\omega_s\to\infty} E^{\textrm{twist-three}}_p(\omega_s) = 2\Gammacusp \log\omega_s + C + E(p) + \OO(1/\omega_s)\,. \label{GKP_twist_3}
\ee
All the objects discussed in this paper arise implicitly from this limit.
The quantum number $p$ is related to the discrete mode number of the excitations, which becomes dense in the large spin limit (see~\cite{Belitsky:2006en} for a one-loop discussion);
we use the subscript $s$ to distinguish the color-singlet quantum numbers from the background-subtracted ones used elsewhere in this paper.

We wish to argue that a similar embedding into gauge invariant operators can be made on the BFKL side.
We begin with the reggeized gluon, which in the planar limit is a single null infinite Wilson line.
This can be naturally viewed as a limit of a very large dipole, wherein a ``spectator'' Wilson
line is inserted far away in order to cancel the charge.
Identifying this distance with the infrared cutoff of the theory, $\mu_{IR}\sim 1/L$,
and using the simple form of dipole eigenfunctions in Mellin space,
$(L+\delta L)^{i\nu_s} \approx L^{i\nu_s} e^{i \delta L \frac{\nu_s}{L}}$, we obtain the map between
the plane wave momentum $q$ of the reggeized gluon and the dipole quantum number: $\nu_s\approx \frac{q}{\mu_{\rm IR}}$.
Now the gluon Regge trajectory in conformal field theory is universally controlled by the cusp anomalous dimension \cite{Korchemskaya:1996je}:
$\omega_g(q)\sim -\frac{\Gammacusp}{2}\log(q^2/\mu_{\rm IR}^2)$.
Viewing the large dipole as two independent Reggeized gluons thus yields the prediction:
\be
 \lim_{\substack{\nu_s\to\infty\\ m\,{\rm fixed}}} \omega^{\rm Pomeron} (\nu_s,m) = -2\Gammacusp \log\frac{|\nu_s|}{2} - C' +\OO(1/\nu_s) \,.
\ee
This can be verified against the two-loop expressions of \cite{Kotikov:2000pm,Kotikov:2002ab}.
(The constant term in the logarithm is somewhat arbitrary, but, curiously, with this choice the finite terms agree to two-loops: $C=C'$.)
By the same argument, a color-adjoint dipole should arise as a low-lying excited state of  the Odderon%
\footnote{
In the traditional BFKL language the Odderon is a (CPT-odd) composite state of three Reggeized gluons.
In the Wilson line description, the Odderon ground state is a color \emph{dipole} with odd angular momentum,
while its excited states are realized as special states of the ``zig-zag" quadrupoles described in ref.~\cite{Caron-Huot:2013fea}.}
at large-$\nu_s$:
\be
 \lim_{\nu_s\to\infty} \omega^{\rm Odderon}_{\nu} (\nu_s,m) \stackrel{?}{=} -2\Gammacusp \log \frac{|\nu_s|}{2} - C'  + \omega(\nu,m) + \OO(1/\nu_s)\,.  \label{excited_Odderon}
\ee
The Odderon is known to have a discrete spectrum of excitations \cite{Derkachov:2002wz,deVega:2002im,Korchemsky:2003rc}, with spacing $\sim g^2$ at weak coupling,
which we expect to become dense at large $\nu_s$.
The adjoint-dipole quantum number $\nu$ used in the main text 
should presumably arise as the mode number of the excitations, in analogy with how $p$ arises in the twist-three case.
We leave the precise identification to future work.

What we have demonstrated in this paper is that there exists a direct connection between the respective asymptotic limits of twist operators and BFKL,
which by-passes wrapping corrections altogether.
This might prove helpful for constructing the
finite-coupling version of BFKL away from the asymptotic limit, including states with any number of Reggeized gluons.

In this respect, we wish to stress the importance of the first step in our procedure, which was the analytic continuation in the \emph{gauge theory} momenta.
No amount of analytic continuation in the spin chain quantum numbers (spin, dimension, mode number, etc.) could
have produced the Odderon excited states out of the twist-3 data $E(p)$.
Rather, we needed to go through the first step, which replaced $E(p)$ by its ``sister'' version $\check{E}(p)$.
We believe that something similar will also be required away from the asymptotic limit.

More precisely, the analogy with flat space string theory discussed in section \ref{sec:overview} leads us to expect
that there should exist distinct, but equally useful, analytic continuations of twist 3 gauge-invariant operators
(see fig.~5 of ref.~\cite{Braun:1999te} for a closely related observation).
Furthermore, taking the analogy seriously, one should not expect the continuation(s) contributing to four-point correlation functions of local operators
to connect with any excited states of the Odderon in the planar limit, in order to be consistent with the selection rules of ref.~\cite{Caron-Huot:2013fea}.
Excited states, rather, may lie on ``sister'' trajectories, which have nonzero OPE coefficients only starting from the 6-point function.
We hope to see in the future whether these predictions are correct or not.
The problem of organizing local operators into analytic trajectories, remains, at the moment, open.

\section{Conclusion}\la{Conclusion}

The main result of this paper are explicit expressions to all loop orders and strong coupling,
for the BFKL eigenvalues and impact factors in the color-adjoint sector, in planar SYM theory.
Let us summarize the two key ideas of our derivation, which are not specific to this theory,
and briefly comment on the physical interpretation of our result.

A first important observation is that scattering amplitude sometimes exhibit very different high-energy behavior in different kinematic configurations.  We have seen that this is reflected in the existence of qualitatively distinct Regge trajectories, some of which having
vanishing coefficient in some kinematic regions.
We noted the analogy with similar phenomenon in the tree-level scattering of open strings in section \ref{sec:overview}, and also noted in the end of section \ref{discussion} its possible connection with known features of local twist three operators.
In perturbation theory, we thus found ``sister'' trajectories related to the canonical ones by a rather simple mathematical transformation,
which effectively reverses the argument of $\psi$-functions (compare eqs.~(\ref{leadingorder_fE}) and~(\ref{sister_1loop})).  We could also extend that map to finite coupling.

A second important ingredient is the use of analyticity as a means of solving crossing equations.
The relation between the collinear OPE expansion and BFKL (see (\ref{Collinear})--(\ref{BFKL})) is an archetypical example of a crossing equation:
we have two expansions with an overlapping radius of convergence but which resum different physics.
This is formally similar to what one deals with, for example, in the modern conformal bootstrap program \cite{Rattazzi:2008pe}.
Naively, one may think that in order to obtain the leading term of one expansion, for example, it is necessary to ``resum'' infinite towers of terms in the other expansion.
Analyticity offers an alternative way to proceed.

An important way in which the OPE and BFKL expansions differ from other, more Euclidean, expansions, is that they are both
controlled by analytic functions.  
When taking an appropriate double scaling limit, where both expansions can be truncated to their leading term,
the crossing equation reduces to the statement that the two analytic functions agree in a certain open set.
This gives rise to the possibility of obtaining the leading term of one expansion from just the \emph{leading} term of the other, via analytic continuation.
This continuation effectively resums infinitely many terms, without explicitly resumming anything!
The results of this paper demonstrate the feasibility of such an approach. We hope that it will find applications in other problems.

In this work we started from known analytic
expressions on one side (the collinear OPE), and we used the crossing equation to obtain the integrand of the BFKL expansion.
But it is interesting to ask how much can be done when one is in the situation of not knowing
either side.  The analytic continuation step is intrinsically nonperturbative, and to carry it out we relied heavily on the BES equations.
The detailed structure of the BES equations was critical in order for the result to even make sense at all:
for example the obtained BFKL eigenvalue had to satisfy appropriate reality, symmetry and analyticity properties.
This is sufficiently nontrivial that one may speculate about the possibility to derive the BES equations, or maybe some part of them,
from just this requirement.  Since these requirements should hold in any (at least, planar) gauge theory,
one may speculate about the existence of a ``universal core'' of the BES equation
which would possibly lead to powerful constraints on twist operators in other gauge theories.

We have discussed only the leading term in the Regge limit, neglecting corrections suppressed by powers of the energy.
We expect however that there should exist a systematic high-energy expansion with a finite radius of convergence.
There are many questions which remain open in this respect.  For example, do subleading Regge trajectories organize
into a discrete or continuous spectrum (Regge cuts)? How are they related to the collinear OPE?  By how much are they suppressed at strong coupling (e.g. what is the gap)?
Another direction is higher-point amplitudes.  While the seven-point amplitude
should be accessible by a straightforward extension of our methods (only dipoles can be exchanged), beginning with eight-point one can exchange states
with three Wilson lines, so-called BKP states. Understanding these states would be an important step towards exposing
the integrable system which underlies BFKL at finite coupling.

Finally, we would like to conclude with what we find to be a rather remarkable corollary of our results:
that a partonic description of high-energy processes, e.g. involving a finite number of point-like constituents,
appears to be justified even in strongly coupled holographic gauge theories.
This is the physical basis for factorizing the 6-particle amplitude in terms of dipole-dipole scattering:
we find it very difficult to conceive how such a factorization could hold if the transverse coordinates of these dipoles
did not literally correspond to the transverse positions of individual charge carriers.
This identification is further supported by the classical string solution described in section \ref{giant}, which ends on the expected boundary Wilson lines.
Such a partonic interpretation does not contradict the results of ref.~\cite{Polchinski:2002jw} for example, wherein
individual point-like partons were not observed in deep-inelastic scattering type experiments; the essential point is that in the Regge limit
one is not attempting to separate the partons away from each other.
One might simply say that partons at strong coupling are strongly correlated with each others hence difficult to separate.
We look forward to see if such a partonic picture at strong coupling, in the restricted sense advocated here,
can be usefully applied to other situations.

\section*{\it Acknowledgements}
B.B. and S.C.H. are thankful to the participants and organizers of the ``Amplitudes in multi-Regge kinematics'' workshop held in Madrid, February 2014,
for a productive atmosphere during which this work was initiated.
The authors wish to thank Andrei Belitsky, Lance Dixon, Yasuyuki Hatsuda, Romuald Janik, Vladimir Kazakov, Gregory Korchemsky, Juan Maldacena, Jo\~ao Penedones, Jeffrey Pennington, Volker Schomerus, Evgeny Sobko, Martin Sprenger, Pedro Vieira for valuable discussions, and Lev Lipatov in particular for pointing out ref.~\cite{Hoyer:1976xw}.
Work of S.C.H was supported in part by the NSF grant PHY-1314311 and by the Marvin L. Goldberger fund. A.S was supported in part by U.S. Department of Energy grant DE- SC0009988.

\appendix

\section{The impact factor and its derivation}\la{app:impact}

\def\MM{\mathbb{M}}
\def\QQ{\mathbb{Q}}

Here we describe our prediction for the BFKL impact factor to all loop orders, and its derivation.
We proceed in two steps, as for the eigenvalue in the main text. First, we perform the analytic continuation in $\sigma$
which leads to the ``sister'' form factor as defined in eq.~(\ref{LLog_disc}).  Then we analytically continue in $u$ through the cut at ${\rm Im}\,u=-\ell/2$.
The reader not interested in these steps may skip to our final result which is recorded in subsection \ref{ssec:result_impact_factor}.

\subsection{Original measure in the Euclidean region}

We begin with the one-particle contribution obtained in refs.~\cite{Basso:2013aha,Basso:2014koa,Basso:2014nra} in the form
\be
F_\ell(\sigma,\tau) =\int\limits_{-\infty}^{+\infty} \frac{du}{2\pi} \,\mu_\ell(u) \,e^{i\sigma p_\ell(u)-\tau E_\ell(u)}\,,
\ee
with
\be\begin{aligned}
 \mu_\ell(u) &= \frac{g^2 \Gamma(1+\frac{\ell}{2}+iu)\Gamma(1+\frac{\ell}{2}-iu)}{\Gamma(\ell)(x^{[+\ell]}x^{[-\ell]} -g^2)\sqrt{(x^{[+\ell]}x^{[+\ell]}-g^2)(x^{[-\ell]}x^{[-\ell]}-g^2)}}
\\ &\hspace{-2cm}\times \exp\left[\int\limits_0^\infty \frac{dt}{t}\big(J_0(2gt)-1\big) \frac{2\cos(ut)e^{-\ell t/2}-J_0(2gt)-1}{e^t-1}
+f^\ell_3(u)-f^\ell_4(u)\right]\,.
\end{aligned} \label{form_factor}
\ee
Here $x(u)=\frac12\big(u+\sqrt{u^2-4g^2}\big)$ and $x^{[\pm\ell]}(u)=x(u\pm\frac{i}{2}\ell)$.
To describe the functions $f_{3,4}^\ell$ we introduce the matrices:
\be
 \mathbb{K}_{ij} = 2j(-1)^{j(i{+}1)} \int\limits_0^\infty \frac{dt}{t} \frac{J_i(2gt)J_j(2gt)}{e^t-1}\,,
 \qquad
 \MM=(1+\mathbb{K})^{-1}\,,
 \qquad
 \QQ_{ij}= \delta_{ij}(-1)^{i+1}i\,. \label{defQM}
\ee
\def\deltaeven{\delta^{\rm even}}
\def\deltaodd{\delta^{\rm odd}}
We then form the vectors
\be\begin{aligned}
\kappa_j^\ell &\equiv -\int\limits_0^\infty \frac{dt}{t}\frac{J_j(2gt)\left(e^{t\deltaeven_j}\cos(ut)e^{-\ell t/2}-J_0(2gt)\right)}{e^t-1}\,,
\\ \tilde\kappa_j^\ell &\equiv (-1)^{j{+}1} \int\limits_0^\infty \frac{dt}{t}\frac{J_j(2gt)e^{t\deltaodd_j}\sin(ut)e^{-\ell t/2}}{e^t-1}\,,  \label{kappa_original}
\end{aligned}\ee
where $\delta_i^{{\rm even,odd}}\equiv \frac12[1\pm (-1)^i]$.
The symmetrical matrix $\QQ\cdot \MM$, loosely speaking, inverts the BES kernel. Although infinite-dimensional,
in practice its coefficients are suppressed at large order (for example, $K_{ij}\propto g^{i+j}$ at weak coupling),
allowing it to be truncated to a finite size.
The functions $f_{3,4}^\ell$ are then given as
\be
f_3^\ell(u) = 2\tilde \kappa^\ell(u)\cdot\QQ\cdot\MM\cdot\tilde\kappa^\ell(u)\,,
\qquad f_4^\ell(u) = 2\kappa^\ell(u)\cdot\QQ\cdot\MM\cdot\kappa^\ell(u)\,. \label{def_f34}
\ee
It will be useful to note that $\Gammacusp=4g^2(\QQ\cdot\MM)_{11}$ and that
\be
 E_\ell(u) = \ell + 4g\big(\QQ\cdot\MM\cdot\kappa^\ell(u)\big)_1\,,\qquad
 p_\ell(u) = 2u - 4g\big(\QQ\cdot\MM\cdot\tilde \kappa^\ell(u)\big)_1\,. \label{dispersion_in_appendix}
\ee

\subsection{Continuation in $\sigma$ and the sister measure}

To obtain the ``sister'' measure which governs the $\sigma$-space discontinuity, we follow the procedure of section \ref{analyticcontinuationsec}.
Since the procedure is friendly with $\sigma$-space convolutions, it can be applied to each factor separately.
It amounts to the following: for each factor which has an infinite sequence of singularities in the lower-half $u$-plane,
we flip the sign of the Fourier-conjugate $t$ variable.
For the $\Gamma$-functions the result can be deduced using that $\psi(x)=\partial_x\log\Gamma(x)$,
together with the known rule for the $\psi$ function in eq.~(\ref{sister_1loop}): $\psi(a-iu)\mapsto \psi(1-a+iu)+i\pi$. This gives
\be
 \Gamma(1+\ell/2-iu) \mapsto \frac{C e^{\pi u}}{\Gamma\left(-\ell/2+iu\right)} = \Gamma(1+\ell/2-iu) \times -2e^{\pi u+i\pi \ell/2}\sinh(\pi u+i\pi\ell/2)\,.
\ee
The $u$-independent constant $C$ was fixed by requiring that the function be unmodified at large negative $u$, see comment below
eq.~(\ref{expansion24}).
Rational factors represent convolutions against pure power laws and thus go simply to themselves, and thus so do the algebraic factors in eq.~(\ref{form_factor}) (as is clear from e.g. series expanding them at weak coupling).

For the more complicated functions $f_{3,4}$, we proceed by flipping the sign of $t$
in the $e^{+iut}$ terms in eqs.~(\ref{kappa_original}):
\be\begin{aligned}
\sister{\kappa}_j^\ell &\equiv -\int\limits_0^\infty \frac{dt}{t}\frac{J_j(2gt)}{2(e^t-1)}
 \left[e^{t\deltaeven_j}e^{-iut-\ell t/2}  - (-1)^{j} e^{t\deltaodd_j}e^{-iut+\ell t/2}-2J_0(2gt)\right]\,,
\\ \sister{\tilde\kappa}_j^\ell & \equiv  \int\limits_0^\infty \frac{dt}{t}\frac{J_j(2gt)}{2i(e^t-1)}
 \left[e^{t\deltaeven_j}e^{-iut+\ell t/2}+ (-1)^{j}e^{t\deltaodd_j}e^{-iut-\ell t/2}\right]\,.
\end{aligned}\label{sister_kappa} \ee
In defining these ``sister'' sources we have omitted the anomalies (\ref{anomaly}), which affect only the $j=1$ terms: $\delta \sister{\kappa}_1^\ell =i\pi g/2$, $\delta \sister{\tilde\kappa}_1^\ell =-\pi g/2$.
These anomalies are simple to deal with thanks to the identities (\ref{dispersion_in_appendix}).
Including them, we find that $f_{3,4}^\ell$ continue to the same expressions (\ref{def_f34}) with the $\kappa$'s replaced by their
sister versions (\ref{sister_kappa}), plus some simple shifts:
\be
 f_3^\ell(u)\mapsto \sister{f}_3^\ell(u)+ \frac{\pi}{2}(\sister{p}_\ell(u)-2u)-\frac{\pi^2}{8}\Gammacusp\,,\qquad
 f_4^\ell(u)\mapsto \sister{f}_4^\ell(u)+ \frac{i\pi}{2}(\sister{E}_\ell(u)-\ell)+\frac{\pi^2}{8}\Gammacusp\,.\nonumber
\ee
These shifts neatly cancel various phases coming from the shifts $\sigma\to \sigma+i\pi/2$ and $\tau\to\tau-i\pi/2$.
Collecting these ingredients, we obtain the discontinuity of the amplitude in the $3\to 3$ region in the form (\ref{LLog_disc}), with
the ``sister'' measure:
\be\begin{aligned}
 \mu^\updownarrow_\ell(u) &=\frac{-2\pi i (-1)^\ell g^2 \Gamma(1+\frac{\ell}{2}+iu)}
 {\Gamma(\ell)\Gamma(-\frac{\ell}{2}+iu) (x^{[+\ell]}x^{[-\ell]} -g^2)\sqrt{(x^{[+\ell]}x^{[+\ell]}-g^2)(x^{[-\ell]}x^{[-\ell]}-g^2)}}e^{-\frac{\pi^2}{4}\Gammacusp}
\\ &\hspace{-1.5cm}\times\exp\left[\int\limits_0^\infty \frac{dt}{t}\big(J_0(2gt)-1\big) \frac{e^{-iut}(e^{-\ell t/2}-e^{(\ell/2+1)t})-J_0(2gt)-1}{e^t-1}+\sister{f}_3^\ell(u)-\sister{f}^\ell_4(u)\right]\,.
\end{aligned}
\label{sister_form_factor}
\ee
This is defined for ${\rm Im}\,u<-\ell/2$. For $\ell=1$, we have compared this extensively against the discontinuity
of the analytic expressions given in refs.~\cite{Basso:2013aha,Papathanasiou:2013uoa}.

\subsection{Passing through the cut}\label{pass-cut}

The form factor (\ref{sister_form_factor}) describes the hexagon amplitude after the analytic continuation in $\sigma$
which takes us to the $3\to 3$ region, but we have not yet taken the Regge limit.
To reach the BFKL region it remains to analytically continue through the cut at $u=-i\ell/2\pm 2g$.
The identities which allow us to do so are a simple generalization of eq.~(\ref{BES}):
\be
 \left\{ \begin{array}{l}
 \left(\QQ\cdot\MM\cdot\kappa^0(u)\right)_j \\
\left(\QQ\cdot\MM\cdot\tilde\kappa^0(u)\right)_j\end{array}\right\}
= \frac12\left[\left(\frac{x}{g}\right)^j+\left(\frac{x}{g}\right)^{-j}\right]\times
\left\{\begin{array}{l}
(-1)^{j/2}\deltaeven_j \\
(-1)^{\frac{j-1}{2}}\deltaodd_j\end{array}\right\}\,, \label{BES_appendix}
\ee
which again hold precisely on the cut, $-2g<u<2g$. By adding suitable multiples of $\kappa_0(u+i\ell/2)$
or $\tilde\kappa_0(u+i\ell/2)$ to eqs.~(\ref{sister_kappa}) as done in eqs.~(\ref{goingthrough}),
all functions can be easily continued across the cut.
The comparatively simpler terms produced by the right-hand sides may then be dealt with using the identities
\def\ellhalf{\frac{\ell}{2}}
\be
 \sum_{j=1}^\infty J_j(2gt)\left[\left(\frac{x}{g}\right)^j+\left(\frac{x}{g}\right)^{-j}\right] \times
 \left\{\begin{array}{l} 
 (-1)^{j/2}\deltaeven_j \\
(-1)^{\frac{j-1}{2}}\deltaodd_j\end{array}\right\}
= \left\{\begin{array}{c}
 \cos(ut)-J_0(2gt)\\ \sin(ut) \end{array}\right\}\,.
\ee
The resulting integrals may then be done analytically, the point being that they do not involve
the complicated matrix $\QQ\cdot\MM$.  After some straightforward algebra we get to use the following representation of unity:
\be\begin{aligned}
 1&=
  \frac{-\Gamma(1+\frac{\ell}{2}+iu)}{g\Gamma(1+\ell)\Gamma(-\frac{\ell}{2}+iu)}
\times
\left(\frac{u+i\sqrt{4g^2-u^2}}{u-i\sqrt{4g^2-u^2}}\right)^{[+\ell]}
\\&\hspace{-0.5cm}
\times\exp\left[\int_0^\infty\frac{dt}{t}\left\{
e^{-iut+\ell t/2}\big(1-2J_0(2gt)\big)+J_0(2gt)+ \frac{e^{-\ell t}+e^{-iut+\ell t/2}-e^{-iut-\ell t/2}-1}{e^t-1}\right\}\right]\,.
\end{aligned}\nonumber\ee
The parenthesis with the $[+\ell]$ superscript is evaluated with $u$ shifted by $i\ell/2$.
The identity is valid inside the cut and using the positive branch for the square root, e.g. for $u+i\ell/2$ real and smaller in absolute value than $2g$;
it is similar to the algebraic identities used in appendix~C.4 of ref.~\cite{Basso:2014koa}.
Thanks to it, the $\Gamma$ functions disappear from eq.~(\ref{sister_form_factor}) and the expression
simplifies dramatically after going through the cut.

\subsection{Final result}
\label{ssec:result_impact_factor}

\def\m{m}
\def\A{A}

Our finite-coupling prediction for the measure is:
\be
\mu^{\rm BFKL}_{\m}(u) = \frac{g^2(x^{[+\m]}x^{[-\m]}-g^2)}{x^{[+\m]}x^{[-\m]}\sqrt{\big(x^{[+\m]}x^{[+\m]}-g^2\big)\big(x^{[-\m]}x^{[-\m]}-g^2\big)}}\times e^{\A+2f_{\m,3}(u)-2f_{\m,4}(u)}\, . \label{bfkl_measure}
\ee
As before, $x^{[\pm \m]} = x(u\pm i\tfrac{\m}{2})$ and $x(u) = \tfrac{1}{2}(u+\sqrt{u^2-4g^2})$.
This measure enters the remainder function in the $3\to 3$ kinematics through
\be
 \mathcal{R}_{\rm hex}^{3\to 3}e^{-i\delta} =
 -2\pi i \sum_{m=-\infty}^{\infty} (-1)^m \left(\frac{w}{w^*}\right)^{\frac{m}{2}} \int\limits_{-\infty}^\infty \frac{du}{2\pi}
 \,\mu^{\rm BFKL}_{\m}(u)
 \,|w|^{i\nu_{\m}(u)} \left(\sqrt{|u_2u_3|}\right)^{-\omega_{\m}(u)}\, ,
\ee
with $\delta=\frac{\pi\Gammacusp}{2}\log\frac{|w|}{|1+w|^2}$. In the Regge limit the finite cross-ratio $w$ is related to the Mandelstam invariants through $ww^*=u_3/u_2$ and $(1+w)(1+w^*)=(1-u_1)/u_2$,
with the $u_i$ defined in eq.~(\ref{crossratios}).

The most complicated quantities involve the source terms $\kappa_{\m}(u), \tilde{\kappa}_{\m}(u)$ for the BFKL mode $(u, \m)$:
\be\begin{aligned}
 \kappa_{\m,j} &= 
  -\int\limits_0^\infty \frac{dt}{t}\frac{J_i(2gt)}{e^t-1}\left( \frac{e^{t\deltaeven_j}-(-1)^j e^{t\deltaodd_j}}{2}\cos(ut)e^{-\m t/2}-J_0(2gt)\right)\,,\\
 \tilde\kappa_{\m,j} &= -\int\limits_0^\infty \frac{dt}{t}\frac{J_i(2gt)}{e^t-1}\frac{e^{t\deltaeven_j}+(-1)^j e^{t\deltaodd_j}}{2}\sin(ut)e^{-\m t/2}\,.
\end{aligned}\label{kappa_BFKL}\ee
Then
\be\label{f34}
f_{\m,3}(u) = 2\tilde{\kappa}_{\m}(u)\cdot \mathbb{Q}\cdot \mathbb{M}\cdot \tilde{\kappa}_{\m}(u)\, , \qquad f_{\m,4}(u) = 2\kappa_{\m}(u)\cdot \mathbb{Q}\cdot \mathbb{M}\cdot \kappa_{\m}(u)\, ,
\ee
with the $\mathbb{Q}$ and $\mathbb{M}$ matrices as defined in eq.~(\ref{defQM}).
Finally, the constant $\A$ (independent on $(u, \m)$) is given as
\be
\A = 2\int_0^{\infty}\frac{dt}{t}\frac{1-J_0(2gt)^2}{e^t-1} - \frac{\pi^2}{4}\Gamma_{\textrm{cusp}} \,.
\ee
For completeness, we record the corresponding prediction for the $2\to 4$ region, which differs by a simple phase \cite{Bartels:2010tx}:
\be
 \mathcal{R}_{\rm hex}^{2\to 4}e^{i\delta} =
 2\pi i \sum_{m=-\infty}^{\infty} (-1)^m \left(\frac{w}{w^*}\right)^{\frac{m}{2}} \int\limits_{-\infty}^\infty \frac{du}{2\pi}
 \,\mu^{\rm BFKL}_{\m}(u)
 \,|w|^{i\nu_{\m}(u)} e^{-i\pi\omega_{\m}(u)} \left(\sqrt{|u_2u_3|}\right)^{-\omega_{\m}(u)}\,.\nonumber
\ee
One can also show that the BFKL eigenvalues (\ref{BFKL_exact}) can be expressed in terms of the same sources,
\be
 -\omega_{\m}(u) = 4g\big(\QQ\cdot\MM\cdot\kappa_{\m}(u)\big)_1\,,\qquad
 \nu_{\m}(\ell) = 2u -4g\big(\QQ\cdot\MM\cdot\tilde{\kappa}_{\m}(u)\big)_1\,. \label{bfkl_disk_appendix}
\ee
This completes the description of our predictions for the BFKL expansion.

Its expansions at weak and strong coupling
are described in the next appendices.
As argued in the main text, even though our derivations are performed for all $m\neq 0$, there is good evidence
that the same formula applies to the $m=0$ case as well. The $u$-plane contour is more subtle in this case, and is described
in appendix~\ref{meq0appendix}. This contour implements a prescription $1/(\nu^2-\frac{\pi^2}{4}\Gammacusp^2\mp i0)$ for the real-axis poles at $(\nu,m)=(\pm \frac{\pi\Gammacusp}{2},0)$,
the sign depending on whether one considers the $3\to 3$ or $2\to 4$ region, and which fully accounts for the omitted ``$\cos\omega_{ab}$'' term between
our formulas and those of ref.~\cite{Bartels:2010tx}.

\subsection{NMHV form factor}

The measure (or impact factor) for the NMHV 6-gluon amplitude can also be straightforwardly obtained from its collinear cousin.%
\footnote{The 6-gluon amplitude of interest here is dual to the ``gluonic" NMHV super-Wilson-loop component considered in~\cite{Basso:2013aha,Basso:2014nra}.}
On the OPE side, the prescription for passing from the MHV to the NMHV amplitude
is to multiply the measure for the gluonic flux tube excitation by the so called NMHV form factor~\cite{Basso:2013aha,Basso:2014nra}%
\footnote{One should distinguish between positive and negative angular momenta in the sum~(\ref{collinear_2}) by writing $\sum_{\ell = \pm 1, \ldots} e^{i\ell \phi}\times (\ldots)$ in place of $\sum_{\ell\geqslant 1}2\cos{(\ell \phi)}\times (\ldots)$.}
\beq\label{NMHVff}
h_{\ell}(u) =\left( \frac{x^{[+\ell]}(u)x^{[-\ell]}(u)}{g^2}\right)^{\textrm{sign}\, \ell}\, ,
\eeq
whose effect at weak coupling is to suppress the contribution from the modes with $\ell<0$ and, simultaneously, enhance the role of the positive modes.

The transformation of the form factor~(\ref{NMHVff}) to its BFKL counterpart is easily achieved. The first step of the procedure is just trivial; the sister version of~(\ref{NMHVff}) being precisely itself, since it only involves algebraic quantities at weak coupling which map to themselves, as mentioned earlier. Only the second step does something to~(\ref{NMHVff}). Because this one corresponds to crossing the cut in the lower half plane, it turns $x^{[+|\ell|]}$ into $g^2/x^{[+|\ell |]}$ and leaves $x^{[-|\ell |]}$ invariant. The BFKL measure for the NMHV amplitude is therefore given by
\beq\label{NMHVff2}
\mu^{\textrm{NMHV}}_{m}(u) = \frac{x(u-\frac{im}{2})}{x(u+\frac{im}{2})}\times \mu^{\textrm{MHV}}_{m}(u)\, ,
\eeq
where $\mu^{\textrm{MHV}}_{m}(u)$ is the BFKL measure constructed in the previous subsection. We immediately notice that the BFKL outcome~(\ref{NMHVff2}) is more analytic as a function of the angular momentum $m$ than its collinear relative~(\ref{NMHVff}) was as a function of $\ell$. This feature supports that our above analysis of the $m\neq 0$ cases can be lifted to $m=0$ by simply setting $m=0$ in~(\ref{NMHVff2}). Interestingly, the NMHV and MHV measures are then predicted to be identical when $m=0$.

Our finding~(\ref{NMHVff2}) is easily seen to reproduce the recipe of ref.~\cite{Lipatov:2012gk} for converting among MHV and NMHV amplitudes (in the BFKL regime) to leading order at weak coupling (i.e.~when $x(u)\rightarrow u$). It also reproduces the higher-loop results obtained through the hexagon function bootstrap up to 3 loops in ref.~\cite{Dixon:2014iba}.%
\footnote{We are very thankful to Lance Dixon and Matt von Hippel for sharing their findings with us prior to publication.}

\section{Expansion at weak coupling}\label{app:expansion}

The leading order expression for the eigenvalues at weak coupling was given in~(\ref{loBFKL}).
It is given in terms of the function $K(t)$, to be obtained by solving the BES equation.
The solution, to order $g^6$, is given by
the Bessel expansion (\ref{Bessel_expansion}) with the coefficients
\be\label{gamma-coeff}
\gamma_2= 4g^4\zeta_3 -g^6\left(8\zeta_2\zeta_3+40\zeta_5\right)+\OO(g^{8})\,,\quad
\gamma_3= -4\zeta_4g^5+\OO(g^7)\,,\quad \gamma_4 = 4g^6\zeta_5+\OO(g^{8})\,,
\ee
with $\gamma_1$ given already in eq.~(\ref{cusp}).
Using the integrals\footnote{These formulas will give the correct result for convergent combinations of integrals
when $n=0$ provided one sets $\zeta_1\to -\psi(1)$.}
\be
 \int_0^\infty \frac{t^n dt}{e^t-1}e^{iut} = (-1)^{n+1} \psi^{(n)}(1-iu)\,, \qquad \int_0^\infty \frac{t^n\,dt}{e^t-1} = n!\,\zeta_{n{+}1}\,,
\ee
one can then readily evaluate the eigenvalue (\ref{loBFKL}) to order $g^6$.

An equivalent, but perhaps more systematic, way to proceed is to use the $\QQ\cdot \MM$ matrix defined in eq.~(\ref{defQM}),
together with eqs.~(\ref{bfkl_disk_appendix}). This is the method implemented in the attached notebook.
Its main advantage is that it applies uniformly both to the eigenvalue and impact factor (\ref{bfkl_disk_appendix}).

We simplify the notation by introducing the $E,V,N,D$ alphabet as in ref.~\cite{Dixon:2012yy}:
\be\begin{aligned}\label{BuildingBlocks}
 E&=-\frac12\frac{|m|}{u^2+\frac{m^2}{4}}+\psi\left(1+iu+\frac{|m|}{2}\right) +\psi\left(1-iu+\frac{|m|}{2}\right)-2\psi(1)\,,
\\
V&\equiv \frac{iu}{u^2+\frac{m^2}{4}}\,,\qquad N = \frac{m}{u^2+\frac{m^2}{4}}\,,
\qquad D_u=-i\partial/\partial u\,.
\end{aligned}\ee
Then our result for $\nu$ and $\omega$, up to order $g^6$, can be written
\be
\begin{aligned}\label{WeakCouplingApp}
\nu = & \, \, 2u +2ig^2 V-ig^4\left(D^2V+4\zeta_2V\right)\\&
+ig^6\left(\frac{1}{6}D^4V+2\zeta_2 D^2V-4\zeta_3 DE+44\zeta_4 V\right) 
+\OO(g^8)\,, \\
-\omega = &\, \,  2g^2 E-g^4\left(D^2E+4\zeta_2E+12\zeta_3\right)\\
&+g^6\Bigg(\frac{1}{6}D^4E+2\zeta_2D^2E+4\zeta_3DV+44\zeta_4E+80\zeta_5+16\zeta_2\zeta_3\Bigg)
+\OO(g^8)\,.
\end{aligned}
\ee
Here all functions are evaluated at rapidity $u$ and angular momentum $m$ (as specified in~(\ref{BuildingBlocks})). One would notice the parallelism between these two expressions upon exchanging the role of $E$ and $V$, up to ($u$-independent) constants. It is observed to work at any loop order. The expressions are always linear combinations of $V$ and $E$ and derivatives thereof with overall weight fixed by the loop order (with $E, V, N, D$ having weight $1$ by definition). All coefficients are
products of simple Riemann $\zeta$ values, $\zeta_{z} = \zeta(z)$, to any loop order. 

Equipped with~(\ref{WeakCouplingApp}) it is now straightforward to derive the eigenvalues $\omega$ as a function of $\nu$. Eliminating the Bethe rapidity $u$ we obtain
\be
\begin{aligned}\label{omeganu}
-\omega  =&\, \,  2g^2 E - g^4\left(D^2E-2VDE+4\zeta_2 E+12\zeta_3\right) \\
&+g^6\Bigg(\frac{1}{6}D^4 E-VD^3E+(V^2+2\zeta_2)D^2E-(N^2+8\zeta_2)VDE \\
&\qquad \qquad +(4V^2+N^2)\zeta_3 + 44\zeta_4 E+16(\zeta_2\zeta_3+5\zeta_5)\Bigg) + g^8 E_{N^3LLA} + \mathcal{O}(g^{10})\, .
\end{aligned}
\ee
The terms up to order $g^6$ follow from the preceding equations, while the N$^3$LLA term, which we obtained by working out the expansion for $\omega$ to one order higher, is:
\be
\begin{aligned}\label{N3LLA}
E_{N^3LLA} &= -\frac{1}{72}D^6E +\frac{1}{6} V\,D^5E-\left(\frac{1}{2}V^2+\frac13\zeta_2\right)D^4E+
\left(\frac13V^2+\frac12N^2+4\zeta_2\right)V\,D^3E
\\
& -\left(N^2V^2+6\zeta_2V^2+24\zeta_4\right)D^2E -4\zeta_3(DE)^2- \left(438\zeta_6+16\zeta_3^2\right)E
\\
&+\left( \frac58N^4+\frac32N^2V^2+\big(3N^2-4V^2\big)\zeta_2+108\zeta_4\right)V\,DE 
\\
& -\zeta_3N^2\left(\frac12N^2+6V^2\right)-(10\zeta_5+2\zeta_2\zeta_3)(N^2+4V^2)-168\zeta_4\zeta_3-80\zeta_2\zeta_5-700\zeta_7\,.
\end{aligned}
\ee
All functions are now evaluated at $\nu/2$ (which is often referred to as $\nu$ in the literature). It is easy to see that our result~(\ref{omeganu}) precisely matches the known expressions for the BFKL eigenvalues up to N$^2$LLA, see~\cite{Bartels:2008ce,Bartels:2008sc,Fadin:2011we,Dixon:2012yy,Dixon:2014voa}.
The formula~(\ref{N3LLA}) is a prediction for the next order.

In order to compare the measure (\ref{bfkl_measure}) with the literature,
one needs to convert the $u$ integral to a $\nu$-integral,
which comes at the cost of a Jacobian. Furthermore,
the $\nu$-space version of the lowest order result,
$\mu^{\rm BFKL}_{\m}(u) = \frac{g^2}{u^2+\frac{m^2}{4}}+\OO(g^4)$ with $\nu= 2u+\OO(g^2)$, is usually factored out.
To compare with the literature we thus define the function $\Phi$ through
\be
\mu^{\rm BFKL}_{\m}(u) = 2g^2\frac{d\nu}{du}\frac{\Phi(\nu, \m)}{\nu^2+m^2},\qquad \Phi(\nu,m) = 1+\mathcal{O}(g^2)\, .
\ee
Apart from our different convention for the variable $\nu$, our function $\Phi$ is the same as $\Phi_{\rm reg}$ in refs.~\cite{Dixon:2012yy}: $ \Phi_{\textrm{reg}}(\nu, \m) = \Phi(2\nu, m).$  At the lowest nontrivial order,
\be
 \Phi(\nu,m) = 1- g^2\left( E^2+\frac34N^2+2\zeta_2\right) + \OO(g^4)\,.
\ee
This agrees with ref.~\cite{Dixon:2012yy} (see also ref.~\cite{Lipatov:2010ad}).  We have performed the expansion to higher orders and found perfect agreement
with the four-loop (i.e., $\OO(g^6)$) result of ref.~\cite{Dixon:2012yy,Dixon:2014voa}. Since the formulas are somewhat lengthy we do not reproduce them here.
We have also obtained new predictions to order $g^8$ and $g^{10}$.
These are attached in an ancillary file with the arXiv submission of this paper,
together with a Mathematica notebook which implements the formula to arbitrary loop order.

\section{BFKL eigenvalue at $m=0$}\la{meq0appendix}

A peculiarity of the $m=0$ eigenvalue~(\ref{BFKL_exact}) is that it displays square-root singularities at $u = \pm 2g$ in rapidity space. This contrasts with the $|m|>0$ eigenvalues which are analytical within the strip $|\textrm{Im} \,  u| < \tfrac{1}{2}|m|$ and for which the full real $u$ line maps into the full real $\nu$ line. What happens for $m=0$ is that the integral representation~(\ref{BFKL_exact}) only holds for $u^2 > (2g)^2$ and then only covers the domain $|\nu|>\nu_0$ with $\nu_0 = \nu(u=2g)$. The eigenvalue $\omega(\nu, m=0)$ itself is smooth at $\nu=\nu_0$ -- and more generally for any real $\nu$ -- but in order to explore the region where $|\nu| < \nu_0$ we should analytically continue our representation to the second rapidity sheet by going through the cut. This is what is explained in this appendix.

We start by noting that our formula~(\ref{BFKL_exact}) for the $m=0$ eigenvalue can be equivalently written as
\be
\begin{aligned}\label{rep-cut}
&\omega = \int\limits_0^{\infty}\frac{dt}{t}\(\frac{\gamma_{-}(2gt)\cos{(ut)}}{e^{t}-1}+K(t)\) -\frac{1}{2}\int\limits_0^{\infty}\frac{dt}{t}\gamma_{+}(2gt)e^{\pm iut}\, ,\\
&\nu = 2u-\int\limits_0^{\infty}\frac{dt}{t}\frac{\gamma_{+}(2gt)\sin{(ut)}}{e^{t}-1} \pm \frac{i}{2}\int\limits_0^{\infty}\frac{dt}{t}\gamma_{-}(2gt)e^{\pm iut} \, ,
\end{aligned}
\ee
where the functions $\gamma_{\pm}(2gt)$ re-sum the series over the Bessels in~(\ref{Bessel_expansion}), see footnote~\ref{gamma-footnote}. The two representations in~(\ref{rep-cut}) converge for $0<\textrm{Im} \, u<1$ and $-1<\textrm{Im} \, u < 0$ in the $\pm$ case respectively. They merge along the real line if $u^2 > (2g)^2$ where they both equate~(\ref{BFKL_exact}) with $m=0$. This is because the Fourier transform of the even/odd function $\gamma_{+/-}(2gt)$ vanishes outside this interval,
\beq\label{out-form}
\int\limits_0^{\infty}\frac{dt}{t}\gamma_{-}(2gt)\cos{(ut)} = \int\limits_0^{\infty}\frac{dt}{t}\gamma_{+}(2gt)\sin{(ut)} = 0\, , \qquad u^2>(2g)^2\, .
\eeq
The two representations~(\ref{rep-cut}) would, however, disagree if evaluated inside the interval $u^2 < (2g)^2$. This is a manifestation of the presence of the aforementioned square-root cut: one representation is approaching the cut from above and the other one from below. As a result neither $\omega$ nor $\nu$ are real for $u$ within this interval. Instead, if we want to decrease further the value of $|\nu|$, we should continue through this cut and enter the second sheet.
\begin{figure}[t]

\def\svgwidth{13cm}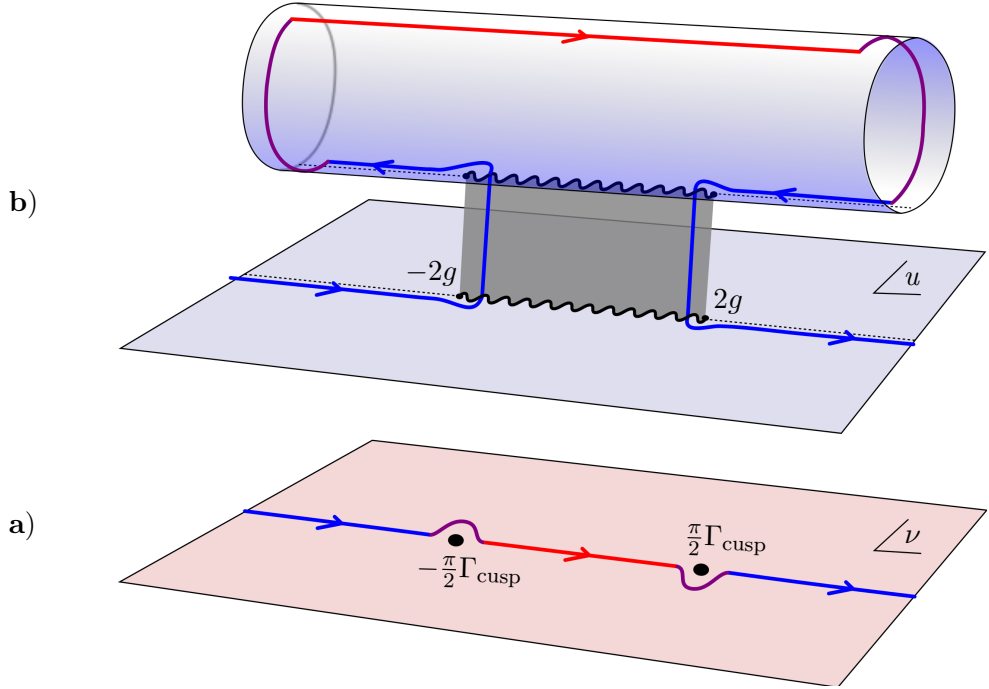
\centering
\caption{Contour of integration for the BFKL mode $m=0$ in $\nu$ space~\textbf{a}) and in $u$ space~\textbf{b}). The contour implements the Feynman-like $i\epsilon$-prescription~\cite{Caron-Huot:2013fea} for integrating around the poles of the $m=0$ impact factor at $\nu=\pm \frac{\pi}{2}\Gamma_{\textrm{cusp}}$. Two copies of the rapidity plane are needed to cover the full integration range. They are depicted by the lower and upper sheet in figure~\textbf{b}). The former sheet, which is associated to the parametric representation~(\ref{BFKL_exact}) of the $m=0$ eigenvalue, only accommodates for $\nu$ in the real intervals $|\nu| > \nu_0 = \nu(u=2g)$. To access to the inner interval $|\nu|< \nu_0$, one should analytically continue eqs.~(\ref{BFKL_exact}) to the upper sheet, by passing through the cut stretching between $u=\pm 2g$. On this sheet the function $\omega(u)$ and $\nu(u)$ are invariant under $u\rightarrow u\pm i$. This periodicity allows us to wrap this $u$ plane into a cylinder of radius $i$, as done in figure~\textbf{b}). The ends of this cylinder map to the points $(\omega, \nu) = (0, \pm \frac{\pi}{2}\Gamma_{\textrm{cusp}})$, respectively. The (red) line, which stands `$\frac{i}{2}$' away from the cut, covers the interval $|\nu| < \frac{\pi}{2}\Gamma_{\textrm{cusp}}$ through the representation~(\ref{m0sdsheet}).}
\la{zerochargepath}
\end{figure}

This analytical continuation is elementary and only makes use of the BES equation. Essentially, once sitting on the cut (that is for $u^2 < (2g)^2$ with a small positive or negative imaginary value), we can add the BES equation~(\ref{BES}) and flip the $\pm$ sign in the exponentials in~(\ref{rep-cut}). This allows us to pass through the cut and obtain a representation on the second sheet. We immediately find
\be
\begin{aligned}\label{rep-cut2}
\omega &= \int\limits_{0}^{\infty}\frac{dt}{t}\frac{\gamma_{+}(2gt)\cos{(ut)}}{e^{t}-1}+\frac{1}{2}\int\limits_0^{\infty}\frac{dt}{t}\gamma_{+}(2gt)e^{\mp iut} \, ,\\
\nu &=\int\limits_{0}^{\infty}\frac{dt}{t}\frac{\gamma_{-}(2gt)\sin{(ut)}}{e^{t}-1} \pm \frac{i}{2}\int\limits_0^{\infty}\frac{dt}{t}\gamma_{-}(2gt)e^{\mp iut} \, ,
\end{aligned}
\ee
which holds within a strip in the lower/upper half plane, respectively. Using the formulae~(\ref{out-form}) one verifies that the two representations~(\ref{rep-cut2}) agree for $u^2 > (2g)^2$. It means that crossing the cut from above or from below boil down to the same operation and lead to the same sheet, as expected.

We succeeded, therefore, in obtaining a representation of the eigenvalue for $|\nu| < \nu_0$. What is not patent, however, is that it allows us to lower $|\nu|$ all the way down to $\nu=0$. If we try to lower $\nu$, assumed positive for definiteness, by cranking up the value of $u$ away from $u=2g$, what we get at its extreme, that is for $u\rightarrow \infty$, is
\beq
\nu = \pi g\gamma_1 + O(e^{-2\pi u}) = \frac{\pi}{2}\Gamma_{\textrm{cusp}} + O(e^{-2\pi u})\, ,
\eeq
where we used the fact that the large rapidity regime of~(\ref{rep-cut2}) is controlled by the small $t$ limit of $\gamma_{-}(2gt) \sim 2g\gamma_{1} t$. In the very same limit we find that $\omega \rightarrow 0$, as a consequence of $\gamma_{+}\propto t^2$ at small $t$. It means that the point $u=\infty$ is the locus of
\beq
\omega(\nu = \frac{\pi}{2}\Gamma_{\textrm{cusp}}, m=0) = 0\, . \label{wcusp}
\eeq
This is precisely the prediction of~\cite{Caron-Huot:2013fea}. (By parity a similar equation holds for $\nu\rightarrow -\nu$.) The interval $2g<u< \infty$ on the second sheet therefore covers the range $\nu_0 < \nu < \frac{\pi}{2}\Gamma_{\textrm{cusp}}$ (assuming $\nu$ is a monotonic function of $u$ along this interval).

To enter inside the interval $|\nu| < \frac{\pi}{2}\Gamma_{\textrm{cusp}}$ we can move along a semi-circle around $\nu = \frac{\pi}{2}\Gamma_{\textrm{cusp}}$ in the lower half $\nu$ plane. As we shall see shortly, this is done by shifting the rapidity by $u\rightarrow u+i/2$. The latter operation can be performed immediately in~(\ref{rep-cut2}) -- picking the representation which is valid in the upper half plane -- and yields
\beq\label{m0sdsheet}
\omega = \int\limits_0^{\infty}\frac{dt}{t}\frac{\gamma_{+}(2gt)e^{t/2}\cos{(ut)}}{e^t-1}\, , \qquad \nu = \int\limits_0^{\infty}\frac{dt}{t}\frac{\gamma_-(2gt)e^{t/2}\sin{(ut)}}{e^t-1}\, .
\eeq
This is the sought representation, which covers all values $|\nu|< \frac{\pi}{2}\Gamma_{\textrm{cusp}}$ as we vary $u$ from $-\infty$ to $\infty$. 

The property we used to arrive at eqs.~(\ref{m0sdsheet}) is that the functions in~(\ref{rep-cut2}), or equivalently~(\ref{m0sdsheet}), are $i$-periodic for $|\textrm{Re}\,  u| > 2g$. This is easily seen in~(\ref{m0sdsheet}) after recalling that in the latter domain we can evaluate the integrals by closing the contour at infinity and taking residues. Namely,
\beq\label{series-omega}
\omega = \frac{1}{4}\int\limits_{\mathbb{R}}\frac{dt}{t}\frac{\gamma_{+}(2gt)e^{iut}}{\textrm{sinh}\,\frac{t}{2}} = \sum_{k\geqslant 1} \frac{(-1)^k}{2k}\gamma_{+}(4i\pi k g) e^{-2\pi k u}\, ,
\eeq
where here we restricted ourselves to $\textrm{Re}\, u > 2g$. The latter requirement is essential to the convergence of the series, since the function $\gamma_{+}(4i\pi k g) \sim e^{4\pi k g}$ at large $k$.
Now from the above representation it is obvious that $\omega (u+i)=\omega(u)$ and also that $\omega \rightarrow 0$ at $u\rightarrow \infty$. The integral for $\nu$ can be handled similarly, though with a bit more care. We get
\beq\label{series-nu}
\nu = \frac{1}{8i} \oint \frac{dt}{t}\frac{\gamma_{-}(2gt)e^{iut}}{\textrm{sinh}\,\frac{t}{2}}  + \frac{1}{4i}\int\limits_{\mathbb{R}+i0}\frac{dt}{t}\frac{\gamma_{-}(2gt)e^{iut}}{\textrm{sinh}\,\frac{t}{2}}
 = \frac{\pi}{2} \Gamma_{\textrm{cusp}} +  \sum_{k\geqslant 1} \frac{(-1)^k}{2ik}\gamma_{-}(4i\pi k g) e^{-2\pi k u}\, ,
\eeq
after using that the contour integral around $t=0$ gives $\pi g \gamma_{1} = \frac{\pi}{2}\Gamma_{\textrm{cusp}}$. We conclude that $\nu(u+i) = \nu(u)$ and, more interestingly, that $\nu\rightarrow \frac{\pi}{2}\Gamma_{\textrm{cusp}}$ for $u\rightarrow \infty$, bringing us back to~(\ref{wcusp}).

One particular application of the representation~(\ref{m0sdsheet}) is found for the intercept, obtained by specifying to $u=0$,
\beq
\omega(\nu=0) = \int\limits_0^{\infty}\frac{dt}{t}\frac{\gamma_{+}(2gt)e^{t/2}}{e^{t}-1}\, .
\eeq
Since $\gamma_+(2gt)=\sum_{k\geqslant 1}2(2k)\gamma_{2k}J_{2k}(2gt)$ and $\gamma_{2k} \sim g^{2k+2}$ at weak coupling, we verify from it that $\omega(\nu=0)$ vanishes to two loops. Using the previously reported weak coupling expression for the leading coefficient $\gamma_2$, see eq.~(\ref{gamma-coeff}), we easily derive that
\beq
\omega(\nu=0) = 4\pi^2\zeta_{3}g^6 - (4\pi^4\zeta_{3}+40\pi^2\zeta_{5})g^8 + O(g^{10})\, ,
\eeq
which agrees with our general result in eqs.~(\ref{omeganu}) and~(\ref{N3LLA}); thus confirming that the analytical continuation was properly done.

A similar analysis can be applied to the measure (i.e.~impact factor). The algebra is straightforward, though a bit more tedious than for the dispersion relation. Starting from the expression~(\ref{bfkl_measure}) with $m=0$, which defines the measure on the bottom sheet in figure~\ref{zerochargepath}~\textbf{b}), one should then pass through the cut, sitting along the interval $u^2 < (2g)^2$, to get to the top sheet in the same figure. The technical tools to perform this step can be adapted from the ones presented in appendix~(\ref{pass-cut}). Once done with this analytical continuation, it is elementary to reach the red line in figure~\ref{zerochargepath}~\textbf{b}) by simply shifing the rapidity as $u\rightarrow u+i/2$. We find in the end that the measure takes the form
\beq\label{measure2nd}
\mu(u)_{2^\textrm{nd}\,\textrm{sheet}} = -\exp{\bigg[-\frac{\pi^2}{4}\Gamma_{\textrm{cusp}}+2f'_{3}(u)-2f'_{4}(u)\bigg]}\, ,
\eeq
where the functions $f'_{3,4}(u)$ are built as in~(\ref{f34}) but with the sources
\beq
\kappa'(u)_{2j} = \int\limits_0^{\infty}\frac{dt}{t}\frac{J_{2j}(2gt)e^{t/2}\cos{(ut)}}{e^t-1}\, , \qquad \tilde{\kappa}'(u)_{2j-1} = -\int\limits_0^{\infty}\frac{dt}{t}\frac{J_{2j-1}(2gt)e^{t/2}\sin{(ut)}}{e^t-1}\, ,
\eeq
and $\kappa'(u)_{2j-1} = \tilde{\kappa}'(u)_{2j} = 0$. The dispersion relation~(\ref{m0sdsheet}) can also be re-written in these terms, through
\beq
\omega(u)_{2^\textrm{nd}\,\textrm{sheet}} = -4g\big(\QQ\cdot \MM\cdot \kappa'(u)\big)_{1}\,, \qquad \nu(u)_{2^\textrm{nd}\,\textrm{sheet}} = -4g\big(\QQ\cdot \MM \cdot \tilde{\kappa}'(u)\big)_{1}\,.
\eeq
We note that all the above quantities are $i$-periodic functions of the rapidity, outside the strip $|\textrm{Re}\, u| < 2g$, since this is the case for the sources, i.e.~$\kappa'(u+i) = \kappa'(u)$ and $\tilde{\kappa}'(u+i) = \tilde{\kappa}'(u)$.

The important property of the measure~(\ref{measure2nd}), which we would like to stress here, concerns its behavior at large rapidity. Namely, we observe that
\beq\label{mu-U}
\mu(u)_{2^\textrm{nd}\,\textrm{sheet}}\rightarrow -1\, , \qquad \textrm{for} \qquad u\rightarrow \pm \infty\, .
\eeq
This immediately follows from the fact that all sources are (exponentially) suppressed at large $u$, except the leading one that goes like $\tilde{\kappa}'(u)_{1}\rightarrow \pm \pi g/2$. It implies that
\beq
f'_{4}(u)\rightarrow 0\, , \qquad f'_{3}(u)\sim 2\tilde{\kappa}'(u)_{1}\MM_{11}\tilde{\kappa}'(u)_{1} \rightarrow \frac{\pi^2}{8}\Gamma_{\textrm{cusp}} \,,
\eeq
after using that $\MM_{11} = \Gamma_{\textrm{cusp}}/(4g^2)$, and eventually yields to eq.~(\ref{mu-U}).
Since at large $u$
\beq
\nu = \pm \frac{\pi}{2} \Gamma_{\textrm{cusp}} \mp  c \, e^{\mp 2\pi u} + \ldots\, ,
\eeq
where $c$ is a coupling dependent constant, we obtain
\beq
du\,  \mu(u)_{2^\textrm{nd}\,\textrm{sheet}} \sim \pm \frac{d\nu}{2\pi} \frac{1}{\nu\mp \frac{\pi}{2}\Gamma_{\textrm{cusp}}}\, ,
\eeq
when $\nu \sim \pm \frac{\pi}{2}\Gamma_{\textrm{cusp}}$. This behavior is in perfect agreement with the general predictions of ref.~\cite{Caron-Huot:2013fea} and constitutes a non-trivial test of our expressions for $m=0$.

\section{Leading behavior at strong coupling}\label{app:strong}

To the leading order at strong coupling $g = \sqrt{\lambda}/(4\pi)\gg 1$, the solution to the BES equation~\cite{Alday:2007qf} reads, in our notations (see~\cite{Basso:2010in} for further detail), as
\be\label{lo-BES}
K(t) = \frac{2gt\sqrt{2}}{\pi(1-e^{-t})}\int\limits_{-1}^{1}d\xi\left(\frac{1+\xi}{1-\xi}\right)^{\tfrac{1}{4}}\sin{(2g\xi t)} - \frac{2gt\sqrt{2}}{\pi(e^t-1)}\int_{-1}^{1}d\xi\left(\frac{1+\xi}{1-\xi}\right)^{\tfrac{1}{4}}\cos{(2g\xi t)} \, .
\ee
As a consequence the integrands in~(\ref{BFKL_exact}) oscillate rapidly, meaning that the integrals are dominated by $t\sim 1/g$. Upon rescaling $t\rightarrow  t/(2g)$ and expanding at large $g$,
it is straightforward to perform the $t$ integrals since the dependence is only through the $\sin$ and $\cos$ terms.
Taking $\tilde{u}=u/(2g)$ fixed, but $m\sim 1$, and performing the $t$ integrals yields
\be\begin{aligned}
 \omega &= \frac{g\sqrt{2}}{\pi} \int\limits_{-1}^{1}d\xi\left(\frac{1+\xi}{1-\xi}\right)^{\tfrac{1}{4}}
 \left[\pi \,\textrm{sign}(\xi) - \log \frac{\left|\xi^2-\tilde{u}^2\right|}{\xi^2}\right] + \OO(g^0)\,,\\
 \nu &= 4g\tilde{u} + \frac{g\sqrt{2}}{\pi} \int\limits_{-1}^{1}d\xi\left(\frac{1+\xi}{1-\xi}\right)^{\tfrac{1}{4}} \log\frac{\left|\xi-\tilde{u}\right|}{\left|\xi+\tilde{u}\right|} + \OO(g^0)\,. 
\end{aligned}
\ee
We observe that all dependence on $m$ dropped out of $\omega$. This is a manifestation of the universality of the (leading) strong coupling limit.
A simple way to evaluate these integrals is to take a derivative with respect to $\tilde{u}$, after which the integrals can be readily performed via contour integration. The result, however, takes different form depending on whether $\tilde{u}^2 \gtrless 1$. As we will see shortly, the inner/outer domain corresponds to $|\nu| \gtrless \pi g$, or, as this turns out to be equivalent, to $\omega \gtrless 0$.
In the inner region ($\tilde{u}^2 < 1$), after re-integrating with respect to $\tilde{u}$ with the appropriate boundary conditions, we obtain
\be
\begin{aligned}\label{scresult-app}
\omega &= \sqrt{2}g\int_{\tilde{u}}^{1}d\xi \bigg[\left(\frac{1+\xi}{1-\xi}\right)^{\tfrac{1}{4}}-\left(\frac{1-\xi}{1+\xi}\right)^{\tfrac{1}{4}}\bigg] + \OO(g^0)\, ,\\ 
\nu &= \sqrt{2}g\int_0^{\tilde{u}}d\xi \bigg[\left(\frac{1+\xi}{1-\xi}\right)^{\tfrac{1}{4}}+\left(\frac{1-\xi}{1+\xi}\right)^{\tfrac{1}{4}}\bigg] + \OO(g^0)\, ,
\end{aligned}
\ee
and immediately verify that $|\nu|$ remains smaller than $\pi g$ and $\omega$ positive as we vary $\tilde{u}^2$ from $0$ to $1$.
We note that the fact that the derivative with respect to $\tilde{u}$ has removed the integrals is essentially the content of the strong-coupling BES equation, which could have been used to directly
write down (\ref{scresult-app}).
Introducing the parametrization $\tilde{u}=\tanh(2\theta)$, we find the formula quoted in the main text (\ref{bfklstrong}).
In the complementary region, picking $\tilde{u}>1$ for definiteness, we get instead that
\be
\begin{aligned}\label{scresult2-app}
\omega &= 2g\int_{1}^{\tilde{u}}d\xi \bigg[\left(\frac{\xi-1}{\xi+1}\right)^{\tfrac{1}{4}}-\left(\frac{\xi+1}{\xi-1}\right)^{\tfrac{1}{4}}\bigg] + \OO(g^0)\, ,\\ 
\nu-\pi g &= 2g\int_{1}^{\tilde{u}}d\xi \bigg[\left(\frac{\xi-1}{\xi+1}\right)^{\tfrac{1}{4}}+\left(\frac{\xi+1}{\xi-1}\right)^{\tfrac{1}{4}}\bigg] + \OO(g^0)\, ,
\end{aligned}
\ee
which shows that $\omega > 0$ and $\nu>\pi g$ for all $\tilde{u}>1$.
Upon the reparameterization $\tilde{u} = \textrm{coth}(2\theta)$, this representation could be re-written equivalently as
\be\la{sctrajectory2}\begin{aligned}
\omega&={\sqrt\lambda\over2\pi}\[{1\over\cosh{\theta}}-\frac{1}{2}\log\({\cosh{\theta}+1\over\cosh{\theta}-1}\)\] \, ,\\
\nu-\frac{\sqrt{\lambda}}{4}&={\sqrt\lambda\over2\pi}\[{1\over \sinh{\theta}}+\frac{i}{2}\log\({\sinh{\theta}+i\over\sinh{\theta}-i}\)\]\,\,.
\end{aligned}\ee
It displays a logarithmic scaling $\omega \sim -\frac{\sqrt{\lambda}}{2\pi}\log{\nu}$ at large $\nu$ (corresponding to small~$\theta$), which persists at any coupling upon $\frac{\sqrt{\lambda}}{2\pi}\rightarrow \Gamma_{\textrm{cusp}}$. 

\begin{figure}[t]
\centering
\includegraphics[scale=1]{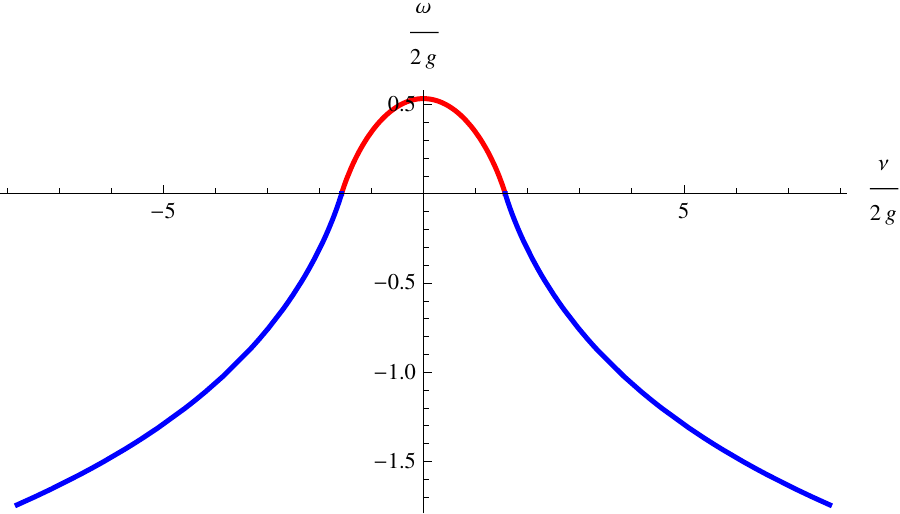}
\caption{The BFKL eigenvalue at strong coupling. The inner part (in red) is controlled by~(\ref{bfklstrong}) and the outer one (in blue) by~(\ref{sctrajectory2}). The former is maximal at $\nu=0$ and minimal at $|\nu| = \pi g$ while the latter is unbounded from below, scaling like $\omega \sim -2g\log{|\nu|}$ at large $\nu$.}\label{branches}
\end{figure}

In short, at strong coupling, the BFKL eigenvalue is described by two (analytical) functions, which are continuously (but not smoothly) connected to each other at $\nu = \pm \pi g$, as depicted in figure~\ref{branches}. Their expressions, given in eqs.~(\ref{gianttoBFKL}) and~(\ref{sctrajectory2}) respectively, are mathematically similar and can both be related to the giant hole dispersion relation, as mentioned in section~\ref{giant}. Namely, we have
\beq\label{matching}
\omega(\theta) = -E_{\textrm{giant}}(\theta)\, , \qquad \nu(\theta) = \frac{\sqrt{\lambda}}{4}\, \textrm{sign}\, \nu + P_{\textrm{giant}}(\theta)\, ,
\eeq
in the outer region, and (\ref{gianttoBFKL}) in the inner one.


The analysis we carried out for $\tilde{u}^2 < 1$ does not apply, strictly speaking, to the particular case $m=0$. This point is easily understood after recalling the observation made in appendix~\ref{meq0appendix} that this eigenvalue has a branch cut along this interval. To obtain the strong coupling eigenvalue for $m=0$ in the regime where $\omega >0$, we should place ourself on the second sheet and analyze the integral representation~(\ref{m0sdsheet}) instead. This one, we have seen, covers the domain $|\nu| < \frac{\pi}{2}\Gamma_{\textrm{cusp}}$ for $|u|<\infty$, which is precisely the range of interest here. In fact, at strong coupling, we reach $|\nu| = \frac{\pi}{2}\Gamma_{\textrm{cusp}}$ at $|u| = 2g$ already (up to exponentially small corrections).
We can therefore focus our attention on the interval $\tilde{u}^2 < 1$. The relevant integrals are easily done and the end result found to be exactly the same as in eq.~(\ref{bfklstrong}). This shows that the universality observed before extends to $m = 0$ as well, to leading order at strong coupling.

The impact factor (\ref{bfkl_measure}) at strong coupling requires more work and will not be discussed here,
but is in principle straightforward.

\bibliographystyle{JHEP}
\providecommand{\href}[2]{#2}\begingroup\raggedright\endgroup

\end{document}

%% file: sistertrajectory.pdf_tex
\begingroup%
  \makeatletter%
  \providecommand\color[2][]{%
    \errmessage{(Inkscape) Color is used for the text in Inkscape, but the package 'color.sty' is not loaded}%
    \renewcommand\color[2][]{}%
  }%
  \providecommand\transparent[1]{%
    \errmessage{(Inkscape) Transparency is used (non-zero) for the text in Inkscape, but the package 'transparent.sty' is not loaded}%
    \renewcommand\transparent[1]{}%
  }%
  \providecommand\rotatebox[2]{#2}%
  \ifx\svgwidth\undefined%
    \setlength{\unitlength}{1157.87431641bp}%
    \ifx\svgscale\undefined%
      \relax%
    \else%
      \setlength{\unitlength}{\unitlength * \real{\svgscale}}%
    \fi%
  \else%
    \setlength{\unitlength}{\svgwidth}%
  \fi%
  \global\let\svgwidth\undefined%
  \global\let\svgscale\undefined%
  \makeatother%
  \begin{picture}(1,0.32885554)%
    \put(0,0){\includegraphics[width=\unitlength]{sistertrajectory.pdf}}%
    \put(0.39175236,0.05014929){\color[rgb]{0,0,0}\makebox(0,0)[lb]{\smash{$\alpha'\,t$}}}%
    \put(0.05882168,0.32135841){\color[rgb]{0,0,0}\makebox(0,0)[lb]{\smash{$j$}}}%
    \put(0.12993061,0.05342197){\color[rgb]{0,0,0}\makebox(0,0)[lb]{\smash{$_1$}}}%
    \put(0.18520431,0.05342197){\color[rgb]{0,0,0}\makebox(0,0)[lb]{\smash{$_2$}}}%
    \put(0.24047801,0.05342197){\color[rgb]{0,0,0}\makebox(0,0)[lb]{\smash{$_3$}}}%
    \put(0.29575171,0.05342197){\color[rgb]{0,0,0}\makebox(0,0)[lb]{\smash{$_4$}}}%
    \put(0.35102541,0.05342197){\color[rgb]{0,0,0}\makebox(0,0)[lb]{\smash{$_5$}}}%
    \put(0.06360216,0.12942331){\color[rgb]{0,0,0}\makebox(0,0)[lb]{\smash{$_1$}}}%
    \put(0.06360216,0.18469701){\color[rgb]{0,0,0}\makebox(0,0)[lb]{\smash{$_2$}}}%
    \put(0.06360216,0.23997071){\color[rgb]{0,0,0}\makebox(0,0)[lb]{\smash{$_3$}}}%
    \put(0.06360216,0.29524442){\color[rgb]{0,0,0}\makebox(0,0)[lb]{\smash{$_4$}}}%
    \put(0.94448938,0.05014929){\color[rgb]{0,0,0}\makebox(0,0)[lb]{\smash{$\alpha'\,t$}}}%
    \put(0.61155861,0.32135841){\color[rgb]{0,0,0}\makebox(0,0)[lb]{\smash{$j$}}}%
    \put(0.68266758,0.05342197){\color[rgb]{0,0,0}\makebox(0,0)[lb]{\smash{$_1$}}}%
    \put(0.73794129,0.05342197){\color[rgb]{0,0,0}\makebox(0,0)[lb]{\smash{$_2$}}}%
    \put(0.79321499,0.05342197){\color[rgb]{0,0,0}\makebox(0,0)[lb]{\smash{$_3$}}}%
    \put(0.84848869,0.05342197){\color[rgb]{0,0,0}\makebox(0,0)[lb]{\smash{$_4$}}}%
    \put(0.90376239,0.05342197){\color[rgb]{0,0,0}\makebox(0,0)[lb]{\smash{$_5$}}}%
    \put(0.61633914,0.12942331){\color[rgb]{0,0,0}\makebox(0,0)[lb]{\smash{$_1$}}}%
    \put(0.61633914,0.18469701){\color[rgb]{0,0,0}\makebox(0,0)[lb]{\smash{$_2$}}}%
    \put(0.61633914,0.23997071){\color[rgb]{0,0,0}\makebox(0,0)[lb]{\smash{$_3$}}}%
    \put(0.61633914,0.29524442){\color[rgb]{0,0,0}\makebox(0,0)[lb]{\smash{$_4$}}}%
    \put(0.19535525,0.00160928){\color[rgb]{0,0,0}\makebox(0,0)[lb]{\smash{$({\bf a})$}}}%
    \put(0.76191065,0.00160928){\color[rgb]{0,0,0}\makebox(0,0)[lb]{\smash{$({\bf b})$}}}%
  \end{picture}%
\endgroup%

%% file: flatstring.pdf_tex
\begingroup%
  \makeatletter%
  \providecommand\color[2][]{%
    \errmessage{(Inkscape) Color is used for the text in Inkscape, but the package 'color.sty' is not loaded}%
    \renewcommand\color[2][]{}%
  }%
  \providecommand\transparent[1]{%
    \errmessage{(Inkscape) Transparency is used (non-zero) for the text in Inkscape, but the package 'transparent.sty' is not loaded}%
    \renewcommand\transparent[1]{}%
  }%
  \providecommand\rotatebox[2]{#2}%
  \ifx\svgwidth\undefined%
    \setlength{\unitlength}{1370.04697266bp}%
    \ifx\svgscale\undefined%
      \relax%
    \else%
      \setlength{\unitlength}{\unitlength * \real{\svgscale}}%
    \fi%
  \else%
    \setlength{\unitlength}{\svgwidth}%
  \fi%
  \global\let\svgwidth\undefined%
  \global\let\svgscale\undefined%
  \makeatother%
  \begin{picture}(1,0.48360864)%
    \put(0,0){\includegraphics[width=\unitlength]{flatstring.pdf}}%
    \put(0.25137668,0.00380918){\color[rgb]{0,0,0}\makebox(0,0)[lb]{\smash{$({\bf a})$}}}%
    \put(0.76522769,0.00380918){\color[rgb]{0,0,0}\makebox(0,0)[lb]{\smash{$({\bf b})$}}}%
    \put(0.41674487,0.05419048){\color[rgb]{0,0,0}\makebox(0,0)[lb]{\smash{$1$}}}%
    \put(0.48717438,0.40124974){\color[rgb]{0,0,0}\makebox(0,0)[lb]{\smash{$6$}}}%
    \put(0.36438267,0.4658629){\color[rgb]{0,0,0}\makebox(0,0)[lb]{\smash{$5$}}}%
    \put(0.15714703,0.46342502){\color[rgb]{0,0,0}\makebox(0,0)[lb]{\smash{$4$}}}%
    \put(0.03706979,0.39754494){\color[rgb]{0,0,0}\makebox(0,0)[lb]{\smash{$3$}}}%
    \put(0.09858222,0.05436993){\color[rgb]{0,0,0}\makebox(0,0)[lb]{\smash{$2$}}}%
    \put(0.93306017,0.04690051){\color[rgb]{0,0,0}\makebox(0,0)[lb]{\smash{$6$}}}%
    \put(0.95733497,0.42659759){\color[rgb]{0,0,0}\makebox(0,0)[lb]{\smash{$1$}}}%
    \put(0.88640725,0.45910979){\color[rgb]{0,0,0}\makebox(0,0)[lb]{\smash{$5$}}}%
    \put(0.66254061,0.45434531){\color[rgb]{0,0,0}\makebox(0,0)[lb]{\smash{$4$}}}%
    \put(0.6231542,0.04490608){\color[rgb]{0,0,0}\makebox(0,0)[lb]{\smash{$3$}}}%
    \put(0.60761684,0.42084925){\color[rgb]{0,0,0}\makebox(0,0)[lb]{\smash{$2$}}}%
    \put(0.20713193,0.27676139){\color[rgb]{0,0,0}\makebox(0,0)[lb]{\smash{$_\text{rapidity gap}$}}}%
    \put(0.55037983,0.41327581){\color[rgb]{0,0,0}\rotatebox{90}{\makebox(0,0)[lb]{\smash{$_\text{time}$}}}}%
  \end{picture}%
\endgroup%

%% file: dipoldipol.pdf_tex
\begingroup%
  \makeatletter%
  \providecommand\color[2][]{%
    \errmessage{(Inkscape) Color is used for the text in Inkscape, but the package 'color.sty' is not loaded}%
    \renewcommand\color[2][]{}%
  }%
  \providecommand\transparent[1]{%
    \errmessage{(Inkscape) Transparency is used (non-zero) for the text in Inkscape, but the package 'transparent.sty' is not loaded}%
    \renewcommand\transparent[1]{}%
  }%
  \providecommand\rotatebox[2]{#2}%
  \ifx\svgwidth\undefined%
    \setlength{\unitlength}{749.60532227bp}%
    \ifx\svgscale\undefined%
      \relax%
    \else%
      \setlength{\unitlength}{\unitlength * \real{\svgscale}}%
    \fi%
  \else%
    \setlength{\unitlength}{\svgwidth}%
  \fi%
  \global\let\svgwidth\undefined%
  \global\let\svgscale\undefined%
  \makeatother%
  \begin{picture}(1,0.59575107)%
    \put(0,0){\includegraphics[width=\unitlength]{dipoldipol.pdf}}%
    \put(0.77994481,0.00230522){\color[rgb]{0,0,0}\makebox(0,0)[lb]{\smash{$({\bf b})$}}}%
    \put(0.16949023,0.00230522){\color[rgb]{0,0,0}\makebox(0,0)[lb]{\smash{$({\bf a})$}}}%
    \put(0.0286161,0.04712881){\color[rgb]{0,0,0}\makebox(0,0)[lb]{\smash{$1$}}}%
    \put(0.31249882,0.04712881){\color[rgb]{0,0,0}\makebox(0,0)[lb]{\smash{$2$}}}%
    \put(0.0286161,0.58501186){\color[rgb]{0,0,0}\makebox(0,0)[lb]{\smash{$4$}}}%
    \put(0.31249882,0.58501186){\color[rgb]{0,0,0}\makebox(0,0)[lb]{\smash{$3$}}}%
    \put(0.61132274,0.04712881){\color[rgb]{0,0,0}\makebox(0,0)[lb]{\smash{$3$}}}%
    \put(0.97418036,0.04712881){\color[rgb]{0,0,0}\makebox(0,0)[lb]{\smash{$6$}}}%
    \put(0.61132274,0.58501186){\color[rgb]{0,0,0}\makebox(0,0)[lb]{\smash{$2$}}}%
    \put(0.97418036,0.58501186){\color[rgb]{0,0,0}\makebox(0,0)[lb]{\smash{$1$}}}%
    \put(0.71804557,0.58501186){\color[rgb]{0,0,0}\makebox(0,0)[lb]{\smash{$4$}}}%
    \put(0.86745753,0.58501186){\color[rgb]{0,0,0}\makebox(0,0)[lb]{\smash{$5$}}}%
  \end{picture}%
\endgroup%

%% file: 3to3kinematics.pdf_tex
\begingroup%
  \makeatletter%
  \providecommand\color[2][]{%
    \errmessage{(Inkscape) Color is used for the text in Inkscape, but the package 'color.sty' is not loaded}%
    \renewcommand\color[2][]{}%
  }%
  \providecommand\transparent[1]{%
    \errmessage{(Inkscape) Transparency is used (non-zero) for the text in Inkscape, but the package 'transparent.sty' is not loaded}%
    \renewcommand\transparent[1]{}%
  }%
  \providecommand\rotatebox[2]{#2}%
  \ifx\svgwidth\undefined%
    \setlength{\unitlength}{611.44243164bp}%
    \ifx\svgscale\undefined%
      \relax%
    \else%
      \setlength{\unitlength}{\unitlength * \real{\svgscale}}%
    \fi%
  \else%
    \setlength{\unitlength}{\svgwidth}%
  \fi%
  \global\let\svgwidth\undefined%
  \global\let\svgscale\undefined%
  \makeatother%
  \begin{picture}(1,0.54559052)%
    \put(0,0){\includegraphics[width=\unitlength]{3to3kinematics.pdf}}%
    \put(0.3074213,0.27984658){\color[rgb]{0,0,0}\makebox(0,0)[lb]{\smash{$_2$}}}%
    \put(0.24047006,0.23111146){\color[rgb]{0,0,0}\makebox(0,0)[lb]{\smash{$_3$}}}%
    \put(0.16273643,0.26581204){\color[rgb]{0,0,0}\makebox(0,0)[lb]{\smash{$_4$}}}%
    \put(0.10794772,0.26561581){\color[rgb]{0,0,0}\makebox(0,0)[lb]{\smash{$_5$}}}%
    \put(0.05088617,0.22434312){\color[rgb]{0,0,0}\makebox(0,0)[lb]{\smash{$_6$}}}%
    \put(0.36272945,0.2652202){\color[rgb]{0,0,0}\makebox(0,0)[lb]{\smash{$_1$}}}%
    \put(0.20514402,0.00205245){\color[rgb]{0,0,0}\makebox(0,0)[lb]{\smash{$({\bf a})$}}}%
    \put(0.35307664,0.15951618){\color[rgb]{0,0,0}\rotatebox{44.07956452}{\makebox(0,0)[lb]{\smash{$_{e^{-2\sigma}}$}}}}%
    \put(0.31905863,0.12549832){\color[rgb]{0,0,0}\rotatebox{44.07956452}{\makebox(0,0)[lb]{\smash{$_0$}}}}%
    \put(0.41849561,0.22231858){\color[rgb]{0,0,0}\rotatebox{44.07956452}{\makebox(0,0)[lb]{\smash{$_\infty$}}}}%
    \put(0.24578927,0.05222894){\color[rgb]{0,0,0}\rotatebox{44.07956452}{\makebox(0,0)[lb]{\smash{$_{-1}$}}}}%
    \put(0.03954934,0.19999551){\color[rgb]{0,0,0}\rotatebox{-45.92043548}{\makebox(0,0)[lb]{\smash{$_{e^{2\tau}}$}}}}%
    \put(0.10496842,0.13457641){\color[rgb]{0,0,0}\rotatebox{-45.92043548}{\makebox(0,0)[lb]{\smash{$_0$}}}}%
    \put(0.12590251,0.1136423){\color[rgb]{0,0,0}\rotatebox{-45.92043548}{\makebox(0,0)[lb]{\smash{$_{-1}$}}}}%
    \put(0.00029788,0.23401339){\color[rgb]{0,0,0}\rotatebox{-45.92043548}{\makebox(0,0)[lb]{\smash{$_\infty$}}}}%
    \put(0.7808319,0.00205245){\color[rgb]{0,0,0}\makebox(0,0)[lb]{\smash{$({\bf b})$}}}%
    \put(0.85066326,0.38705688){\color[rgb]{0,0,0}\makebox(0,0)[lb]{\smash{$_1$}}}%
    \put(0.91217987,0.35837455){\color[rgb]{0,0,0}\makebox(0,0)[lb]{\smash{$_3$}}}%
    \put(0.93690656,0.33809651){\color[rgb]{0,0,0}\makebox(0,0)[lb]{\smash{$_2$}}}%
    \put(0.73571275,0.2678601){\color[rgb]{0,0,0}\makebox(0,0)[lb]{\smash{$_4$}}}%
    \put(0.63100621,0.33641734){\color[rgb]{0,0,0}\makebox(0,0)[lb]{\smash{$_5$}}}%
    \put(0.69373066,0.37978617){\color[rgb]{0,0,0}\makebox(0,0)[lb]{\smash{$_6$}}}%
    \put(0.79449612,0.33527697){\color[rgb]{0,0,0}\makebox(0,0)[lb]{\smash{${\color{applegreen}s_{456}}$}}}%
    \put(0.54760036,0.33397351){\color[rgb]{0,0,0}\makebox(0,0)[lb]{\smash{${\color{applegreen}s_{345}}$}}}%
  \end{picture}%
\endgroup%

%% file: analyticcont.pdf_tex
\begingroup%
  \makeatletter%
  \providecommand\color[2][]{%
    \errmessage{(Inkscape) Color is used for the text in Inkscape, but the package 'color.sty' is not loaded}%
    \renewcommand\color[2][]{}%
  }%
  \providecommand\transparent[1]{%
    \errmessage{(Inkscape) Transparency is used (non-zero) for the text in Inkscape, but the package 'transparent.sty' is not loaded}%
    \renewcommand\transparent[1]{}%
  }%
  \providecommand\rotatebox[2]{#2}%
  \ifx\svgwidth\undefined%
    \setlength{\unitlength}{699.2bp}%
    \ifx\svgscale\undefined%
      \relax%
    \else%
      \setlength{\unitlength}{\unitlength * \real{\svgscale}}%
    \fi%
  \else%
    \setlength{\unitlength}{\svgwidth}%
  \fi%
  \global\let\svgwidth\undefined%
  \global\let\svgscale\undefined%
  \makeatother%
  \begin{picture}(1,0.16043154)%
    \put(0,0){\includegraphics[width=\unitlength]{analyticcont.pdf}}%
    \put(0.3987646,0.14804662){\color[rgb]{0,0,0}\makebox(0,0)[lb]{\smash{$\sigma-t$}}}%
    \put(0.23858154,0.03820682){\color[rgb]{0,0,0}\makebox(0,0)[lb]{\smash{$\sigma-i0$}}}%
    \put(0.15162502,0.07710842){\color[rgb]{0,0,0}\makebox(0,0)[lb]{\smash{$0$}}}%
    \put(0.39602501,0.04091414){\color[rgb]{0,0,0}\makebox(0,0)[lb]{\smash{$\Blue{K_-}$}}}%
    \put(0.00700902,0.04091414){\color[rgb]{0,0,0}\makebox(0,0)[lb]{\smash{$\Red{K_+}$}}}%
    \put(0.94796376,0.14804662){\color[rgb]{0,0,0}\makebox(0,0)[lb]{\smash{$\sigma-t$}}}%
    \put(0.80390417,0.11527355){\color[rgb]{0,0,0}\makebox(0,0)[lb]{\smash{$\sigma+i0$}}}%
    \put(0.7008241,0.07710842){\color[rgb]{0,0,0}\makebox(0,0)[lb]{\smash{$0$}}}%
    \put(0.54934308,0.07752741){\color[rgb]{0,0,0}\makebox(0,0)[lb]{\smash{$\Blue{K_-}$}}}%
    \put(0.54934308,0.13473565){\color[rgb]{0,0,0}\makebox(0,0)[lb]{\smash{$\Red{K_+}$}}}%
    \put(0.19281495,0.00159355){\color[rgb]{0,0,0}\makebox(0,0)[lb]{\smash{$({\bf a})$}}}%
    \put(0.76489733,0.00159355){\color[rgb]{0,0,0}\makebox(0,0)[lb]{\smash{$({\bf b})$}}}%
  \end{picture}%
\endgroup%

%% file: omegaofnu.pdf_tex
\begingroup%
  \makeatletter%
  \providecommand\color[2][]{%
    \errmessage{(Inkscape) Color is used for the text in Inkscape, but the package 'color.sty' is not loaded}%
    \renewcommand\color[2][]{}%
  }%
  \providecommand\transparent[1]{%
    \errmessage{(Inkscape) Transparency is used (non-zero) for the text in Inkscape, but the package 'transparent.sty' is not loaded}%
    \renewcommand\transparent[1]{}%
  }%
  \providecommand\rotatebox[2]{#2}%
  \ifx\svgwidth\undefined%
    \setlength{\unitlength}{609.37695312bp}%
    \ifx\svgscale\undefined%
      \relax%
    \else%
      \setlength{\unitlength}{\unitlength * \real{\svgscale}}%
    \fi%
  \else%
    \setlength{\unitlength}{\svgwidth}%
  \fi%
  \global\let\svgwidth\undefined%
  \global\let\svgscale\undefined%
  \makeatother%
  \begin{picture}(1,0.27848955)%
    \put(0,0){\includegraphics[width=\unitlength]{omegaofnu.pdf}}%
    \put(0.08089967,0.26397983){\color[rgb]{0,0,0}\rotatebox{-30.22595684}{\makebox(0,0)[lb]{\smash{$_\DBlue{|m|=1}$}}}}%
    \put(0.02328141,0.14791227){\color[rgb]{0,0,0}\makebox(0,0)[lb]{\smash{$_{-0.1}$}}}%
    \put(0.02328141,0.09277399){\color[rgb]{0,0,0}\makebox(0,0)[lb]{\smash{$_{-0.2}$}}}%
    \put(0.02328141,0.03501007){\color[rgb]{0,0,0}\makebox(0,0)[lb]{\smash{$_{-0.3}$}}}%
    \put(0.04953774,0.20567619){\color[rgb]{0,0,0}\makebox(0,0)[lb]{\smash{$_{0}$}}}%
    \put(0.20669013,0.19298869){\color[rgb]{0,0,0}\makebox(0,0)[lb]{\smash{$_{2}$}}}%
    \put(0.34847429,0.19298869){\color[rgb]{0,0,0}\makebox(0,0)[lb]{\smash{$_{4}$}}}%
    \put(0.48012658,0.19233649){\color[rgb]{0,0,0}\makebox(0,0)[lb]{\smash{$_\nu$}}}%
    \put(0.02318146,0.26554919){\color[rgb]{0,0,0}\makebox(0,0)[lb]{\smash{$_{\omega(\nu)}$}}}%
    \put(0.0889289,0.19609749){\color[rgb]{0,0,0}\rotatebox{-18.92279015}{\makebox(0,0)[lb]{\smash{$_\Red{m=0}$}}}}%
    \put(0.08287703,0.1739884){\color[rgb]{0,0,0}\rotatebox{-14.09917704}{\makebox(0,0)[lb]{\smash{$_\DBlue{|m|=2}$}}}}%
    \put(0.08214401,0.12286398){\color[rgb]{0,0,0}\rotatebox{-5.10088086}{\makebox(0,0)[lb]{\smash{$_\DBlue{|m|=3}$}}}}%
    \put(0.25657715,0.25600387){\color[rgb]{0,0,0}\makebox(0,0)[lb]{\smash{$g={1\over4}$}}}%
    \put(0.53265414,0.09802525){\color[rgb]{0,0,0}\makebox(0,0)[lb]{\smash{$_{-1}$}}}%
    \put(0.53265414,0.05864076){\color[rgb]{0,0,0}\makebox(0,0)[lb]{\smash{$_{-2}$}}}%
    \put(0.53265414,0.01663064){\color[rgb]{0,0,0}\makebox(0,0)[lb]{\smash{$_{-3}$}}}%
    \put(0.54315667,0.17941986){\color[rgb]{0,0,0}\makebox(0,0)[lb]{\smash{$_{1}$}}}%
    \put(0.54315667,0.22142998){\color[rgb]{0,0,0}\makebox(0,0)[lb]{\smash{$_{2}$}}}%
    \put(0.51802265,0.25735569){\color[rgb]{0,0,0}\makebox(0,0)[lb]{\smash{$_{\omega(\nu)}$}}}%
    \put(0.54315667,0.14003537){\color[rgb]{0,0,0}\makebox(0,0)[lb]{\smash{$_{0}$}}}%
    \put(0.60799713,0.12732577){\color[rgb]{0,0,0}\makebox(0,0)[lb]{\smash{$_{2}$}}}%
    \put(0.66050978,0.12732577){\color[rgb]{0,0,0}\makebox(0,0)[lb]{\smash{$_{4}$}}}%
    \put(0.71302243,0.12732577){\color[rgb]{0,0,0}\makebox(0,0)[lb]{\smash{$_{6}$}}}%
    \put(0.76553509,0.12732577){\color[rgb]{0,0,0}\makebox(0,0)[lb]{\smash{$_{8}$}}}%
    \put(0.81804774,0.12732577){\color[rgb]{0,0,0}\makebox(0,0)[lb]{\smash{$_{10}$}}}%
    \put(0.87056043,0.12732577){\color[rgb]{0,0,0}\makebox(0,0)[lb]{\smash{$_{12}$}}}%
    \put(0.92307308,0.12732577){\color[rgb]{0,0,0}\makebox(0,0)[lb]{\smash{$_{14}$}}}%
    \put(0.96183009,0.12356675){\color[rgb]{0,0,0}\makebox(0,0)[lb]{\smash{$_\nu$}}}%
    \put(0.75544734,0.25600387){\color[rgb]{0,0,0}\makebox(0,0)[lb]{\smash{$g=3$}}}%
  \end{picture}%
\endgroup%

%% file: BFKLintercepts2.pdf_tex
\begingroup%
  \makeatletter%
  \providecommand\color[2][]{%
    \errmessage{(Inkscape) Color is used for the text in Inkscape, but the package 'color.sty' is not loaded}%
    \renewcommand\color[2][]{}%
  }%
  \providecommand\transparent[1]{%
    \errmessage{(Inkscape) Transparency is used (non-zero) for the text in Inkscape, but the package 'transparent.sty' is not loaded}%
    \renewcommand\transparent[1]{}%
  }%
  \providecommand\rotatebox[2]{#2}%
  \ifx\svgwidth\undefined%
    \setlength{\unitlength}{291.32924805bp}%
    \ifx\svgscale\undefined%
      \relax%
    \else%
      \setlength{\unitlength}{\unitlength * \real{\svgscale}}%
    \fi%
  \else%
    \setlength{\unitlength}{\svgwidth}%
  \fi%
  \global\let\svgwidth\undefined%
  \global\let\svgscale\undefined%
  \makeatother%
  \begin{picture}(1,0.52596392)%
    \put(0,0){\includegraphics[width=\unitlength]{BFKLintercepts2.pdf}}%
    \put(0.82168011,0.05181746){\color[rgb]{0,0,0}\makebox(0,0)[lb]{\smash{${\sqrt\lambda\over4\pi}$}}}%
    \put(-0.00111266,0.51251319){\color[rgb]{0,0,0}\makebox(0,0)[lb]{\smash{$_{\omega(\nu=0)}$}}}%
    \put(0.74936665,0.43431507){\color[rgb]{0,0,0}\rotatebox{29.46211448}{\makebox(0,0)[lb]{\smash{$_\DBlue{|m|=1}$}}}}%
    \put(0.74936683,0.3793945){\color[rgb]{0,0,0}\rotatebox{29.46211448}{\makebox(0,0)[lb]{\smash{$_\Red{m=0}$}}}}%
    \put(0.74936676,0.34095001){\color[rgb]{0,0,0}\rotatebox{29.46211448}{\makebox(0,0)[lb]{\smash{$_\DBlue{|m|=2}$}}}}%
    \put(0.74936689,0.25856905){\color[rgb]{0,0,0}\rotatebox{29.46211448}{\makebox(0,0)[lb]{\smash{$_\DBlue{|m|=3}$}}}}%
    \put(0.74835707,0.17775653){\color[rgb]{0,0,0}\rotatebox{27.51846745}{\makebox(0,0)[lb]{\smash{$_\DBlue{|m|=4}$}}}}%
    \put(0.03634563,0.03029166){\color[rgb]{0,0,0}\makebox(0,0)[lb]{\smash{$_{-0.5}$}}}%
    \put(0.0583139,0.10168854){\color[rgb]{0,0,0}\makebox(0,0)[lb]{\smash{$_{0}$}}}%
    \put(0.0583139,0.17308543){\color[rgb]{0,0,0}\makebox(0,0)[lb]{\smash{$_{0.5}$}}}%
    \put(0.0583139,0.24448231){\color[rgb]{0,0,0}\makebox(0,0)[lb]{\smash{$_{1}$}}}%
    \put(0.0583139,0.31587919){\color[rgb]{0,0,0}\makebox(0,0)[lb]{\smash{$_{1.5}$}}}%
    \put(0.0583139,0.38727607){\color[rgb]{0,0,0}\makebox(0,0)[lb]{\smash{$_{2}$}}}%
    \put(0.0583139,0.45867295){\color[rgb]{0,0,0}\makebox(0,0)[lb]{\smash{$_{2.5}$}}}%
    \put(0.21746865,0.08463624){\color[rgb]{0,0,0}\makebox(0,0)[lb]{\smash{$_{0.5}$}}}%
    \put(0.35477037,0.08463624){\color[rgb]{0,0,0}\makebox(0,0)[lb]{\smash{$_{1}$}}}%
    \put(0.47559586,0.08463624){\color[rgb]{0,0,0}\makebox(0,0)[lb]{\smash{$_{1.5}$}}}%
    \put(0.61289756,0.08463624){\color[rgb]{0,0,0}\makebox(0,0)[lb]{\smash{$_{2}$}}}%
    \put(0.73372305,0.08463624){\color[rgb]{0,0,0}\makebox(0,0)[lb]{\smash{$_{2.5}$}}}%
  \end{picture}%
\endgroup%

%% file: stroncouplingsol3.pdf_tex
\begingroup%
  \makeatletter%
  \providecommand\color[2][]{%
    \errmessage{(Inkscape) Color is used for the text in Inkscape, but the package 'color.sty' is not loaded}%
    \renewcommand\color[2][]{}%
  }%
  \providecommand\transparent[1]{%
    \errmessage{(Inkscape) Transparency is used (non-zero) for the text in Inkscape, but the package 'transparent.sty' is not loaded}%
    \renewcommand\transparent[1]{}%
  }%
  \providecommand\rotatebox[2]{#2}%
  \ifx\svgwidth\undefined%
    \setlength{\unitlength}{689.93710938bp}%
    \ifx\svgscale\undefined%
      \relax%
    \else%
      \setlength{\unitlength}{\unitlength * \real{\svgscale}}%
    \fi%
  \else%
    \setlength{\unitlength}{\svgwidth}%
  \fi%
  \global\let\svgwidth\undefined%
  \global\let\svgscale\undefined%
  \makeatother%
  \begin{picture}(1,0.46467062)%
    \put(0,0){\includegraphics[width=\unitlength]{stroncouplingsol3.pdf}}%
    \put(-0.00077869,0.35848229){\color[rgb]{0,0,0}\makebox(0,0)[lb]{\smash{$\text{fold}$}}}%
    \put(0.2100527,0.00219264){\color[rgb]{0,0,0}\makebox(0,0)[lb]{\smash{$({\bf a})$}}}%
    \put(0.76662524,0.00219264){\color[rgb]{0,0,0}\makebox(0,0)[lb]{\smash{$({\bf b})$}}}%
    \put(0.69423498,0.22347415){\color[rgb]{0,0,0}\makebox(0,0)[lb]{\smash{$_1$}}}%
    \put(0.93157672,0.10611158){\color[rgb]{0,0,0}\makebox(0,0)[lb]{\smash{$_2$}}}%
    \put(0.65334874,0.42949928){\color[rgb]{0,0,0}\makebox(0,0)[lb]{\smash{$_3$}}}%
    \put(0.86544567,0.30267566){\color[rgb]{0,0,0}\makebox(0,0)[lb]{\smash{$_4$}}}%
    \put(0.9240323,0.42794378){\color[rgb]{0,0,0}\makebox(0,0)[lb]{\smash{$_5$}}}%
    \put(0.61546258,0.1690102){\color[rgb]{0,0,0}\makebox(0,0)[lb]{\smash{$_6$}}}%
    \put(0.9981811,0.24530519){\color[rgb]{0,0,0}\rotatebox{90}{\makebox(0,0)[lb]{\smash{$\Red{s+ t}$}}}}%
  \end{picture}%
\endgroup%

%% file: CFstrongcoupling2.pdf_tex
\begingroup%
  \makeatletter%
  \providecommand\color[2][]{%
    \errmessage{(Inkscape) Color is used for the text in Inkscape, but the package 'color.sty' is not loaded}%
    \renewcommand\color[2][]{}%
  }%
  \providecommand\transparent[1]{%
    \errmessage{(Inkscape) Transparency is used (non-zero) for the text in Inkscape, but the package 'transparent.sty' is not loaded}%
    \renewcommand\transparent[1]{}%
  }%
  \providecommand\rotatebox[2]{#2}%
  \ifx\svgwidth\undefined%
    \setlength{\unitlength}{439.95615234bp}%
    \ifx\svgscale\undefined%
      \relax%
    \else%
      \setlength{\unitlength}{\unitlength * \real{\svgscale}}%
    \fi%
  \else%
    \setlength{\unitlength}{\svgwidth}%
  \fi%
  \global\let\svgwidth\undefined%
  \global\let\svgscale\undefined%
  \makeatother%
  \begin{picture}(1,0.44652654)%
    \put(0,0){\includegraphics[width=\unitlength]{CFstrongcoupling2.pdf}}%
    \put(0.88465957,0.10507056){\color[rgb]{0,0,0}\makebox(0,0)[lb]{\smash{${-{\nu^2\over4g^2}}$}}}%
    \put(0.42681299,0.43761976){\color[rgb]{0,0,0}\makebox(0,0)[lb]{\smash{${\omega\over2g}$}}}%
    \put(0.6609835,0.1240558){\color[rgb]{0,0,0}\makebox(0,0)[lb]{\smash{$_{10}$}}}%
    \put(0.86100343,0.1240558){\color[rgb]{0,0,0}\makebox(0,0)[lb]{\smash{$_{20}$}}}%
    \put(0.23912327,0.1240558){\color[rgb]{0,0,0}\makebox(0,0)[lb]{\smash{$_{-10}$}}}%
    \put(0.03910334,0.1240558){\color[rgb]{0,0,0}\makebox(0,0)[lb]{\smash{$_{-20}$}}}%
    \put(0.44906152,0.24378802){\color[rgb]{0,0,0}\makebox(0,0)[lb]{\smash{$_{1}$}}}%
    \put(0.44906152,0.34925306){\color[rgb]{0,0,0}\makebox(0,0)[lb]{\smash{$_{2}$}}}%
    \put(0.43451462,0.03285791){\color[rgb]{0,0,0}\makebox(0,0)[lb]{\smash{$_{-1}$}}}%
  \end{picture}%
\endgroup%

%% file: CFplot.pdf_tex
\begingroup%
  \makeatletter%
  \providecommand\color[2][]{%
    \errmessage{(Inkscape) Color is used for the text in Inkscape, but the package 'color.sty' is not loaded}%
    \renewcommand\color[2][]{}%
  }%
  \providecommand\transparent[1]{%
    \errmessage{(Inkscape) Transparency is used (non-zero) for the text in Inkscape, but the package 'transparent.sty' is not loaded}%
    \renewcommand\transparent[1]{}%
  }%
  \providecommand\rotatebox[2]{#2}%
  \ifx\svgwidth\undefined%
    \setlength{\unitlength}{340.9737793bp}%
    \ifx\svgscale\undefined%
      \relax%
    \else%
      \setlength{\unitlength}{\unitlength * \real{\svgscale}}%
    \fi%
  \else%
    \setlength{\unitlength}{\svgwidth}%
  \fi%
  \global\let\svgwidth\undefined%
  \global\let\svgscale\undefined%
  \makeatother%
  \begin{picture}(1,0.60281194)%
    \put(0,0){\includegraphics[width=\unitlength]{CFplot.pdf}}%
    \put(0.605771,0.10690842){\color[rgb]{0,0,0}\makebox(0,0)[lb]{\smash{$_{10}$}}}%
    \put(0.94232705,0.08425355){\color[rgb]{0,0,0}\makebox(0,0)[lb]{\smash{$\Delta^2$}}}%
    \put(0.35416503,0.51202483){\color[rgb]{0,0,0}\makebox(0,0)[lb]{\smash{${\omega(\nu)}$}}}%
    \put(0.78877627,0.10690842){\color[rgb]{0,0,0}\makebox(0,0)[lb]{\smash{$_{20}$}}}%
    \put(0.22099068,0.10690842){\color[rgb]{0,0,0}\makebox(0,0)[lb]{\smash{$_{-10}$}}}%
    \put(0.38156744,0.04035471){\color[rgb]{0,0,0}\makebox(0,0)[lb]{\smash{$_{-0.5}$}}}%
    \put(0.40033722,0.19989777){\color[rgb]{0,0,0}\makebox(0,0)[lb]{\smash{$_{0.5}$}}}%
    \put(0.40033722,0.27497686){\color[rgb]{0,0,0}\makebox(0,0)[lb]{\smash{$_{1}$}}}%
    \put(0.40033722,0.35474837){\color[rgb]{0,0,0}\makebox(0,0)[lb]{\smash{$_{1.5}$}}}%
    \put(0.40033722,0.4345199){\color[rgb]{0,0,0}\makebox(0,0)[lb]{\smash{$_{2}$}}}%
  \end{picture}%
\endgroup%

%% file: contours2.pdf_tex
\begingroup%
  \makeatletter%
  \providecommand\color[2][]{%
    \errmessage{(Inkscape) Color is used for the text in Inkscape, but the package 'color.sty' is not loaded}%
    \renewcommand\color[2][]{}%
  }%
  \providecommand\transparent[1]{%
    \errmessage{(Inkscape) Transparency is used (non-zero) for the text in Inkscape, but the package 'transparent.sty' is not loaded}%
    \renewcommand\transparent[1]{}%
  }%
  \providecommand\rotatebox[2]{#2}%
  \ifx\svgwidth\undefined%
    \setlength{\unitlength}{2225.81171875bp}%
    \ifx\svgscale\undefined%
      \relax%
    \else%
      \setlength{\unitlength}{\unitlength * \real{\svgscale}}%
    \fi%
  \else%
    \setlength{\unitlength}{\svgwidth}%
  \fi%
  \global\let\svgwidth\undefined%
  \global\let\svgscale\undefined%
  \makeatother%
  \begin{picture}(1,0.12597905)%
    \put(0,0){\includegraphics[width=\unitlength]{contours2.pdf}}%
    \put(0.0679837,0.00160475){\color[rgb]{0,0,0}\makebox(0,0)[lb]{\smash{$\text{OPE}$}}}%
    \put(0.32245267,0.00160475){\color[rgb]{0,0,0}\makebox(0,0)[lb]{\smash{$\text{BFKL}$}}}%
    \put(0.5309159,0.00160475){\color[rgb]{0,0,0}\makebox(0,0)[lb]{\smash{$\text{energy suppressed}$}}}%
    \put(0.79976166,0.00160475){\color[rgb]{0,0,0}\makebox(0,0)[lb]{\smash{$\text{twist suppressed}$}}}%
  \end{picture}%
\endgroup%

%% file: zerochargepath3.pdf_tex
\begingroup%
  \makeatletter%
  \providecommand\color[2][]{%
    \errmessage{(Inkscape) Color is used for the text in Inkscape, but the package 'color.sty' is not loaded}%
    \renewcommand\color[2][]{}%
  }%
  \providecommand\transparent[1]{%
    \errmessage{(Inkscape) Transparency is used (non-zero) for the text in Inkscape, but the package 'transparent.sty' is not loaded}%
    \renewcommand\transparent[1]{}%
  }%
  \providecommand\rotatebox[2]{#2}%
  \ifx\svgwidth\undefined%
    \setlength{\unitlength}{629.43198242bp}%
    \ifx\svgscale\undefined%
      \relax%
    \else%
      \setlength{\unitlength}{\unitlength * \real{\svgscale}}%
    \fi%
  \else%
    \setlength{\unitlength}{\svgwidth}%
  \fi%
  \global\let\svgwidth\undefined%
  \global\let\svgscale\undefined%
  \makeatother%
  \begin{picture}(1,0.7021344)%
    \put(0,0){\includegraphics[width=\unitlength]{zerochargepath3.pdf}}%
    \put(0.71945408,0.38980265){\color[rgb]{0,0,0}\makebox(0,0)[lb]{\smash{$2g$}}}%
    \put(0.404393,0.41872664){\color[rgb]{0,0,0}\makebox(0,0)[lb]{\smash{$-2g$}}}%
    \put(0.91380683,0.41351838){\color[rgb]{0,0,0}\makebox(0,0)[lb]{\smash{${\it u}$}}}%
    \put(-0.00130396,0.1593209){\color[rgb]{0,0,0}\makebox(0,0)[lb]{\smash{${\bf a})$}}}%
    \put(-0.00130396,0.48977757){\color[rgb]{0,0,0}\makebox(0,0)[lb]{\smash{${\bf b})$}}}%
    \put(0.91454851,0.14907887){\color[rgb]{0,0,0}\rotatebox{-1.63563391}{\makebox(0,0)[lb]{\smash{$\nu$}}}}%
    \put(0.41555403,0.11587761){\color[rgb]{0,0,0}\rotatebox{-1.63563391}{\makebox(0,0)[lb]{\smash{$-{\pi\over2}\Gamma_\text{cusp}$}}}}%
    \put(0.68940534,0.14767888){\color[rgb]{0,0,0}\rotatebox{-1.63563391}{\makebox(0,0)[lb]{\smash{${\pi\over2}\Gamma_\text{cusp}$}}}}%
  \end{picture}%
\endgroup%